%% file: main.tex
\documentclass[%
	reprint,
	superscriptaddress,
	amsmath,amssymb,amsfonts,
	aps,
	pra,
]{revtex4-2}

\usepackage[utf8]{inputenc}
\usepackage[T1]{fontenc}
\usepackage{lmodern}
\usepackage[english]{babel}
\usepackage[autostyle=true]{csquotes}

\usepackage[dvipsnames]{xcolor}
\usepackage{graphicx}
\graphicspath{{figures/}}

\usepackage{tikz}
\usepackage{tikz-cd}
\usepackage[normalem]{ulem}
\usepackage{mhchem}
\usetikzlibrary{%
  matrix,%
  calc,%
  arrows%
}

\usepackage{amssymb,amsmath}
\usepackage{comment}
\DeclareMathOperator*{\argmax}{arg\,max}
\DeclareMathOperator*{\argmin}{arg\,min}

\usepackage{bm}
\usepackage{mathtools}
\usepackage{braket}
\DeclarePairedDelimiter\abs{\lvert}{\rvert}%

\usepackage{xspace}
\usepackage{extdash}
\usepackage{mleftright}
\mleftright

\usepackage{hyperref}
\hypersetup{
}

\usepackage[%
    separate-uncertainty=false,
    range-phrase=-,
]{siunitx}

\DeclareSIUnit\bohr{\text {\ensuremath {a}}_{0}}

\usepackage[
	capitalise,
]{cleveref}

\usepackage{acro}
\acsetup{%
	first-style=long-short,
}

\DeclareAcronym{magic}{%
	short = MAGIC,
	long = magnetic gradient induced coupling,
}

\definecolor{mypurple}{rgb}{0.49,0.18,0.56}
\definecolor{mygold}{rgb}{0.93,0.49,0.13}
\definecolor{mygreen}{rgb}{0,0.5,0}
\definecolor{myblue}{rgb}{0,0,0.75}
\definecolor{mymagenta}{cmyk}{0,1,0,0.12}
\definecolor{mygray}{rgb}{0.5,0.5,0.5}

\newif\ifcomments
\commentstrue 

\begin{document}
\title{Practical Noise Mitigation for Quantum Annealing via Dynamical Decoupling:\\
Toward Industry-Relevant Optimization using Trapped Ions}

\author{Sebastian Nagies}
\email{sebastian.nagies@unitn.it}
\thanks{Corresponding author}
\affiliation{Pitaevskii BEC Center and Department of Physics, University of Trento, Via Sommarive 14,
38123 Trento, Italy}
\affiliation{INFN-TIFPA, Trento Institute for Fundamental Physics and Applications, Trento, Italy}

\author{Chiara Capecci}
\email{chiara.capecci@unitn.it}
\affiliation{Pitaevskii BEC Center and Department of Physics, University of Trento, Via Sommarive 14,
38123 Trento, Italy}
\affiliation{INFN-TIFPA, Trento Institute for Fundamental Physics and Applications, Trento, Italy}

\author{Marcel Seelbach Benkner}
\affiliation{Department of Electrical Engineering and Computer Science, 
University of Siegen, Hölderlinstraße 3, 57076 Siegen, Germany
}
\affiliation{eleQtron GmbH }

\author{Javed Akram}
\email{research@eleqtron.com}
\affiliation{eleQtron GmbH }

\author{Sebastian Rubbert}
\affiliation{eleQtron GmbH }

\author{Dimitrios Bantounas}
\affiliation{eleQtron GmbH }

\author{Michael Moeller}
\affiliation{Department of Electrical Engineering and Computer Science, 
University of Siegen, Hölderlinstraße 3, 57076 Siegen, Germany
}

\author{Michael Johanning}
\affiliation{eleQtron GmbH }

\author{Philipp Hauke}
\email{philipp.hauke@unitn.it}
\affiliation{Pitaevskii BEC Center and Department of Physics, University of Trento, Via Sommarive 14,
38123 Trento, Italy}
\affiliation{INFN-TIFPA, Trento Institute for Fundamental Physics and Applications, Trento, Italy}

\begin{abstract}

Quantum annealing is a framework for solving combinatorial optimization problems. While it offers a promising path towards a practical application of quantum hardware, its performance in real-world devices is severely limited by environmental noise that can degrade solution quality.
We investigate the suppression of local field noise in quantum annealing protocols through the periodic application of dynamical decoupling pulses implementing global spin flips. As test problems, we construct minimal Multiple Object Tracking QUBO instances requiring only five and nine qubits, as well as cutting stock instances of five and six qubits. Moreover, using the Sherrington--Kirkpatrick model, we demonstrate the robustness of our protocol to problem structure and size. To further place our results in a practical context, we consider a trapped-ion platform based on magnetic gradient-induced coupling as a reference architecture, using it to define experimentally realistic noise and coupling parameters.
We show that external magnetic field fluctuations, typical in such setups, significantly degrade annealing fidelity, while moderate dynamical decoupling pulse rates, which are achievable in current experiments, restore performance to near-ideal levels. Our analytical and numerical results reveal a universal scaling behavior, with fidelity determined by a generalized parameter combining noise amplitude and dynamical decoupling pulse interval. While our analysis is grounded in the trapped-ion platform, the proposed noise mitigation strategy and resulting performance improvements are applicable to a broad range of quantum annealing implementations and establish a practical and scalable route for error mitigation in near-term devices.

\end{abstract}

\maketitle

\section{Introduction}\label{sec:intro}
\input{Intro}

\section{Quantum annealing}\label{sec:annealing}

\input{annealing}

\section{Test QUBO problem: Multiple object tracking}\label{sec:MOT}

\input{MOT}

\section{Reference platform}
\label{sec:platform}
\input{platform}

\section{Error analysis on cost function}
\label{sec:disorder}
\input{disorder}

\section{Dynamical decoupling}\label{sec:dd}

\input{dd}

\section{Conclusions}\label{sec:conclusion}
\input{Conclusion}

\input{Acknowledgments}

\appendix
\input{appendix}

\bibliographystyle{apsrev4-2}
\bibliography{trappedions}
\end{document}

%% file: Intro.tex
Quantum computing has emerged as a transformative technology for addressing classically intractable computational problems \cite{Ladd2010,Fedorov2022,Beck2023, Hoefler2023, Scholten2024, Troyer2024}, with quantum algorithms broadly categorized into gate-based and analog approaches. Within this landscape, quantum annealing occupies a distinctive position as an analog quantum algorithm especially suitable for combinatorial optimization problems \cite{Santoro2002, Martonak2004, Hauke2020, Rajak2022, Yarkoni2022}. Unlike gate-based algorithms that rely on discrete quantum operations, quantum annealing leverages a continuous time evolution to steer a system towards a quantum state encoding the solution to an optimization problem. 

Hardware considerations play a central role in shaping quantum annealing implementations, as different platform architectures provide unique strengths and functionalities. Superconducting qubits \cite{Arute2019,Kjaergaard2020, Bravyi2024}, central to early annealing demonstrations \cite{Boixo2014}, enable nanosecond-scale gate operations \cite{Werninghaus2021} with large numbers of physical qubits. Neutral atom systems \cite{Brodoloni2025, GonzalezCuadra2025}, leveraging optical tweezers and Rydberg interactions, provide highly scalable architectures with flexible, programmable coupling. Trapped-ion platforms \cite{Wineland2003, Leibfried2003a, HAFFNER2008, Bruzewicz2019, Srinivas2021, Chen2024, Liu2025, Meth2025} offer all-to-all qubit connectivity and extended coherence times. These characteristics could potentially make the latter well-suited for the implementation of quantum annealing protocols which typically require long coherence windows and complex coupling topologies.

Within the trapped-ion paradigm, systems employing magnetic-gradient-induced coupling (MAGIC) represent a promising architecture for quantum information processing \cite{Mintert2001, Wunderlich2002, Johanning2009, Zippilli2014, Piltz2016, Bassler2023, Weidt2016, Arrazola2018, Huber2021, Leu2023, Nagies2025}. In MAGIC-based trapped-ion processors, a static magnetic field gradient mediates spin--spin interactions through collective motional modes of the ion crystal, enabling native two-body interactions that are both tunable and naturally suited to optimization problems. This approach offers additional practical advantages, including the use of microwave fields for qubit control---eliminating the need for complex laser systems---and the ability to implement problem-specific Hamiltonians directly in hardware.

However, the practical realization of quantum annealing on any physical platform must contend with the inevitable presence of noise and systematic imperfections that can compromise algorithmic performance. 
Typically, among the most problematic sources of errors are local field fluctuations, which in trapped-ion platforms arise from temporal and spatial fluctuations of external magnetic fields. These fluctuations can directly distort the cost Hamiltonian that encodes the optimization problem, potentially leading to suboptimal solutions.

To mitigate these detrimental effects while preserving the integrity of the encoded optimization problem, we investigate the application of dynamical decoupling strategies \cite{Viola1998,Khodjasteh,Souza2011, Cai2023, Piltz2013,Genov2017, Valahu2021, Morong2023, Barthel2023, Nuennerich2024} specifically tailored to quantum annealing protocols \cite{Lidar2008,Quiroz2012}. We discuss how dynamical decoupling pulses implementing periodic qubit rotations can average out the impact of time-dependent local field fluctuations without disrupting the quantum annealing protocol. We demonstrate this noise suppression technique using minimal five-qubit and nine-qubit Multiple Object Tracking QUBO instances \cite{Zaech2022}, cutting stock instances with five and six qubits, and instances of the Sherrington--Kirkpatrick model \cite{Sherrington1975, Panchenko2013} up to 12 qubits. These problem sizes were chosen to be small enough for detailed numerical analysis and thus suitable for first proof-of-concept demonstrations. At the same time, their modest qubit counts ensure that they remain compatible with the limited qubit numbers available on current experimental platforms.

Through our numerical simulations employing a realistic noise spectrum and experimentally achievable coupling parameters, specifically referencing MAGIC-based trapped-ion systems, we establish that moderate dynamical decoupling pulse rates, achievable on current hardware, are sufficient to restore annealing fidelity to near-ideal levels across different noise frequencies and amplitudes at least up to an order of magnitude larger than the energy scale of the annealing Hamiltonian. Moreover, our analysis reveals a universal scaling behavior governed by a generalized parameter that combines noise amplitude with the dynamical decoupling pulse interval, providing a compact figure of merit for optimizing error mitigation strategies across diverse noise environments. 
Finally, by exploiting the discretization of time into steps given by consecutive dynamical decoupling pulses, the scheme can be adapted with little overhead to also permit one to program desired interaction parameters that encode the optimization problem.  
These findings establish a practical and scalable framework for implementing quantum annealing on near-term devices, with direct applicability extending beyond trapped-ion platforms to other quantum annealing architectures.

This paper is structured as follows: Section \ref{sec:annealing} provides a brief review of quantum annealing and introduces the specific annealing algorithm considered throughout this work. Section \ref{sec:MOT} presents Multiple-Object Tracking, an industrially relevant computer vision problem that can be formulated as a Quadratic Unconstrained Binary Optimization (QUBO) problem suitable for quantum annealing. We construct a minimal problem instance that serves as our test case for numerical simulations and discuss the scaling of required qubit number with problem size. 
The initial Multiple-Object Tracking problem instance for this purpose is taken from the MOTChallenge publication from 2015 \cite{leal2015motchallenge}.
Section \ref{sec:platform} presents an overview of trapped-ion quantum computers with magnetic gradient induced coupling, which serves as our reference platform. We examine experimental parameters, including realizable coupling strengths on current hardware, relevant noise sources, and the feasibility of implementing quantum annealing on this platform. Section \ref{sec:disorder} provides a numerical investigation of how different noise sources affect quantum annealing protocol performance, emphasizing the critical need for local field noise mitigation.
Section \ref{sec:dd} addresses active noise mitigation through dynamical decoupling pulses, providing both analytical justification for the underlying mechanism and numerical simulations of annealing sweeps with varying pulse counts. Section~\ref{sec:conclusion} presents our conclusions. Technical details and complementary results for the cutting stock problem and different system sizes are delegated to several Appendices.

%% file: annealing.tex
Quantum annealing is a common technique for solving optimization problems on quantum computers. It exploits quantum fluctuations to guide a system towards the ground state of a cost Hamiltonian $H_{\text{cost}}$, which encodes the solution to a combinatorial optimization problem \cite{Santoro2002, Martonak2004, Hauke2020, Rajak2022, Yarkoni2022}. In the standard framework of quantum annealing, the system is initialized in the ground state of a driving Hamiltonian $H_\text{driving}$, which introduces quantum fluctuations, whose strength is gradually reduced while simultaneously the strength of the cost Hamiltonian is ramped up. At the end of the annealing sweep, when the driving Hamiltonian has been switched off, the system ideally reaches the problem-encoding ground state. This procedure can be performed either adiabatically, where the system remains close to the instantaneous ground state over the entire evolution time, quantified by the adiabatic theorem \cite{Jansen2007,Lidar2009,Amin2009,Cheung2011}, or diabatically \cite{Cepaite2023,GarciaPintos2024, Bottarelli2024, Choi2021, Feinstein2025}, where higher excited states can be strategically populated during the evolution, to potentially decrease the total required annealing time.

\textit{Modified annealing protocol ---} 
In this work, we employ a slightly modified annealing protocol, where the strength of the cost Hamiltonian is kept constant. This form is potentially better suited to the trapped-ion devices based on MAGIC, which will be introduced in Sec.~\ref{sec:platform} as our reference hardware platform. We define the total Hamiltonian during the annealing sweep as
\begin{align}
    H(t) &=  H_{\text{cost}} + C(t) H_\text{driving}\,,\label{eq:modified_annealing}\\
    H_{\text{cost}} &=\sum_{ij} J_{ij} \sigma_i^z \sigma_j^z + \sum_i h_i^z \sigma_i^z \,,\label{eq:H0}\\
    H_{\text{driving}} &= - h^x \sum_i \sigma_i^x\,, \label{eq:Hdrive}
\end{align}
where $H_{\text{cost}}$ contains two-body interactions and local fields (corresponding to a QUBO problem), and $H_{\text{driving}}$ is the standard choice for a driver with homogeneous transverse local fields. The coefficient is set to $C(0)=1$ at the beginning of the protocol and then linearly ramped down to zero. 

The system is initialized in the ground state of the driving Hamiltonian, which is the equal superposition of all computational basis states. However, in contrast to the standard quantum annealing protocol, where $H_{\mathrm{standard}}(t=0) = H_{\text{driving}}$, the equal superposition is not the true ground state of the total Hamiltonian in the modified protocol, as $H(0) = H_{\text{cost}} + H_{\text{driving}}$. To ensure that the system still has a high overlap with the true ground state at the beginning of the protocol, the energy scale of the driver has to be chosen sufficiently larger than the energy scale of the cost Hamiltonian to ensure an approximately adiabatic evolution. In contrast, for the conventional protocol, it is usually optimal to choose the energy scales of driving and cost Hamiltonian to be roughly equal \cite{Nagies2025a}. 

\textit{Encoding scheme ---} 
As we will discuss in Sec.~\ref{sec:platform}, a significant source of errors in the trapped-ion platform is fluctuating local field noise. This noise can interfere with the specific local field strengths necessary to faithfully encode an optimization problem in $H_{\text{cost}}$. To be able to treat noise and cost Hamiltonian on separate footings, it  can be useful to encode the problem purely into two-body interactions. To that end, one can introduce an additional ancilla qubit and write the cost Hamiltonian as
\begin{align}\label{eq:ancilla_encoding}
    H_{\text{cost}} =\sum_{ij} J_{ij} \sigma_i^z \sigma_j^z + \sigma_a^z \sum_i J_{ai} \sigma_i^z, 
\end{align}
where the local biases of the original cost Hamiltonian (Eq.~\ref{eq:H0}) are identified as $h_i^z \to \sigma_a^z J_{ai}$, with index $a$ referring to the ancilla qubit. The driving Hamiltonian now also addresses the ancilla qubit, and the system is initialized in the equal superposition of all qubits, including the ancilla.

As a direct effect of this reformulation of the problem, the ground state of the cost Hamiltonian becomes two-fold degenerate. Measuring either one of the two ground states at the end of the annealing sweep gives the solution to the problem, as the measurement result of the ancilla qubit implies whether the rest of the qubits have to be flipped or not. In the main text of this work, we always assume that the problem is encoded solely in the two-body interactions with this scheme, and all fidelities we report correspond to the sum of the two overlaps with the two respective ground states. 

An important consequence of this encoding scheme is that the noise in the local field fluctuations gets isolated from the problem itself and can be effectively compensated by applying dynamical decoupling sequences during the annealing sweep, which we will discuss in detail in Sec.~\ref{sec:dd}. Note that it is not strictly necessary to introduce an ancilla qubit as above to perform quantum annealing with dynamical decoupling to mitigate local field noise. In Appendix \ref{app:alt_protocol}, we discuss a similar annealing protocol that does not require the ancilla qubit and has comparable performance.

%% file: MOT.tex
Multiple Object Tracking (MOT) is a fundamental computer vision problem that involves identifying and following multiple objects of interest across consecutive video frames \cite{Luo2021}. The task requires solving two key challenges: first, detecting objects in each frame, and second, associating these detections across frames to form coherent trajectories or tracks. A detection refers to a localized object instance in a single frame (typically represented by a bounding box) \cite{zhao2019object}, while a track represents the temporal sequence of detections corresponding to the same object as it moves through the scene.
The association step (determining which detections in consecutive frames belong to the same object) naturally leads to complex combinatorial optimization problems. These problems become particularly challenging in crowded scenes with occlusions, appearance changes, and false positive detections, making MOT a computationally demanding task that often requires solving NP-hard assignment problems \cite{ganian2021complexity}.

We investigate quantum annealing as a solver for MOT problems, focusing on the challenge of associating given detections to coherent tracks. A QUBO formulation offers a suitable framework to compare detections across frames. To construct the QUBO instance, we follow the approach discussed in Ref.~\cite{Zaech2022}, which scales linearly with the number of detections, tracks, and frames and could, in principle, scale to industry-relevant problem sizes. As this serves as an initial testbed for quantum annealing experiments, we introduce simplifying assumptions to reduce the problem size. A related work, which focuses on fast, online algorithms, is presented in Ref.~\cite{ihara2025enhancing}. 

\subsection{General QUBO formulation}

The considered QUBO problem consists of two parts: a cost term that evaluates the similarity between detections from different frames, and a penalty term that enforces constraints on the solution vector. The output of the annealing protocol is a binary vector indicating the track assignment for each detection. For a given frame $f$, a fixed number of tracks $T$, and $D_f$ detections in the frame, we denote a binary vector as

\begin{equation} \label{eq:sol_vector}
x^{(f)} \in \{0,1\}^{D_fT},
\end{equation}

which encodes detections as

\begin{equation}
    x^{(f)}_{k\cdot T+ l}=1     \quad \longleftrightarrow  \quad   \textnormal{  
    \\Detection $k$ in frame $f$ is in track $l$.
    }
\end{equation}

To ensure a valid assignment, we impose two constraints, both expressible as a linear system ($Ax = b$): each detection $k$ must be assigned to exactly one track $l$, and each track must receive exactly one detection per frame: 

\begin{align}
    \sum_{l=0}^{T-1}  x^{(f)}_{k\cdot T+ l}= 1  \quad \forall k \in \{0,...,D-1 \},
    \label{eq:constraint1}
    \\
    \sum_{k=0}^{D-1}  x^{(f)}_{k\cdot T+ l}= 1  \quad \forall l \in \{0,...,T-1 \}.
\label{eq:constraint2}
\end{align}

\begin{figure}[!tbp]
  \centering
  \includegraphics[width=0.49\columnwidth]{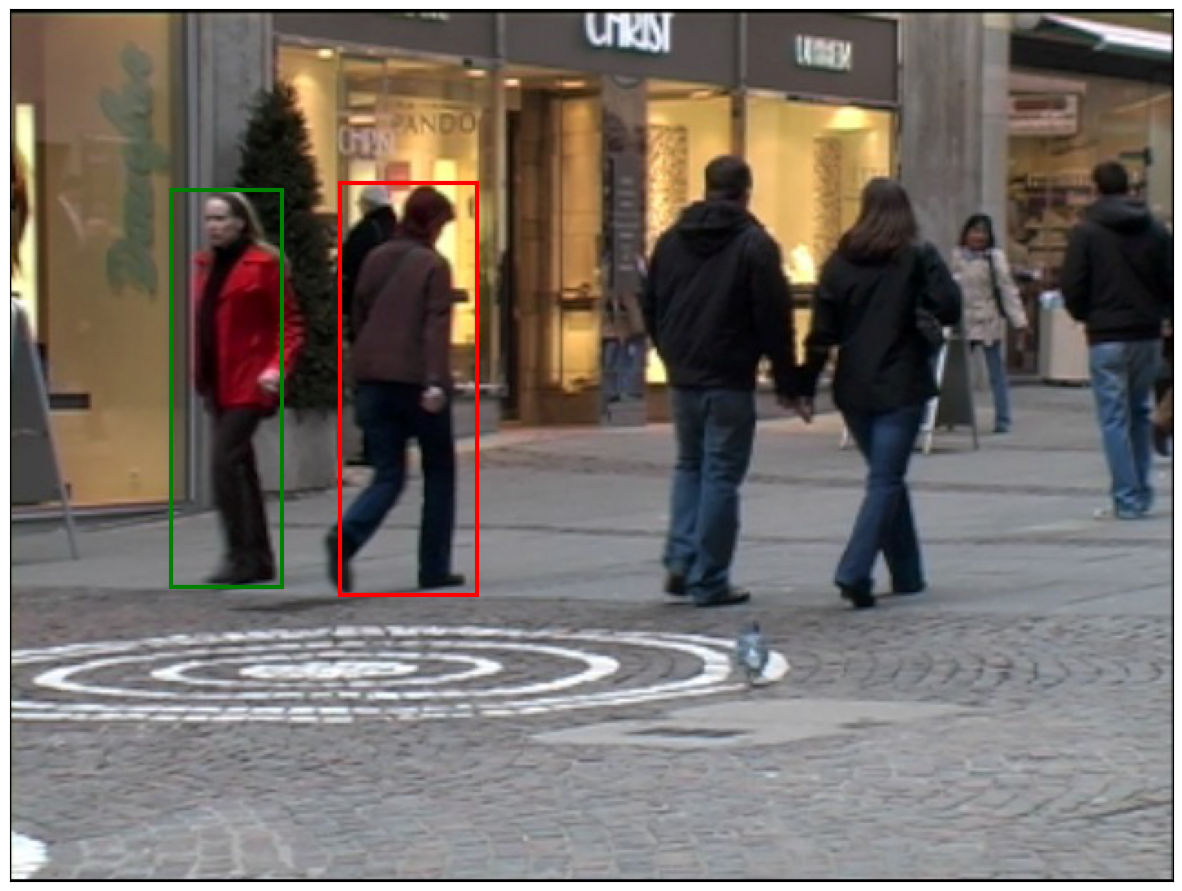}  
 \includegraphics[width=0.49\columnwidth]{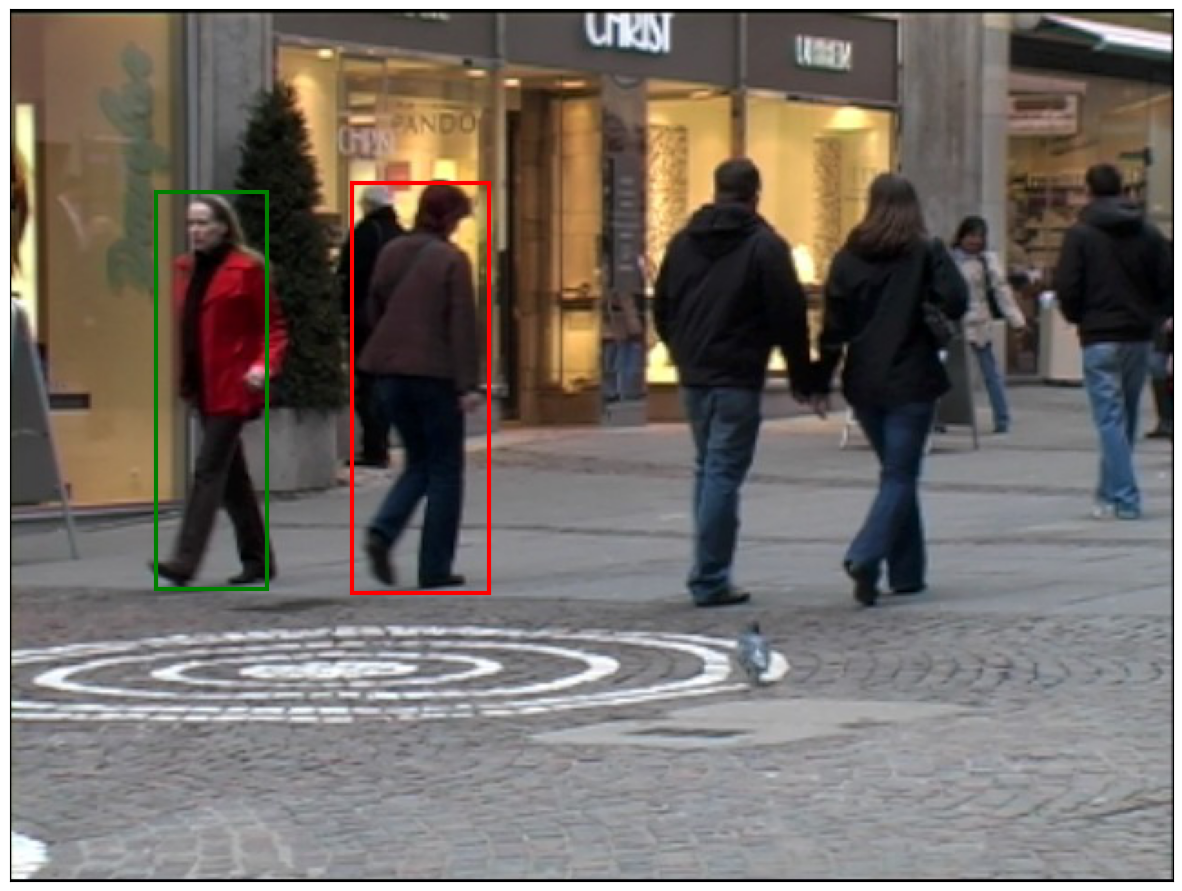}
  \caption{Two frames of a video from \cite{andriluka2010monocular,leal2015motchallenge} with pedestrians walking in front of the camera. The color of the bounding boxes identifies the persons. Two frames with two detections each can be formulated as a QUBO problem with four binary variables.}
  \label{fig:MOTCityCenter}
\end{figure}

Figure~\ref{fig:MOTCityCenter} illustrates this encoding for pedestrian tracking. This formulation does not yet account for hidden objects or false detections. These can be handled by introducing dummy tracks and detections; additional variables that are excluded from the constraints by increasing $T$ and $D$ accordingly. 

Since the objective term addresses how likely detections from different frames correspond to the same object, a geometric cost can be introduced by assuming limited pedestrian speed, making nearby bounding boxes in consecutive frames more likely to match. This is quantified using the ratio between the area of intersection and the area of union for two bounding boxes $B_1$ and $B_2$:
\begin{equation}
G(B_1,B_2):= \frac{\abs{B_1 \cap B_2}}{\abs{B_1 \cup B_2}}. 
\label{eq:geometric}
\end{equation}

Furthermore, a measure of visual similarity $C$ can be obtained by taking the cosine similarity between features $f_{k_1},f_{k_2} \in \mathbb{R}^{1024} $ that are extracted from a pre-trained neural network \cite{zheng2019joint}:
\begin{equation}
    C(f_{k_1},f_{k_2}):=  \frac{f_{k_1}^T f_{k_2}}{||f_{k_1}||\cdot||f_{k_2}||}.\label{eq:visual}
\end{equation}
The indices $k_1,$ and $ k_2 $ refer to different detections. 
The associated cost is considered only if both detections are included in the same track. This similarity cost is also used in \cite{hornakova2021making}.

To combine geometric and visual similarity, we take a weighted sum. In our implementation, both components contribute equally. Calling $m$ and $l$ the indices corresponding to sorting detections
$k_1$ and  $k_2$ to track $t$, respectively, the considered cost function is
\begin{equation}\label{eq:mot_objective}
    W_{m,l}=  \frac{1}{2} C(f_{k_1},f_{k_2}) + \frac{1}{2} G(k_1, k_2).
\end{equation}

To combine the objective and the constraints (Eqs. \eqref{eq:constraint1} and \eqref{eq:constraint2}) into a single QUBO problem, we apply the penalty method. For a sufficiently large penalty parameter $\lambda \in \mathbb{R}_{+}$, the constrained optimization problem is equivalent to the unconstrained QUBO problem 

\begin{align} \label{eq:MOT_qubo}
 &    \argmin_{\{ x\in \{0,1\}^n|  Ax = b \}} x^T W x    \nonumber \\
&= \argmin_{\{ x\in \{0,1\}^n \}} x^T W x    + \lambda || Ax-b ||^2.
\end{align}

While $\lambda$ can be chosen arbitrarily large in theory, in practice, smaller values that still enforce the constraints are preferred, as large penalties may effectively amplify hardware noise and reduce precision. Furthermore, the problem exhibits a symmetry: in pedestrian tracking videos, the identities (or colors) of the tracked objects can be permuted; the track numbering is arbitrary. To break this symmetry, 
we fix the assignment in the first frame. This is done by setting the corresponding binary entries in the solution vector, which has the added benefit of reducing the QUBO size.

\subsection{Minimal example with four qubits}\label{subsec:min_qubo_problem}

We construct the smallest non-trivial example by considering the MOT problem with only two frames, two tracks ($T=2$),  and two detections ($D=2$). 
The binary solution vector would be $8$-dimensional ($\Vec{x} = (\Vec{x}^{(1)}, \Vec{x}^{(2)})^T$, c.f. Eq. \eqref{eq:sol_vector}). By exploiting the symmetry under permutation of tracking labels, we reduce the problem to four dimensions by fixing the first four entries of the solution vector. This breaks the degeneracy and ensures a unique solution. The fixed assignment --- consistent with the constraints in Eqs. \eqref{eq:constraint1} and \eqref{eq:constraint2} --- is chosen arbitrarily as $x^{(1)}_1 = x^{(1)}_4 = 1$ and $x^{(1)}_2 = x^{(1)}_3 = 0$.  We then solve only for the remaining four-dimensional vector $\Vec{x}^{(2)}$. Formulating the constraints as a linear system of equations $A \Vec{x}^{(2)} = b$, we get
\begin{equation}
        \begin{pmatrix}
         1 & 1 & 0 & 0\\
         0 & 0 & 1 & 1 \\
         1 & 0 & 1 & 0 \\
         0 & 1 & 0 & 1 
        \end{pmatrix} \Vec{x}^{(2)} = \begin{pmatrix}
         1 \\
         1\\
         1 \\
          1
         \end{pmatrix}.
\end{equation}

Since we fixed the assignments of detections to tracks on the first frame, the objective part $W$ of the cost function in Eq.~\eqref{eq:MOT_qubo} reduces to only local field terms which depend on the chosen scene. For this minimal problem with four qubits, all two-body interactions in the total QUBO instance are only defined by the constraints and consequently identical for all chosen scenes and frames. 
In the MOT instance corresponding to Fig.~\ref{fig:MOTCityCenter}, the objective is given by $W \approx (-2.18, -0.95, -0.58, -2.26)$. 

To write Eq.~\eqref{eq:MOT_qubo} as a cost Hamiltonian for quantum annealing, we need to translate the Boolean variables to quantum spins via the transformation $x_i \to (1 + \sigma_i^z) / 2$. The penalty factor $\lambda$ in Eq.~\eqref{eq:MOT_qubo} has a large effect on the minimal energy gap in the corresponding quantum annealing protocol and thus has to be chosen carefully \cite{Nagies2025a,benkner2020adiabatic}. For this small problem instance, we can simulate the annealing sweep numerically and choose a value which is close to optimal ($\lambda = 2.5$). For larger systems, there are different heuristics available for choosing a sensible value (e.g., see \cite{Alessandroni2023}).
Finally, for our numerical simulations in Sections \ref{sec:disorder} and \ref{sec:dd}, we always consider a quadratized cost Hamiltonian with only two-body interactions and no biases, which can be obtained by introducing an ancilla qubit as discussed in Sec.~\ref{sec:annealing}. 

For the minimal problem described above, we then obtain the following five-qubit cost Hamiltonian
\begin{equation}\label{eq:5qubit_MOT}
H_{\mathrm{cost}} / J  = 
        \begin{pmatrix}
         0 & -0.87 & -0.38 & -0.23 & -0.91\\
         0 & 0 & 1 & 1 & 0\\
         0 & 0 & 0 & 0 & 1\\
         0 & 0 & 0 & 0 & 1\\
         0 & 0 & 0 & 0 & 0
        \end{pmatrix},
\end{equation}
which we write in units of the characteristic energy scale, i.e., the maximal realizable two-body coupling strength $J:=\max_{ij} |J_{ij}| $. This energy scale depends on the considered hardware platform. We discuss details for trapped-ion hardware using magnetic gradient-induced couplings in the next section. In Appendix \ref{app:MOT8qubit}, we provide the explicit QUBO formulation of the same MOT problem given in Fig.~\ref{fig:MOTCityCenter}, but with one additional frame, which results in a QUBO cost Hamiltonian with $9$ qubits. 
In the same Appendix, we also discuss how the number of qubits scales when considering more complex MOT problems. Furthermore, in Appendix \ref{app:cuttingstock}, we discuss the cutting stock problem---another industrially relevant optimization problem---and how to construct minimal instances.

%% file: platform.tex
Trapped-ion systems with their native all-to-all connectivity \cite{Wineland2003, Leibfried2003a, HAFFNER2008, Bruzewicz2019, Srinivas2021, Chen2024, Liu2025, Meth2025}  are a promising candidate for quantum annealing applications~\cite{Hauke2015}. In this work,  we consider a specific variant that uses microwave fields to address individual ions, whose resonance frequencies are shifted relative to each other by an external magnetic field gradient \cite{Mintert2001, Wunderlich2002, Johanning2009, Zippilli2014, Piltz2016, Bassler2023, Weidt2016, Arrazola2018, Huber2021, Leu2023, Nagies2025}. In this setup, there exists an always-on, all-to-all magnetic gradient induced coupling (MAGIC) between ions, which could be utilized for encoding cost Hamiltonians of optimization problems for quantum annealing algorithms. We start by giving a short summary of the MAGIC scheme, after which we provide specific details of the considered experimental platform. In particular, we discuss realizable coupling strengths on current hardware and realistic noise spectra that are detrimental for quantum annealing.

\subsection{Magnetic Gradient Induced Coupling}

While traditional trapped-ion setups use optical lasers to address individual ions and implement two-qubit gates (e.g., M\o lmer--S\o rensen type interactions \cite{Moelmer1999, Soerensen1999}), the Magnetic Gradient Induced Coupling (MAGIC) scheme utilizes microwave fields \cite{Mintert2001,Wunderlich2002,Johanning2009}. To preserve individual addressability of ions, an external magnetic field gradient is applied, creating position-dependent Zeeman shifts for each ion in the trap. A beneficial side effect of this magnetic field gradient is the automatic generation of two-body interactions between all ions, even without external driving fields, which can be harnessed to implement entangling many-body gates \cite{Khromova2012,Bassler2023}. We provide a brief overview of how these interaction strengths arise and direct readers to the literature for further details \cite{Mintert2001,Wunderlich2002,Johanning2009,Nagies2025}. 

For the qubit states, a magnetically sensitive transition is chosen (which is typically a transition in a hyperfine ground state manifold with a resonance frequency in the microwave regime \cite{Balzer2006,Piltz2016, Weidt2016}). In the trap, when choosing the longitudinal and radial trapping frequencies appropriately, the ions equilibrate into a one-dimensional chain \cite{Schiffer1993}. When the ions are then subjected to a magnetic field gradient $\partial_z B$ along the chain axis, the resonance frequencies $\omega(z)$ obtain a Zeeman shift. 
The position-dependent resonance frequencies then result in spin-dependent forces on the ions, which couple internal spin degrees of freedom with the collective motion of the ion chain (phonons). Switching to a rotating frame and applying a polaron transformation, one can decouple internal and external degrees of freedom and show that the system (without driving fields) is described by the Hamiltonian \cite{Mintert2001, Wunderlich2002, Johanning2009}
\begin{align}
\label{eq:Hmagic}
    H = &-\frac{\hbar}{2} \sum_{n=1}^N \omega_n \sigma^z_{n} + \hbar \sum_{n=1}^N \nu_n a^\dagger_n a_n \nonumber\\
    &- \frac{\hbar}{2}\sum_{i<j}^N J_{ij} \sigma^z_{i}\sigma^z_{j} \,, 
\end{align}
where $\omega_n$ is the resonance frequency of ion $n$ in its equilibrium position and $a^\dagger_n$ and $a_n$ are bosonic creation and annihilation operators for the collective phonon modes of the ion chain with respective frequencies $\nu_n$. The only approximations for arriving at Eq.~\eqref{eq:Hmagic} are assuming a quadratic external potential, given by the combination of trapping potential and Coulomb repulsion between ions, and a vanishing curvature of the magnetic field (see Ref.~\cite{Nagies2025} for a discussion of higher-order effects). The native two-body coupling strength $J_{ij}$ in the MAGIC scheme is given by
\begin{align} \label{eq:J2}
    J_{ij} = \sum_{n=1}^N \nu_n \epsilon_{in} \epsilon_{jn} \, ,
\end{align}
with the effective Lamb--Dicke parameter $\epsilon_{nl}$, which determines how strongly ion~$n$ couples to phonon mode $l$ and which is proportional to the applied magnetic field gradient. Denoting the width of the ground-state wave packet of the harmonic collective mode $l$ as $\Delta z_l = \sqrt{\hbar / 2m\nu_l}$ ($m$ is the mass of the ions), and using the orthogonal matrix $S$ that translates between individual ion coordinates and the collective normal modes, the Lamb--Dicke parameter is given by
\begin{align}\label{eq:lambdicke}
    \epsilon_{nl} &= \frac{\Delta z_l \partial_z \omega_n}{\nu_l} S_{nl}.
\end{align}

The native spin--spin coupling strength $J_{ij}$ in the MAGIC setup is directly influenced by the magnetic field gradient and the strength of the external harmonic longitudinal trapping potential $\nu$, e.g., $J_{ij} \sim (\partial_zB/\nu)^2$ for a harmonic trap of strength $\frac{1}{2}m\nu^2$ \cite{Nagies2025a}. 

In contrast to the two-qubit gates, single-qubit rotations are implemented in the MAGIC setup analogously to other trapped-ion platforms. In particular, when tuning a driving field with Rabi frequency $\Omega$ to the resonance frequency of an ion, the Hamiltonian $H_{\text{drive}}$ describing the coupling to the field, in a rotating frame, can be expressed as (see, e.g., \cite{Nagies2025})

\begin{align}\label{eq:1q_gate}
    H_{\text{drive}} = \frac{\hbar\Omega}{2} \left(\cos( \phi) \sigma_{n}^{x} - \sin (\phi) \sigma_n^y \right),
\end{align}

where $\phi$ is the phase of the driving field. By leaving the driving field on for a finite time and tuning the phase appropriately, rotations around an arbitrary axis in the XY plane can be realized. Rotations around the Z-axis are typically implemented virtually \cite{McKay2017}.

\subsection{Quantum annealing with trapped ions}\label{subsec:annealing_with_ions}

As discussed in Sec.~\ref{sec:annealing}, we choose to consider a modified version (Eq.~\eqref{eq:modified_annealing}) of the standard quantum annealing protocol, where the cost Hamiltonian remains constant throughout the evolution, while the driving Hamiltonian is ramped down linearly. The MAGIC-based trapped-ion platform lends itself naturally to this protocol: as we have explained above, the presence of the magnetic gradient induces a native all-to-all two-body coupling that is always active, even without any additional driving fields. If one manages to directly encode an optimization problem into this coupling matrix, one only needs to implement the single-qubit rotations of the driving Hamiltonian, which can be easily achieved with microwave fields tuned to the individual resonance frequencies of the ions. To match the two-body couplings to a specific problem instance, there are multiple possibilities: by employing different, specifically tailored and not necessarily harmonic trapping potentials along the chain, coupling strengths between different pairs of ions can be enhanced or reduced \cite{Zippilli2014}. Furthermore, by applying suitable dynamical decoupling sequences, effective two-body interaction strengths can be continuously tuned (see Appendix \ref{app:dd}).
Finally, arbitrary subsets of the ions can be excluded from the all-to-all interaction by transferring their state populations to magnetically insensitive states \cite{Piltz2016}.

If the problem cannot be directly encoded into the MAGIC couplings, one can always perform a Trotterized version of the modified annealing protocol \cite{Sack2021}. In our numerics in Sec.~\ref{sec:dd}, we consider a continuous-time protocol, though the same dynamical decoupling strategy can also be incorporated into Trotterized annealing protocols.

\subsection{Experimental setup} \label{subsec:exp_parameters}
While our numerical simulations in the following sections employ dimensionless units throughout, we provide concrete timescale estimates based on realistic device parameters. Our analysis references experimental parameters and coupling strengths previously demonstrated in the MAGIC setup \cite{Piltz2016}.

Specifically, we consider a linear chain of five \ce{^{171}_{}Yb+} ions (matching our 5-qubit MOT problem constructed in Sec.~\ref{subsec:min_qubo_problem}), with an axial trapping frequency of $\omega_z = 2\pi \times 130$ kHz and a magnetic field gradient of $\partial_z B = \SI{19}{\tesla\per\meter}$. Under these conditions, the system achieves a maximal two-body coupling strength of $\max_{ij} |J_{ij}| \approx 26.5$~Hz between ions. We refer to this value when we set the characteristic energy scale to $J = 26$~Hz (see Eq.~\eqref{eq:5qubit_MOT}), which determines the fundamental timescale for the quantum annealing sweeps analyzed in Sec.~\ref{sec:dd}. Next-generation devices offer the potential for substantially enhanced performance through increased magnetic field gradients. For instance, with a stronger gradient of $\partial_z B = \SI{150}{\tesla\per\meter}$, the maximal coupling strength increases dramatically to $\max_{ij} |J_{ij}| \approx 1650.6$~Hz (see also Ref.~\cite{Nagies2025}). This approximately 60-fold enhancement would correspondingly reduce the required annealing timescales and improve the overall protocol efficiency.

For atomic qubits encoded in magnetically sensitive states, as employed in the MAGIC scheme, the dominant source of noise arises from fluctuations in the externally applied magnetic field. The qubit splitting depends in first order linearly on the field strength through the Zeeman effect, making it highly susceptible to such variations. These fluctuations typically originate from ambient laboratory fields, most prominently at 50 Hz and its harmonics, as well as in the tens-of-kilohertz range due to power-supply noise, instabilities in current or voltage sources, and slow thermal drifts in coils or permanent magnets. In the absence of magnetic shielding, the noise at 50 Hz typically dominates and in the worst case produces fluctuations with amplitudes up to the order of kilohertz, directly causing dephasing and reducing gate fidelity. For example, Ref.~\cite{Merkel2019} reports magnetic field rms noise of around 100 nT without stabilization. For \ce{^{171}_{}Yb+} ions, when considering $\ket{F=0, m_F = 0}$ and $\ket{F=1, m_F =  + 1}$ in the hyperfine ground state manifold as the qubit states, this corresponds to local-field noise on the order of $\delta h_i^z \approx 1.4$ kHz.
Additional noise channels further degrade performance. Microwave drive amplitude and phase noise introduce Rabi-rate fluctuations and ac Stark shifts, while motional noise from trap-voltage instabilities can modulate the effective magnetic field experienced by the atoms in the gradient. 

Similar issues arise in other quantum-computing platforms. Broadly, fast technical noise sets the limit on short-time coherence, whereas slow drifts accumulate over minutes to hours, necessitating compensation through recalibration and active feedback. 
It is the purpose of the rest of the paper to, first, study how such salient error terms impact the possible solution (Sec.~\ref{sec:disorder}) and, second, how dynamical decoupling can be applied to counteract one of the most prominent noise sources (Sec.~\ref{sec:dd}), fluctuations of the longitudinal field (equivalent to fluctuations of the qubit frequencies).

%% file: disorder.tex
This section examines two significant error sources for any quantum annealing experiment: fluctuations in the two-body couplings (Sec.~\ref{subsec:trap_freq_disorder}) and undesired local field terms (Sec.~\ref{subsec:local_field_disorder}).
To assess the impact of these errors on annealing performance, we calculate the probability that random fluctuations of each type alter the ground state of the cost Hamiltonian. 
The quantum annealing protocol fundamentally relies on encoding the problem solution into the ground state of the cost Hamiltonian: if fluctuations cause the ground state to change, the protocol can no longer identify the optimal solution.
Our analysis reveals contrasting behaviors for these two error sources when applied to our test QUBO problem given in Eq.~\eqref{eq:5qubit_MOT}. Under realistic experimental conditions the impact of fluctuations in the two-body couplings are negligible for annealing performance. However, local field fluctuations present a substantial challenge: the noise amplitudes typically encountered in trapped-ion setups are sufficient to prevent successful quantum annealing without active intervention. Consequently, effective noise mitigation strategies are essential for protocol success, which we will examine in detail in Sec.~\ref{sec:dd}.

\subsection{Fluctuations in the two-body couplings}\label{subsec:trap_freq_disorder}

\begin{figure}[htbp]
    \centering
\includegraphics{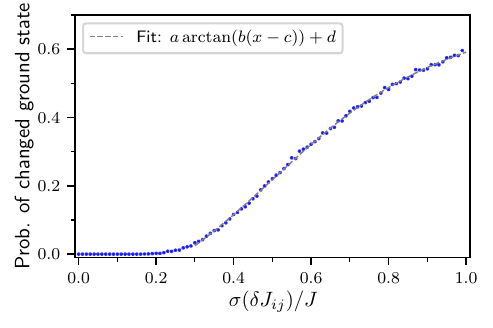}
\caption{Probability of a changed ground state of the cost Hamiltonian for the quadratized minimal MOT problem (five qubits, see Eq.~\eqref{eq:5qubit_MOT}) as a function of the standard deviation of uncorrelated fluctuations in the two-body couplings $\sigma(\delta J_{ij}) / J$. Each point corresponds to the mean of $10\,000$ samples and has a variance of $\leq 2.5\cdot 10^{-5}$. The grey dashed line corresponds to a fit with an arctan function, starting at $\sigma (\delta J_{ij})) / J = 0.3$. Fit parameters: $a, b, c, d \approx 0.41, 2.59, 0.50, 0.22$.}
    \label{fig:Jcouplings_disorder}
\end{figure}

We consider two types of two-body coupling fluctuations: First, in the MAGIC scheme variations of the overall trapping frequency cause correlated fluctuations in the coupling strengths. Second, we discuss the general case of uncorrelated fluctuations of the two-body couplings, which could for example be caused by erroneous two-body gate implementations in a Trotterized annealing sweep.

The trapping frequencies in a trapped-ion system typically exhibit slow temporal drifts. 
To model this effect, we consider variations in the harmonic axial trap frequency $\omega_z$ according to
\begin{align}
\omega_z \rightarrow \omega_z + \delta \omega_z,
\end{align}
where we assume that the fluctuation $\delta \omega_z$ remains constant on the timescale of a single annealing sweep.

Since the two-body coupling strengths $J_{ij}$ in the MAGIC scheme are directly proportional to $1/\omega_z^2$, any fluctuations in the trap frequency induce correlated fluctuations in all of the two-body couplings.
From Eq.~\eqref{eq:J2}, we calculate the two-body coupling variations to first order in $\delta \omega_z$ as
\begin{align}\label{eq:trap_fluctuations}
     \delta J_{ij} \approx \sum_{n=1}^N \frac{-\hbar \partial_z \omega_i \partial_z \omega_j}{\nu_n^2 \omega_z  m} S_{in} S_{jn} \delta \omega_z = -2 \frac{J_{ij}}{\omega_z} \delta \omega_z,
\end{align}
where we used the fact that the axial phonon modes $\nu_n$ are proportional to the axial trap frequency $\omega_z$. Consequently, the standard deviations of the coupling and frequency fluctuations are related by $\sigma (\delta J_{ij}) \approx |2 J_{ij} / \omega_z |  \sigma(\delta\omega_z)$. 
Numerically, we verified that this first-order approximation remains accurate for fluctuation amplitudes up to $\sigma(\delta \omega_z) / \omega_z \approx 0.1$. For larger amplitudes, quadratic corrections become significant.

All $J_{ij}$ are thus scaled by the same factor when the axial trap frequency $\omega_z$ fluctuates. For the annealing protocol we consider in this work, where the problem is encoded only in the two-body interactions without utilizing local biases (see Sec.~\ref{sec:annealing}), this corresponds to effectively rescaling the entire cost Hamiltonian while preserving its structure.
Importantly, such a global rescaling shifts only the eigenvalues, but leaves the eigenstates unaltered. As a result, the ground state, which encodes the solution to the optimization problem, remains unaffected.
Therefore, fluctuations in $\omega_z$ and the resulting correlated fluctuations in $J_{ij}$ do not alter the computational outcome of the annealing protocol. However, for other annealing protocols that also utilize local biases to encode the problem (e.g., see Appendix \ref{app:alt_protocol}), these fluctuations could in principle become relevant, as they change the relative energy scales of the two-body interactions with respect to the local biases.

We now discuss the case of uncorrelated fluctuations in the two-body coupling terms $J_{ij}$. Unlike the correlated disorder induced by fluctuations in the axial trap frequency $\omega_z$, uncorrelated coupling noise does not scale all interaction terms uniformly and can, in principle, modify the structure of the cost Hamiltonian and thus affect its ground state.

To assess the impact of such uncorrelated disorder, we numerically compute the probability that random fluctuations $\delta J_{ij}$ change the ground state of the cost Hamiltonian, choosing the minimal MOT instance with 5 qubits defined in Eq.~\eqref{eq:5qubit_MOT} as a test problem. Each data point in Fig.~\ref{fig:Jcouplings_disorder} represents the mean probability obtained from 10\,000 independent samples. For each sample, the coupling perturbations $\delta J_{ij}$ are drawn independently from a Gaussian distribution with zero mean and specified standard deviation. The figure shows the probability of a ground-state change as a function of the relative disorder strength $\sigma(\delta J_{ij})/J_{ij}$. The variance of each data point $p$ can be calculated as $p(1-p)/10000 \leq 2.5 \cdot 10^{-5}$. We observe that the ground state remains stable for disorder levels up to approximately $20\%$, beyond which the probability for a changed ground state begins to increase. Fortunately, such large fluctuations are atypical in realistic experimental conditions. 
Moreover, in the MAGIC setup (see Sec.~\ref{sec:platform}), uncorrelated coupling disorder can only arise from imperfections in the trap potential or inhomogeneities in the magnetic field gradient, which are unlikely to occur over the length scale of the ion chain. Otherwise, uncorrelated disorder in $J_{ij}$ could in principle stem from imperfect two-body gate operations in the case of Trotterized annealing protocols. However, two-body gate fidelities of $\sim 99.9\%$ fidelity can already be achieved for current trapped-ion quantum computers \cite{Ballance2016}. 
Although we cannot exclude that the picture may change when scaling up the size of the optimization problem, the large robustness encountered for the benchmark models gives optimism that there is still significant room before the small realistic levels of uncorrelated coupling fluctuations can become detrimental. 
We therefore conclude that, while uncorrelated coupling fluctuations could in principle affect solution fidelity, they are not expected to play a significant role in practice under typical conditions of current experiments.

\subsection{Local field fluctuations } \label{subsec:local_field_disorder}

Magnetic field gradient fluctuations in the MAGIC setup (see Sec. \ref{sec:platform}), along with unwanted external magnetic fields, induce corresponding fluctuations in the ion resonance frequencies. These perturbations manifest as additional longitudinal fields $\sim \delta h^z \sigma^z$ in the cost Hamiltonian of the annealing protocol (Eq.~\eqref{eq:H0}). We model these disorder terms by modifying the local field components:
\begin{align}
    h_i^z \rightarrow h_i^z+\delta h_i^z\,,
\end{align}
where $\delta h^z_i$ represents the field fluctuation for ion $i$. In our considered encoding scheme there are no local fields in the cost Hamiltonian ($h^z_i = 0$, see Sec.~\ref{sec:annealing}). In this section, we assume that the fluctuations remain constant on the timescale of a quantum annealing sweep. We will consider field fluctuations with colored noise spectra in Sec.~\ref{sec:dd}. 

To assess the quantum annealing protocol's resilience to local field variations, we add disordered diagonal terms $\delta h^z_i$ to the cost Hamiltonian given in Eq.~\ref{eq:5qubit_MOT}. We sample the $\delta h^z_i$ from Gaussian distributions with varying standard deviations. For each standard deviation, we generate $10\,000$ samples and determine numerically whether the ground-state configuration changes.

\begin{figure}[htbp]
    \centering
\includegraphics{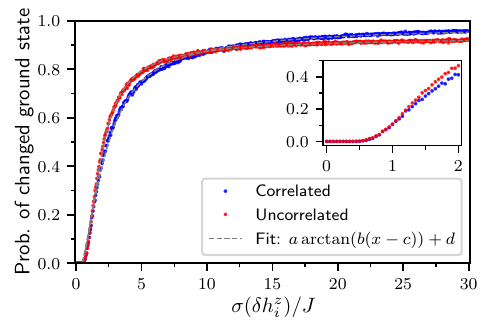}
    \caption{Probability of a changed ground state of the cost Hamiltonian for the quadratized minimal MOT problem (5 qubits, see Eq. \ref{eq:5qubit_MOT}) as a function of the standard deviation of the local field disorder $\delta h_i^z$ in units of $J$ for correlated (blue) and uncorrelated (red) disorder. The data points show the mean probabilities of a changed ground state for $10\,000$ simulations each (variance of $\leq 2.5\cdot 10^{-5}$), where $\delta h_i^z$ is sampled from a Gaussian distribution with zero mean. The grey dashed lines represent fits with the arctangent function $f(x) = a \arctan(b(x-c)) + d$, starting from $\delta h_i^z / J = 1$. For correlated disorder: $a, b, c, d \approx 0.74, 0.63, 0.39, -0.17$. For uncorrelated disorder: $a, b, c, d \approx 0.57, 0.81, 0.99, 0.10 $.}
    \label{fig:local_field_disorder}
\end{figure}

Figure \ref{fig:local_field_disorder} presents our results for two distinct disorder scenarios: correlated fluctuations ($\delta h^z_i = \delta h^z_j$ for all $i,j$) and uncorrelated fluctuations (all $\delta h^z_i$ independently sampled). In realistic experimental setups, the case of correlated disorder is typically more relevant, as external magnetic fields are unlikely to significantly vary on the length scale of the ion chain. Uncorrelated variations could, however, effectively occur by erroneous calibrations of frequencies addressing the individual qubit transitions. Both cases exhibit a sharp transition in ground-state stability at approximately $\delta h^z_i / J \approx 0.5$, see inset, where $J$ represents the energy scale set by the maximal two-body coupling strength. The grey dashed lines show heuristic arctan fits starting at $\delta h^z_i / J = 1$.

The exact value where local field variations change the optimization result will depend on the precise problem considered. However, the realistic experimental implementations face significantly larger local field fluctuations than the order $\delta h^z_i / J \approx 0.5$ found for the MOT benchmark problem.
Using experimental parameters that have been previously demonstrated for the MAGIC setup \cite{Piltz2016}, a five-ion chain reaches a maximal two-body coupling strengths of $J_{ij} \approx 26$ Hz \cite{Nagies2025}. In the same setup, local field fluctuations can reach magnitudes on the order of $10^3$ Hz in the worst-case scenario without any magnetic shielding and stabilization \cite{Merkel2019}, corresponding to $\delta h^z_i / J \sim 40$. This places the system well within the regime where the desired solution of the optimization problem is no longer the ground state. As mentioned above, improved setups can reach $\max_{ij} |J_{ij}| \approx 1650.6$~Hz~\cite{Nagies2025}, which, however, is still not sufficient to safely ignore variations of the local fields. 
To avoid degradation of the quantum annealing performance dedicated mitigation strategies are thus required. 
The next section demonstrates that local fields can be completely compensated through simple dynamical decoupling sequences, even when they fluctuate substantially and when they are time dependent.

%% file: dd.tex
This section explores how simple dynamical decoupling protocols can be implemented during annealing sweeps to completely suppress the effect of local field fluctuations and recover ideal annealing performance, both under slow drifts that are approximately constant within each experimental run (equivalent to random disorder) and under time-dependent noise. 
We begin in Sec.~\ref{subsec:analytics} with a general explanation of dynamical decoupling (DD). We also derive an effective time evolution operator for annealing protocols subject to varying fields and incorporating DD pulses, which provides further analytical insights. 
Our numerical results are presented in Sec.~\ref{subsec:results}, validating our theoretical discussion and providing concrete estimates for the required pulse counts in experimental implementations across different noise spectra and amplitudes. Finally, Sec.~\ref{subsec:collapse} discusses a universal scaling behavior we observe in the number of dynamical decoupling pulses needed for effective noise suppression.

\subsection{Applying dynamical decoupling pulses during a quantum annealing protocol}
\label{subsec:analytics}

Dynamical decoupling represents a powerful error mitigation technique designed to extend the coherence times of quantum systems by suppressing environmental decoherence \cite{Viola1998,Khodjasteh,Souza2011, Cai2023, Piltz2013, Valahu2021, Morong2023, Barthel2023, Nuennerich2024}. The fundamental principle involves applying periodic control pulses that rotate all qubits around specific axes by predetermined angles, effectively averaging out unwanted error terms. 

When implementing dynamical decoupling within quantum annealing protocols, a fundamental challenge arises: the applied pulses generally do not commute with the quantum annealing Hamiltonian given in Eq.~\eqref{eq:modified_annealing}. This means that a control pulse can excite the system from its instantaneous ground state into higher energy eigenstates, potentially compromising the annealing process. Consequently, careful corrections must be implemented to ensure that the annealing sweep continues to converge towards the correct solution.

A general control pulse, assumed to be applied nearly instantaneously (i.e., much faster than the relevant time scales of the annealing sweep), can be mathematically represented as a collective rotation of all qubits around an arbitrary axis: $U_\mathrm{pulse} = \prod_k \exp(-i\bold{v^{(k)}} \cdot \bold{\sigma^{(k)}})$, where $\bold{\sigma^{(k)}} = (\sigma_x^{(k)}, \sigma_y^{(k)}, \sigma_z^{(k)})$ denotes the vector containing the Pauli matrices acting on qubit $k$. 

In this work, motivated by the specific form of the annealing Hamiltonian and salient error term, we focus on the simplest case: $\pi$-pulses that rotate qubits around the x-axis, equivalent to spin-flip operations. Such pulses can be implemented in the MAGIC setup by turning on driving fields, tuned to the qubit resonances, for a finite time interval (Eq.~\eqref{eq:1q_gate}). This choice of simple $\pi$-pulses is sufficient for mitigating the local field fluctuations considered in our analysis (see Sec. \ref{subsec:local_field_disorder}). The action of such pulses on the driver and cost Hamiltonians of the annealing protocol, as given in Eq.~\eqref{eq:modified_annealing}, yields
\begin{align}
   e^{i \frac{\pi}{2} \sigma_x} H_{\text{driving}} e^{-i \frac{\pi}{2} \sigma_x} &= H_{\text{driving}} \label{eq:xpulse_driving},\\
   e^{i \frac{\pi}{2} \sigma_x} H_{\text{cost}} e^{-i \frac{\pi}{2} \sigma_x} &= \sum_{ij} J_{ij} \sigma_i^z \sigma_j^z - \sum_i h_i^z \sigma_i^z. \label{eq:xpulse_cost}
\end{align}
Notably, the spin-flip operations leave the driver Hamiltonian unchanged and only modify the signs of the local fields $h_i^z$ in the cost Hamiltonian. 
In the case where the fields are undesired error terms, this sign flip will enable alleviating their effect, as detailed below. 
In the case where the fields are part of the problem cost Hamiltonian, the sign reversal is a crucial perturbation that needs to be compensated.

To address this issue, we employ here the ancilla qubit approach described in Sec.~\ref{sec:annealing}. By encoding all local fields into additional two-body interactions through the introduction of an ancilla qubit, see Eq.~\eqref{eq:ancilla_encoding}, we render the complete annealing protocol Hamiltonian invariant under dynamical decoupling pulses. This ensures that the system remains in the instantaneous ground state throughout the entire annealing sweep.
An alternative approach involves actively compensating for the sign flip with each pulse application, resulting in the annealing protocol described in Appendix \ref{app:alt_protocol}. 

Practical experimental implementations often employ more sophisticated dynamical decoupling sequences. For instance, universally robust sequences \cite{Souza2011, Piltz2013, Genov2017} provide enhanced protection against a broader spectrum of noise sources and compensate for imperfections in the pulse sequences themselves. These advanced protocols generally incorporate rotations around multiple axes with varying angles.
The integration of such general pulse sequences within quantum annealing frameworks has been investigated in Refs.~\cite{Lidar2008, Quiroz2012}.

\subsection{Time evolution operator}\label{subsec:dd_time_evolution}

We now demonstrate how periodically applied dynamical decoupling pulses in the form of global spin flips can cancel local field fluctuations during an annealing protocol. 
To this end, we consider the quantum annealing Hamiltonian discussed in Sec.~\ref{sec:annealing} with noise, 
\begin{align}\label{eq:annealing_noise}
    H(t) &= C(t) H_{\text{driving}} + H_{\text{cost}} + H_{\text{noise}} \nonumber\\
    &= - C(t)\sum_i h^x \sigma_i^x + \sum_{ij} J_{ij} \sigma_i^z \sigma_j^z + \sum_i \delta h_i^z(t) \sigma_i^z,
\end{align}
where the problem is encoded only in the two-body interactions as discussed in Sec.~\ref{sec:annealing}, while the local fields $\delta h_i^z(t)$ in $H_{\text{noise}}$ are time-dependent error terms that we aim to effectively cancel via dynamical decoupling.

To make analytic progress, we calculate the time evolution operator of this annealing protocol for a small time interval $2 \Delta t \equiv t_+ - t_-$. During this interval, we apply two dynamical decoupling control pulses, the first at the center of the interval at time $t_0 = t_-+\Delta t$ and the second pulse at time $t_+$. Before and after the first pulse, the system evolves according to Eq.~\ref{eq:annealing_noise} for the respective time interval $\Delta t$.

We assume that $ 2\Delta t$ is small enough so that the function $C(t)$ can be approximated as linear during the corresponding annealing segment. At a given time $t$ in the small time interval $2\Delta t$, it is then defined as
\begin{equation}\label{eq:Ct_approx}
    C(t) = C(t_-) - v(t_-) t\quad, \qquad t\in [t_-,t_+].
\end{equation}
Here, $v(t_-)$ is a generally time-dependent function that determines the form of the annealing protocol. In our numerical simulations in the following section, the driving ramp protocol is chosen to be linear during the entire annealing sweep, which is often the standard choice. This means that the slope is constant ($v(t_-) \equiv v$ for all $t_-$) during the entire sweep. Further, we assume that the disorder terms fluctuate sufficiently slowly so that we can take them as constant during the short time interval $2 \Delta t$: $\delta h_i^z(t) \equiv \delta h_i^z$, $\forall t\in[t_-,t_+]$.

The unitary time-evolution operator in the first half of the interval from $t_-$ to $t_0$ can be written using the Magnus expansion \cite{Blanes2009}:
\begin{align}\label{eq:t1_to_t2_teo}
    U_{t_- \to t_0} &= \mathcal{T} e^{-i\int_{t_-}^{t_0} dt^\prime H(t^\prime)} \nonumber \\ 
    &= e^{-i\int_{t_-}^{t_0} dt^\prime H(t^\prime) - \frac{1}{2}\int_{t_-}^{t_0}dt^\prime \int_{t_-}^{t^\prime} dt^{\prime \prime} [H(t^\prime), H(t^{\prime\prime})] + \mathcal{O}(\Delta t^4)}.
\end{align}
Evaluating the commutators, integrating $C(t)$ (using Eq.~\eqref{eq:Ct_approx}) and keeping only terms up to third order in $\Delta t$ gives
\begin{align}
U_{t_- \to t_0} \label{eq:t1_to_t2}
&\approx \exp\left( -i [C(\bar{t}_{-}) H_{\text{driving}}  + H_{\text{cost}} + H_{\text{noise}}] \Delta t \right. \nonumber \\
 &-\frac{i}{6}h^x  v(t_-) \bigg[\sum_i \delta h_i^z \sigma_i^y + 2 \sum_{ij} J_{ij}  \sigma_i^z \sigma_j^y\bigg]  \Delta t^3 \big)\,,
\end{align}

where we defined $\bar{t}_{-} = (t_- + t_0)/2$. The time-evolution operator $U_{t_0 \to t_+}$ for the interval from $t_0$ to $t_+$ has an analogous structure. To calculate the total time evolution operator for the annealing sweep of duration $2\Delta t$ from time $t_-$ to $t_+$, we can use the Baker--Campbell--Hausdorff formula (again neglecting terms of fourth and higher order in $\Delta t$), 

\begin{align}
     \sigma^x U_{t_0 \to t_+} \sigma^x U_{t_- \to t_0} &= e^{-i [\Hat{A}_1 +  \Hat{A}_2 
    + \Hat{A}_3 
    ] 2\Delta t+ \mathcal{O}(\Delta t^4)}, 
\end{align}
with

\begin{align}
    \Hat{A}_1 &= \sum_{ij} J_{ij} \sigma_i^z \sigma_j^z - h^x C(t_0)\sum_i \sigma_i^x, \\
    \Hat{A}_2 &= \bigg( h^x  C(t_0)\sum_i \delta h_i^z \sigma_i^y \bigg) \Delta t \label{eq:A2},\\
    \Hat{A}_3 &= \bigg( -\frac{2}{3} h^x v \sum_{ij}J_{ij} \sigma_i^z \sigma_j^y + \frac{2}{3} h^x  C(t_0) \sum_i \delta{h_i^z}^2 \sigma_i^x  \bigg) \Delta t^2. \label{eq:A3}
\end{align}

Here, we write $\sigma^x  = \bigotimes_i \sigma^x_i$ as shorthand notation denoting the global spin flip due to the dynamical decoupling pulses. Further, we assume that the slope of the driving protocol remains nearly constant during the time interval $2\Delta t$, so that we can set $v(t_-) = v(t_+) \equiv v$.

The first-order terms recover the ideal quantum annealing evolution with the local field disorder being completely canceled, i.e., $\Hat{A}_1  = C(t) H_{\text{driving}} + H_{\text{cost}}$. The leading-order corrections, given in Eq.~\eqref{eq:A2}, to this ideal evolution can be interpreted as a modification of the driving Hamiltonian, as they act in a basis direction orthogonal to the cost Hamiltonian. Their strength is proportional to the amplitude of the local field disorder terms and the interval between the dynamical decoupling pulses ($\sim \delta h_i^z \Delta t$). Our numerical simulations in the following section support this observation, showing that the local field noise can always be suppressed by choosing the interval between dynamical decoupling pulses to be sufficiently small. Further, in Section \ref{subsec:collapse}, we demonstrate that the annealing protocol with dynamical decoupling follows a universal scaling behavior, solely determined by the product $\delta h_i^z \Delta t$.

The corrections to the ideal annealing protocol of third order in $\Delta t$, given in Eq.~\eqref{eq:A3}, are composed of two parts: First, there is a two-body interaction term whose strength is proportional to $\sim h^x v J_{ij} \Delta t^2$. This correction is independent of the local field noise and vanishes in the limit of adiabatic driving ($v \to 0$). Second, there is another correction to the driving Hamiltonian that scales as the square of the first-order correction ($\sim \delta {h_i^z}^2 \Delta t^2$).

All corrections to the ideal annealing sweep considered so far rely on the assumption of perfectly instantaneous dynamical decoupling pulses applied at uniform intervals. In practice, however, pulse operations introduce additional sources of error: they have finite duration and their spacing may deviate from uniformity. These control imperfections are well documented, and several dynamical decoupling schemes with intrinsic robustness to such errors have been proposed, e.g., in Ref.~\cite{Genov2017}. Additional errors can become relevant at higher pulse rates which can necessitate a higher Rabi frequency in experiments. If this frequency gets too close to the trapping frequency, this can produce further errors due to sideband excitations, which need to be compensated, e.g., via sympathetic cooling \cite{Rohde2001, Sriarunothai2017, Sosnova2021}.

As an example of a control error, we calculate the effect of a small pulse timing error $\epsilon$ on the time evolution operator for a small time interval $2\Delta t = t_+ - t_-$ during an annealing sweep, analogous to the discussion above: We let the system evolve with the annealing Hamiltonian (Eq.~\eqref{eq:annealing_noise}) for a time interval $\Delta t + \epsilon = t_0^\epsilon - t_-$ ($t_0^\epsilon = t_0 + \epsilon$), after which we apply a global spin flip. Afterwards the system evolves for a time interval $\Delta t - \epsilon =  t_+ - t_0^\epsilon $ and we apply a second global spin flip at time $t_+$. The total time evolution operator to first order in $\Delta t$ is then given as
\begin{align}\label{eq:magnus_timing_error}
     \sigma^x U_{t_0^\epsilon \to t_+} \sigma^x U_{t_- \to t_0^\epsilon} &= e^{-i\hat{A}_1^\epsilon 2\Delta t + \mathcal{O}(\Delta t^2)}, 
\end{align}

where the higher-order terms also include contributions proportional to $\epsilon \Delta t$. The first-order terms can be written as the original annealing Hamiltonian with a rescaled local-field noise contribution:
\begin{align}
    \hat{A}_1^\epsilon &=  H_{\text{cost}} + C(t_0) H_{\text{driving}} + \frac{\epsilon}{\Delta t} H_{\text{noise}}.
\end{align}

To get a rough estimate of how large $\epsilon$ can be before affecting the performance of the annealing sweep, we can consider the numerical result in Sec.~\ref{subsec:local_field_disorder} for the maximal local-field disorder strength the protocol is robust to: $\epsilon /  \Delta t \approx 0.5  J / \delta h^z_i$, which corresponds to a required accuracy in pulse timing of around 99\%  for $\delta h^z_i = 1$ kHz and $J = 26$ Hz.
In our numerical simulations of annealing sweeps with time-dependent local-field noise, we observe that moderately larger fluctuations in pulse spacing of the order of 98\% fidelity also have negligible impact on performance, provided they do not exhibit a systematic pattern that would enable coherent error accumulation over the course of the annealing sweep (see Appendix \ref{app:methods}).

In summary, one can always recover the ideal annealing protocol and obtain the correct solution by choosing the time interval $\Delta t$ between consecutive dynamical decoupling pulses of uniform length and sufficiently small with respect to the amplitude of the local field noise $\delta h_i^z$. The salient deviations for slow sweeps scale as powers of $\delta {h_i^z} \Delta t$. 
Finally, we note that the above scheme of periodically applying dynamical decoupling pulses can also be used to effectively modulate the strength of the two-body coupling weights $J_{ij}$, which can be useful for encoding arbitrary optimization problems into the two-body couplings. We discuss this method in Appendix~\ref{app:dd}.

\subsection{Numerical results for noise mitigation}
\label{subsec:results}

\begin{figure*}[htb!]
    \centering        \includegraphics{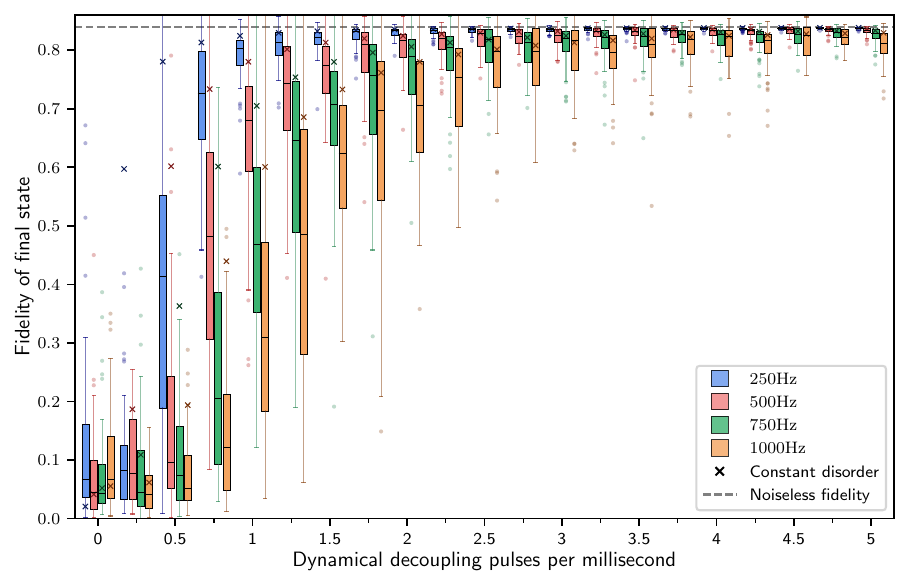}
    \caption{Fidelity of the final state, compared to the state encoding the problem solution, at the end of a quantum annealing sweep versus dynamical decoupling pulse rate, using the minimal MOT test problem with five qubits and an annealing time of $2.6/J$ (corresponding, for an energy scale of $J = 26$ Hz, to a sweep duration of $100$ms). Box plots compare results for correlated noise ($\delta h_i^z(t) = \delta h^z(t)$) using identical noise spectra with two Lorentzian frequency peaks at 50 Hz and 150 Hz but different amplitudes of 250 Hz (blue), 500 Hz (red), 750 Hz (green), and 1000 Hz (orange). The crosses in each boxplot give a comparison with the median value of annealing sweeps subject to constant disorder ($\delta h^z(t) = \delta h^z$) of the same respective amplitude. The horizontal dashed line shows the achievable fidelity of the ideal noiseless protocol ($\approx 0.84$). Data for each boxplot corresponds to 50 random independent noise realizations (see Appendix \ref{app:methods}). We observe that the achieved final fidelity converges to the noiseless case for a sufficient rate of dynamical decoupling pulses, with higher noise amplitudes requiring higher rates.}
    \label{fig:2peaks_all_corr}
\end{figure*}

\begin{figure*}[htb!]
    \centering        \includegraphics{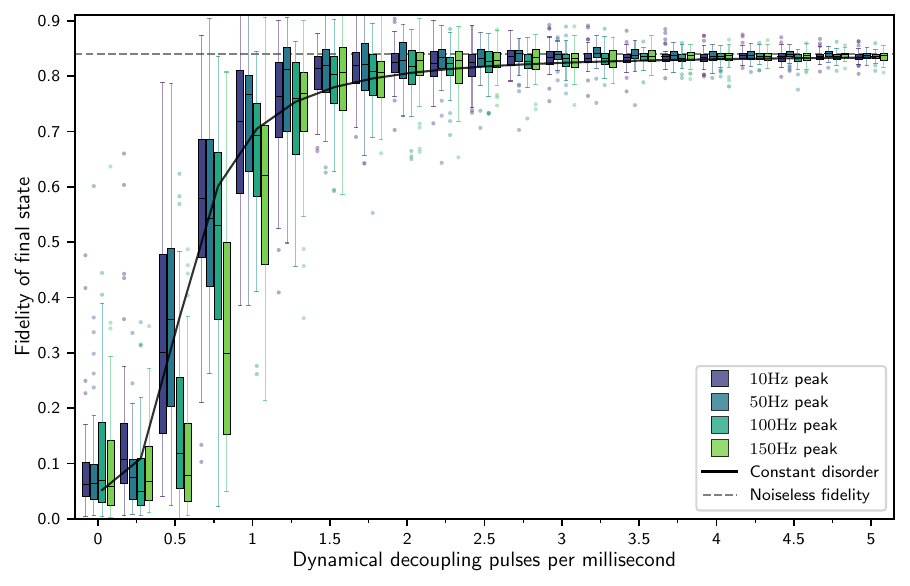}
    \caption{
    Impact of noise peak frequency on dynamical-decoupling performance. As Fig.~\ref{fig:2peaks_all_corr}, but using different noise spectra with a single Lorentzian peak centered at varying frequencies, all with the same amplitude of 750Hz. The solid line gives a comparison with the median value of annealing sweeps with constant disorder ($\delta h^z(t) = \delta h^z$) and the same amplitude of 750Hz. The horizontal dashed line shows the achievable fidelity of the ideal noiseless protocol. Data for each boxplot corresponds to 50 random independent noise realizations (see Appendix \ref{app:methods}). Similar to Fig.~\ref{fig:2peaks_all_corr}, the final fidelity converges to the noiseless, case for sufficient rates of dynamical decoupling pulses. The required rates are only slightly dependent on the noise frequency. }
    \label{fig:1peak_all_corr}
\end{figure*}

In this section, we present numerical simulations of the dynamical decoupling protocol introduced in the previous section. We employ the minimal MOT instance with five qubits, defined in Eq.~\eqref{eq:5qubit_MOT} as our test problem, though our findings regarding noise mitigation through dynamical decoupling extend to different problem instances and system sizes. In Appendix \ref{app:alt_protocol}, we discuss an alternative annealing protocol solving the same problem but using only four qubits and Appendix \ref{app:MOT8qubit} contains an analogous treatment of a MOT problem instance encoded into 9 qubits. Additionally, in Appendices \ref{app:cuttingstock} and \ref{app:sk_model} we show numerical results for noise mitigation in quantum annealing with five and six qubit cutting stock problems, as well as Sherrington-Kirkpatrick model instances up to 12 qubits, further demonstrating the universality of the dynamical decoupling protocol. 

Our simulations solve the time-dependent Schrödinger equation using the annealing Hamiltonian defined in Eq.~\eqref{eq:annealing_noise}. In the numerics below, we implement a linear annealing sweep over a fixed duration of $2.6/J$, which provides a good fidelity in the absence of any error terms. To put this value into context of the timescales in a typical laboratory experiment, the energy scale $J = 26$~Hz demonstrated in the MAGIC setup \cite{Piltz2016} corresponds to an annealing time of 100 ms for our chosen parameters.
We set the driving strength to $h_x = 3J$ as a compromise to avoid overly large separations of energy scales while keeping the initialization into a fully $x$-polarized state a good approximation of the initial ground state (see discussion in Sec.~\ref{sec:annealing}). In our numerics, we discretize the annealing sweep into 50\,000 steps. For each discrete step, we set a value of the local field $\delta h_i^z$, sampled from a given noise spectrum, thus approximating the continuous fluctuations of an external magnetic field.
Here, we consider two distinct types of noise spectra: First, we model an experimentally realistic spectrum featuring two dominant Lorentzian peaks at frequencies of 50 Hz and the leading next harmonic at 150 Hz. As discussed in Sec.~\ref{subsec:exp_parameters}, such a spectral structure is typical of external magnetic field fluctuations arising from the electricity grid. Second, we also examine spectra with individual Lorentzian peaks at various frequencies to systematically investigate our protocol's dependence on noise characteristics. All noise realizations are generated using the approximate frequency domain method detailed in Ref.~\cite{percival1992}.
At approximately equidistant intervals across the 50\,000 discrete time steps (see Appendix \ref{app:methods}), we apply dynamical decoupling pulses as instantaneous global spin flips affecting all qubits simultaneously.  

To evaluate the success of the protocol, we calculate the overlap between the final state after the annealing sweep and the two-fold degenerate true ground state encoding the problem solution. 
Figure \ref{fig:2peaks_all_corr} presents this protocol fidelity as a function of the number of DD pulses for a noise spectrum featuring two Lorentzian peaks centered at 50 Hz and 150 Hz. We consider four noise amplitudes---250 Hz (blue), 500 Hz (red), 750 Hz (green), and 1000 Hz (orange). These noise amplitudes are chosen as a range going from typical values to a worst-case scenario: In an experimental setup without any magnetic shielding, local-field noise can realistically reach values around 1 kHz. Using stabilization techniques, these values can potentially be reduced by an order of magnitude \cite{Merkel2019}. 

The considered DD pulse rates vary from 0 to 5 pulses per millisecond, which for the chosen sweep duration of 100 ms correspond to a total of 0 to 500 dynamical decoupling pulses applied throughout the annealing process.
For each pulse rate, we simulate 50 independent annealing runs, each subject to a distinct, correlated noise realization (i.e., identical across all qubits). Results are visualized using boxplots, where the boxes indicate the interquartile range and horizontal lines denote median fidelities. The dashed grey line at $\approx 84$\% represents the fidelity achieved  under the same annealing schedule in the absence of noise.

We observe that increasing the dynamical decoupling pulse rate systematically improves the final fidelity, with all noise amplitudes showing convergence towards the ideal (noiseless) performance. For each amplitude, a pulse rate of approximately 2.5 pulses per ms is sufficient to yield fidelities above 70\% in the majority of runs. This rate is readily achievable by current trapped-ion platforms: the experimental upper limit of the pulse rate is set by the width of the dynamical decoupling pulses (since they should not overlap), which in turn depends on the accessible Rabi frequency. For a large Rabi frequency, the pulse width to implement the same $\pi$-pulse is smaller. The Rabi frequency is effectively limited in the experiment by the trapping frequency, as it should be set below that frequency in order to avoid heating via unwanted sideband excitations. Typical trapping frequencies in an experiment are well above 100 kHz. When setting the Rabi frequency to 100 kHz, the pulse width would be approximately 0.06 ms  (using Eq.~\eqref{eq:1q_gate} for single-qubit rotations), which implies an upper limit for the pulse rate of around 15 pulses per millisecond. In our numerical examples, we find that significantly lower pulse rates are sufficient to restore the final fidelity to noise-free levels.

For comparison, we also include results from simulations with static (time-independent) disorder in Fig.~\ref{fig:2peaks_all_corr}, marked by crosses. To this end, we compute the median fidelity from 50 annealing runs where each qubit experiences a constant local field sampled from a Gaussian distribution with a standard deviation matching the corresponding noise amplitude. The resulting fidelities lie a little higher and show a similar convergence trend, indicating that time-dependent noise and static disorder can be mitigated with similar dynamical decoupling pulse rates for the parameters considered.

To examine the role of spatial correlations, we also ran simulations with \emph{uncorrelated} noise, where each qubit is subject to an independent noise realization. The results (not shown) are qualitatively very similar to those in Fig.~\ref{fig:2peaks_all_corr}, with convergence to ideal fidelity occurring slightly faster.

Figure \ref{fig:1peak_all_corr} explores the dependence of annealing performance on the spectral shape of the noise. We fix the noise amplitude to 750 Hz (corresponding to the green boxplots from Fig.~\ref{fig:2peaks_all_corr}) and consider single Lorentzian peaks at 10, 50, 100, and 150 Hz, respectively (from darkest to lightest green). A constant disorder comparison is shown as a solid line. 
We find no significant qualitative sensitivity to the exact peak frequency in the considered regime: all spectra exhibit similar convergence behavior, with slightly improved performance for lower noise frequencies.

As these results show, dynamical decoupling can render the performance of quantum annealing in realistic devices robust against field fluctuations of large strength and various spectral properties. 
As we show in Appendices \ref{app:cuttingstock} and \ref{app:sk_model}, when normalizing to the noise-free fidelity that is achievable for a given sweep speed, both MOT, cutting stock and Sherrington-Kirkpatrick model problems of different sizes follow very similar curves, showing the generality of this protocol. 
The performance can be improved with enhanced two-body interaction strengths, achievable in the MAGIC setup with existing technology. As mentioned previously, magnetic field gradients on the order of 150 T/m are feasible \cite{Weidt2016, SiegeleBrown2022}, enabling coupling strengths up to 
$J \approx 1650$ Hz \cite{Nagies2025}. In such scenarios, the annealing time required to achieve fidelities comparable to those in our simulations would reduce from 100 ms to approximately 1.5 ms. Moreover, the ratio of noise amplitude to coupling strength would become more favorable, enabling high-fidelity sweeps with even fewer dynamical decoupling pulses.

\subsection{Universal scaling behavior}
\label{subsec:collapse}

\begin{figure}
    \centering
    \text{(a) Constant disorder} \par
    \includegraphics[width=\linewidth]{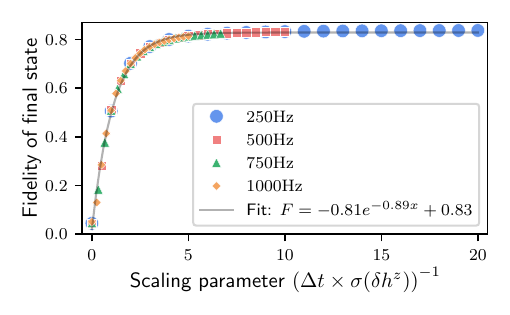}
    
    \text{(b) Noise spectrum with two Lorentzian peaks} \par
    \includegraphics[width=\linewidth]{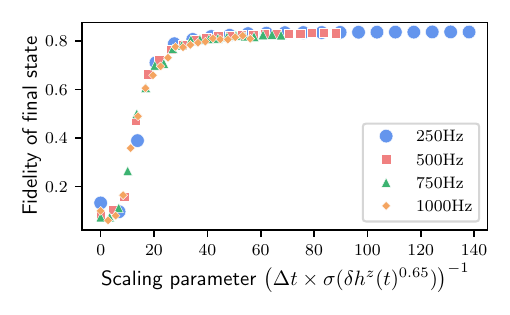}
    
    \caption{Fidelity of the final state, compared to the state encoding the problem solution, at the end of a quantum annealing sweep versus a universal scaling parameter $(\Delta t \times \sigma(\delta h^z)^c)^{-1}$, which depends on the time $\Delta t$ between two dynamical decoupling pulses, the standard deviation of local field fluctuations $\sigma(\delta h^z)$ (correlated noise), and a noise-dependent exponent $c$. 
    (a) For constant disorder ($\delta h^z (t) = \delta h^z$), an excellent data collapse is obtained for $c=1$, in agreement with analytic derivations (see Sec.~\ref{subsec:analytics}). The grey line in (a) shows an exponential fit considering all data points for the different noise amplitudes together.  
    (b) For time-dependent noise spectra with two Lorentzian peaks at 50 Hz and 150 Hz, an approximate data collapse can still be obtained for a (fitted) value of $c\approx 0.65$. 
    In both cases, we show the mean value of 50 random independent noise realizations. }
    \label{fig:collapse}
\end{figure}

In the previous section, we demonstrated that quantum annealing remains effective under local field fluctuations when dynamical decoupling pulses are applied at regular intervals during the protocol. As shown in Fig.~\ref{fig:2peaks_all_corr}, the final fidelity converges to the noiseless limit as the dynamical decoupling pulse rate increases, with the required pulse rate depending on the noise amplitude and to a lesser degree the noise spectrum. In this section, we reveal a unifying trend across these cases by identifying a generalized scaling parameter that causes all fidelity curves to collapse onto a single universal curve.

To this end, we build upon the analytical treatment of the time evolution operator of the noisy annealing protocol with dynamical decoupling pulses derived in Sec.~\ref{subsec:analytics}. There, the leading-order corrections due to local field fluctuations appear as perturbations to the driver Hamiltonian proportional to $\delta h_i^z \times \Delta t$, see Eq.~\eqref{eq:A2}, where $\delta h_i^z$ is the local field fluctuation at the position of qubit $i$ and $\Delta t$ is the time interval between two consecutive dynamical decoupling pulses (corresponding to the inverse of the pulse rates on the horizontal axis in Figs.~\ref{fig:2peaks_all_corr} and \ref{fig:1peak_all_corr}). 
This implies that the deviation from the ideal (noiseless) annealing performance should be governed by this product. 

To test this prediction, we plot in Fig.~\ref{fig:collapse} (a) the mean final fidelities for the case of static correlated disorder, now as a function of the inverse parameter $(\sigma(\delta h^z) \times \Delta t)^{-1}$, where $\sigma(\delta h^z)$ denotes the standard deviation of the static local field disorder. This rescaling causes the data from all four disorder amplitudes to collapse onto a single, smooth curve. The collapse confirms our analytical prediction and demonstrates that the key quantity controlling fidelity loss is indeed the product of noise amplitude and time between dynamical decoupling pulses. The gray line in Fig.\ref{fig:collapse} (a) shows an exponential fit to the combined data.

However, the perturbative argument in Sec. \ref{subsec:analytics} assumes the field fluctuations $\delta h^z$ remain constant between dynamical decoupling pulses. This assumption breaks down once the noise acquires a significant time dependence. Consistent with this, we find that applying the same rescaling to annealing sweep results with time-dependent noise fails to produce a similarly precise data collapse.
To recover a universal description, we generalize the scaling parameter to the form $\sigma(\delta h_i^z)^c \times \Delta t$ with $c$ being a fitting exponent between zero and one. 

In Fig.~\ref{fig:collapse}(b), we plot the mean fidelities for time-dependent correlated noise, corresponding to the spectrum considered in the previous section with two Lorentzian peaks (at 50 Hz and 150 Hz), as a function of this generalized parameter. For $c\approx 0.65$ we achieve an approximate collapse of the data, suggesting a sublinear dependence of the fidelity degradation on noise amplitude in the considered noise regime.

We further analyzed noise spectra with single Lorentzian peaks at various frequencies: the optimal value of $c$ consistently decreases as the dominant noise frequency increases. 

It will be interesting to understand the microscopic origin of this generalized scaling behavior. A rigorous analytical derivation of the generalized parameter and its dependence on the noise spectral density could provide deeper insight into the mechanisms of dynamical decoupling in time-dependent environments and represents a promising avenue for future research.

%% file: Conclusion.tex
To summarize, we have discussed how local field noise in quantum annealing can be robustly mitigated by interleaving periodic dynamical decoupling pulses into the annealing schedule. Starting from a minimal, five‑qubit Multiple‑Object‑Tracking QUBO instance, chosen to capture industrially relevant optimization challenges while remaining numerically tractable, we demonstrated that current small‑scale devices can already encode and solve practical problems as a proof of principle. Focusing on a microwave‑controlled trapped‑ion platform with magnetic‑gradient–induced coupling (although our findings generalize straight-forwardly to other architectures and optimization problems), we explained how such a platform could be especially suitable for quantum annealing applications due to the native all-to-all two-body interactions, which could potentially be tuned to directly encode a QUBO optimization problem. We argued that two-body coupling‑strength fluctuations play a negligible role under realistic experimental conditions and identified external magnetic‑field fluctuations as the dominant error channel on such platforms.

As an error mitigation technique for such local field noise, we explained in detail the periodic application of dynamical decoupling pulses implementing global spin flips during an annealing protocol. We discussed how the annealing protocol needs to be modified to accommodate such an error mitigation scheme, and analytically calculated the time evolution operator of the annealing sweep with noise and dynamical decoupling pulses.

We performed numerical simulations demonstrating the success of this protocol for a realistic noise spectrum and different noise amplitudes. We found that around 2.5 dynamical decoupling pulses per millisecond are already enough to efficiently counteract the noise, for all considered regimes with noise amplitudes up to 1000 Hz (chosen as a worst-case scenario) and when considering already demonstrated experimental two-body coupling strengths of 26 Hz. 
Furthermore, we verified that the exact form of the noise spectrum has little effect on the required number of dynamical decoupling pulses (using different spectra with single Lorentzian peaks located at different frequencies up to 150 Hz). 
Moreover, we found that different system sizes and both MOT and cutting stock problems, as well as Sherrington--Kirkpatrick model instances, show quantitatively similar behavior when normalizing the fidelity to the respective noiseless value. 
We commented on the improvements to be expected for near-future devices with much higher interaction strengths, which enable much faster annealing sweeps with fewer dynamical decoupling pulses to obtain a similar performance.

Finally, we identified a universal scaling behavior for the achieved final fidelities with different noise amplitudes and dynamical decoupling pulse rates. For time-independent local field disorder, we found that the achieved fidelity is determined by the product of noise amplitude and the time interval between two dynamical decoupling pulses, matching our analytical calculations. For time-dependent noise, we find a dependence on a generalized parameter where we observe a sublinear dependence on the noise amplitude, with an exponent that decreases for noise spectra containing larger frequencies. This universal scaling provides a compact figure of merit for designing dynamical decoupling sequences across diverse noise environments.

Our work opens several immediate research directions. For example, it will be important to investigate additional noise sources, more sophisticated dynamical-decoupling protocols, and a wider range of optimization problems. 
Motivated by the universal behavior suggested by our analytical and numerical results, it will also be interesting to understand whether a product of error strength and decoupling pulse rate can serve as a general guiding principle to design practical annealing protocols. Furthermore, incorporating  dynamical decoupling sequences within quantum optimization algorithms provides a means to experimentally tune the impact of noise, allowing to controllably study the performance of quantum optimization and the role of quantum effects therein \cite{Santra2025,Capecci2025}, under various degrees of decoherence \cite{Hauke2015}.  
Finally, we anticipate that our results will motivate near‑term experimental quantum annealing demonstrations on trapped-ion platforms, and that they will inform error‑mitigation strategies for quantum annealing algorithms in other architectures.

%% file: Acknowledgments.tex
\begin{acknowledgments}
We thank Kevin T. Geier, Patrick Huber, Florian Köppen and Christof Wunderlich for helpful discussions.
The work reported in this publication is based on a project that was funded by the German Federal Ministry for Education and Research under the funding reference number 13N16437.
The authors are solely responsible for the content of this publication. We acknowledge the CINECA award under the ISCRA initiative, for the availability of high performance computing resources and support.
This work has benefited from Q@TN, the joint lab between University of Trento, FBK---Fondazione Bruno
Kessler, INFN---National Institute for Nuclear Physics, and CNR---National Research Council.
We acknowledge support by Provincia Autonoma di Trento.
\end{acknowledgments}

%% file: Appendix.tex
\section{Annealing protocol without additional ancilla qubit}\label{app:alt_protocol}

In Sec.~\ref{sec:annealing}, we presented a quantum annealing protocol that encodes QUBO problems exclusively through two-body interactions. This approach eliminates local field terms by encoding them as interactions with an additional ancilla qubit (see Eq.~\eqref{eq:ancilla_encoding}). The key advantage of this encoding scheme is that it decouples the problem representation from local field noise, enabling effective noise suppression through dynamical decoupling control pulses (Sec.~\ref{sec:dd}). 
However, the encoding requires one additional ancilla qubit for any problem instance. Given the limited qubit resources available in current quantum hardware, it may be more practical to retain the original QUBO formulation that directly incorporates local field terms in the cost Hamiltonian. In this Appendix, we demonstrate how the annealing sweep with dynamical decoupling pulses can be adapted to mitigate unwanted local field noise while preserving the desired local fields that encode the optimization problem.

This alternative annealing protocol Hamiltonian with local field noise $\delta h_i^z (t)$ is defined as
\begin{align}\label{eq:alt_protocol}
    H &=  C(t) H_\text{driving} + \sum_{ij} J_{ij} \sigma_i^z \sigma_j^z + \sum_i \big(h_i^z + \delta h_i^z (t)\big) \sigma_i^z.
\end{align}
Implementing the same dynamical decoupling scheme as in Sec.~\ref{sec:dd} would cancel both the noise $\delta h_i^z (t)$ and the local field terms $h_i^z$ encoding the problem. However, this can be easily avoided by applying a sign change to the problem-encoding local fields ($h_i^z \to - h_i^z$), simultaneously with each application of a dynamical decoupling control pulse. This additional sign change cancels the one due to the global dynamical-decoupling spin flip, such that only the noise terms $\delta h_i^z(t)$ get effectively canceled out. 

The additional sign flips are easily implementable in experiments. E.g., when the problem-encoding local fields are implemented via detuning of a microwave signal from resonance of the corresponding ion transition (see for example discussion in Ref.~\cite{Nagies2025}), it suffices to quickly switch the sign of the detuning simultaneously with the action of the dynamical decoupling pulse.  
Other implementations rely on virtual Z rotations \cite{McKay2017} or single-qubit gates in a Trotterized annealing protocol, in both of which one can just choose opposite rotation angles. 

\begin{figure}
    \centering
    \includegraphics{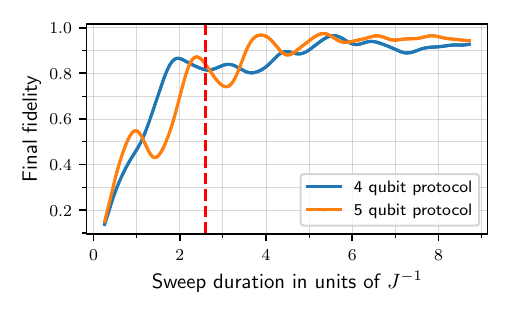}
    \caption{Final noiseless fidelities achieved for different total linear annealing sweep durations for the minimal MOT QUBO problem (Sec. \ref{subsec:min_qubo_problem}). We compare the performance of the annealing protocol considered in the main text (Eq. \ref{eq:modified_annealing}), encoding the problem exclusively into the two-body interactions using 5 qubits overall (orange), versus the direct encoding utilizing local fields with 4 qubits overall (blue). The annealing time is given in units of the inverse characteristic energy scale $J$. The vertical red dashed line denotes the annealing time considered for all numerical simulations in the main text ($2.6 J^{-1}$).}
    \label{fig:alt_protocol_comp}
\end{figure}

We now compare the performance of this alternative protocol against the method employed in the main text when applied to the minimal MOT problem (Sec.~\ref{subsec:min_qubo_problem}). The original protocol requires five qubits, including an additional ancilla qubit for the encoding, while the alternative approach uses only four qubits with local field terms.

Figure~\ref{fig:alt_protocol_comp} presents the final noiseless fidelities achieved for both protocols across varying annealing sweep durations, expressed in units of the inverse characteristic energy scale $J$. In the main text we considered a sweep duration of $2.6 J^{-1}$ for numerical calculations, corresponding to 100~ms when assuming $J = 26$~Hz (indicated by the vertical red dashed line). At this annealing time, both protocols achieve nearly identical final fidelities ($\approx 84\%$). 
More generally, the final fidelities exhibit oscillatory behavior with annealing time, resulting in alternating advantages between the two protocols at different sweep durations, but without significant qualitative differences.

We also investigated whether the dynamical decoupling requirements differ between the two protocols by examining the number of control pulses needed to mitigate the noise of a given amplitude. Simulations of the four-qubit protocol yield results very similar to those obtained with the five-qubit approach (Fig.~\ref{fig:2peaks_all_corr}). In both cases, approximately 2.5 control pulses per millisecond suffice to achieve final fidelities above 70\% for the majority of annealing runs. This consistency in dynamical decoupling requirements suggests that the choice between protocols can be made primarily based on qubit resource constraints rather than noise mitigation and performance considerations.

\section{MOT problem instance with 9 qubits}\label{app:MOT8qubit}

\begin{figure*}[!t] 
\centering
\begin{equation*}
H_{\mathrm{cost}} / J = 
        \begin{pmatrix}
         0 & -0.96 & -0.55 & -0.43 & -0.99 & -0.84 & -0.89 & -0.38 & -0.97\\
         0 & 0 & 1 & 1 & 0 & -0.19 & 0 & -0.04 & 0\\
         0 & 0 & 0 & 0 & 1 & 0 & -0.19 & 0 & -0.04\\
         0 & 0 & 0 & 0 & 1 & -0.05 & 0 & -0.18 & 0\\
         0 & 0 & 0 & 0 & 0 & 0 & -0.05 & 0 & -0.18\\
         0 & 0 & 0 & 0 & 0 & 0 & 1 & 1 & 0\\
         0 & 0 & 0 & 0 & 0 & 0 & 0 & 0 & 1\\
         0 & 0 & 0 & 0 & 0 & 0 & 0 & 0 & 1\\
         0 & 0 & 0 & 0 & 0 & 0 & 0 & 0 & 0
        \end{pmatrix}
\end{equation*}
\caption{Cost Hamiltonian matrix $H_{\mathrm{cost}}$ for the MOT QUBO problem corresponding to an added third frame in Fig.~\ref{fig:MOTCityCenter}. Encoding this problem only into two-body interactions requires 9 qubits.}
\label{fig:8qubit_matrix}
\end{figure*}

\begin{figure}
    \centering
    \includegraphics[width=1\linewidth]{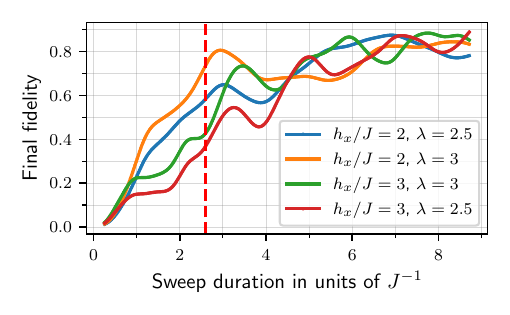}
    \caption{Final fidelities achieved for different total linear annealing sweep durations for the MOT QUBO problem with 9 qubits. We compare the (noiseless) performance of the annealing protocol when choosing different hyperparameters, namely the strength of the penalty term $\lambda$ in the QUBO problem (Eq.~\eqref{eq:MOT_qubo}) as well as the strength of the driver Hamiltonian $h_x/J$. The vertical red dashed line denotes the sweep time $2.6 J^{-1}$ used in our numerical simulations in the main text. 
    In an intermediate regime of medium-slow sweeps, it can be beneficial to carefully choose these hyperparameters, either through a variation of test runs or heuristic methods for finding optimal penalty parameters. 
    }
    \label{fig:8qubit_optimal_parameters}
\end{figure}

\begin{figure*}[htb!]
    \centering        \includegraphics{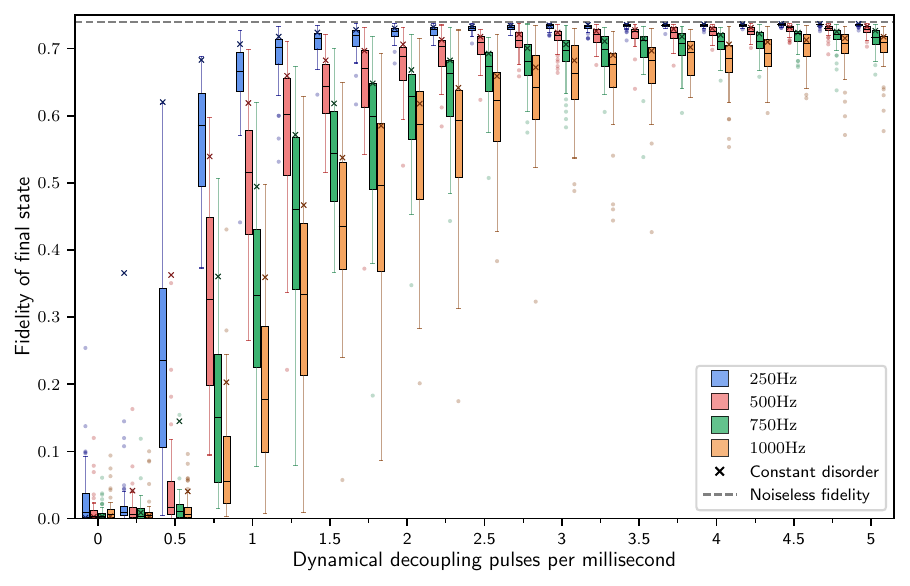}
    \caption{Fidelity of the final state, compared to the state encoding the problem solution, after a quantum annealing sweep versus the applied dynamical decoupling pulse rate during the protocol (see Sec. \ref{sec:annealing}), using the MOT test problem with $9$ qubits (see Fig. \ref{fig:8qubit_matrix}) and an annealing time of $2.6/J$ (corresponding to a sweep duration of $100$ms for an energy scale of $J = 26$ Hz). Box plots compare results for correlated noise using identical noise spectra with two Lorentzian peaks at $50$~Hz and $150$~Hz (see Appendix \ref{app:methods}) but different amplitudes of $250$ Hz (blue), $500$ Hz (red), $750$ Hz (green), and $1000$ Hz (orange). The crosses in each boxplot give a comparison with the median value of annealing sweeps subject to constant disorder ($\delta h^z(t) = \delta h^z$) of the same respective amplitude. The horizontal gray dashed line shows the achievable fidelity of the ideal noiseless protocol ($\approx 0.74$). Data for each boxplot corresponds to $50$ random independent noise realizations (see Appendix \ref{app:methods}). We observe that the achieved final fidelity converges to the noiseless case for a sufficient rate of dynamical decoupling pulses, with higher noise amplitudes requiring higher rates.}
    \label{fig:8qubits_2peaks_all_corr}
\end{figure*}

The main text employed a minimal 5-qubit MOT instance as a reference problem (Sec.~\ref{subsec:min_qubo_problem}) to evaluate our dynamical decoupling noise mitigation protocol (Sec.~\ref{sec:dd}). To demonstrate the scalability of this approach, we now extend the analysis to the next largest MOT QUBO instance, which requires 9 qubits in our encoding (Sec.~\ref{sec:annealing}). We present the explicit QUBO matrix structure and show that the number of dynamical decoupling pulses required for noise suppression remains similar to that for the smaller instance.

\textit{Problem construction ---}  The 9-qubit instance is generated by adding a third frame to the video sequence used for the minimal problem (Fig.~\ref{fig:MOTCityCenter}). Following the same pattern as before, we consider the first, third, and fifth frame of the video, which yields a 9-qubit QUBO formulation when encoded exclusively through two-body interactions. In the construction of the QUBO problem we rescale the off-diagonal elements of the objective by a factor of $0.5$, which corresponds to giving a larger weight to the overlap with the detections in the first fixed frame encoded in the diagonal elements (Eq.~\eqref{eq:mot_objective}). For this particular instance we observe slightly increased annealing performance with this rescaling and the remaining chosen parameters (see below). Figure~\ref{fig:8qubit_matrix} displays the resulting $9 \times 9$ QUBO matrix, where entries equal to one correspond to constraint terms and the remaining entries encode the objective function.

\textit{Parameter optimization ---}  The optimal penalty factor $\lambda$ for the QUBO problem (see Eq.~\eqref{eq:MOT_qubo}) varies between problem instances and must be determined individually. For this 9-qubit problem, we select $\lambda = 3$, which provides good annealing performance at the reference time of $2.6 J^{-1}$ used throughout the main text. Similarly, the optimal transverse driving strength $h_x$ (Eq. \eqref{eq:Hdrive} differs from the minimal problem: while the 5-qubit instance performed well with $h_x/J = 3$, we use $h_x/J = 2$ here to achieve a final noiseless fidelity of approximately 0.74.

Figure~\ref{fig:8qubit_optimal_parameters} shows the final noiseless fidelities for various combinations of driving strength and penalty factor across different sweep durations (with $2.6 J^{-1}$ marked by the red vertical dashed line). The parameter landscape reveals that optimal parameter choices can vary with annealing time, and no single combination remains optimal across all durations. In experimental practice, multiple annealing runs with varied parameters would be performed, potentially guided by heuristic methods for penalty factor estimation~\cite{Alessandroni2023}.

For the present analysis, we employ $h_x/J = 2$ and $\lambda = 3$ without further optimization, as these parameters yield sufficiently good results for our primary objective: demonstrating the efficacy of the dynamical decoupling protocol for larger problem instances.

\textit{Dynamical decoupling ---} Figure~\ref{fig:8qubits_2peaks_all_corr} presents our numerical results for the 9-qubit problem under the same correlated local field noise conditions discussed in Sec.~\ref{sec:dd}. The time-dependent noise is sampled from a spectrum with two dominant frequency peaks at 50~Hz and 150~Hz, analogous with the 5-qubit analysis (Fig.~\ref{fig:2peaks_all_corr}). We assume the same energy scale of $J = 26$~Hz to enable direct comparison with the main text results, though we note that in trapped ion implementations, coupling strengths typically decrease for longer ion chains, leading to lower effective energy scales for larger problems \cite{Nagies2025}. 

The $9$-qubit problem behaves similar to its 5-qubit counterpart. Final fidelities converge toward the ideal noiseless performance (gray dashed line) as the number of control pulses increases. Importantly, pulse rates of approximately $2.5$ pulses per millisecond again prove sufficient to achieve good performance, with final fidelities exceeding 60\% for the majority of annealing runs across all considered noise amplitudes. These results demonstrate that our dynamical decoupling protocol is robust and shows little dependence on the specific problem instance and system size (further supported by our results on the cutting stock problem and the Sherrington-Kirkpatrick model in Appendices \ref{app:cuttingstock} and \ref{app:sk_model}). The pulse requirements depend primarily on the noise amplitude and, to a lesser extent, the noise spectrum characteristics. This scaling behavior aligns with our analytical predictions from Sec.~\ref{subsec:analytics}, where corrections to the ideal annealing protocol are shown to be independent of the specific coupling strengths that encode the optimization problem.

\textit{Scaling to larger instances ---} 
It is interesting to study how the number of required qubits increases when considering a larger number of frames and detections in the MOT problem.
The number of binary variables for larger object tracking problems in this formulation can be obtained from Eq.~\eqref{eq:sol_vector}. If we have a number of $T$ tracks on a sequence of $F$ frames, each with the same number of detections $D_f$ that belong to the same persons, one would need $FD_fT$ binary variables, i.e., the number of qubits needed scales linearly with the problem parameters. 

For larger sequences, it is also beneficial to have a maximal frame gap, so that we stop calculating similarity scores for frames that are far apart from each other. 
An advantage of this method is that if a track for a person terminates and at some frame, which is further away than the maximal frame gap, a new person enters the frame, then one can use the same ID \cite{Zaech2022}.
The two persons are then separated in a post-processing phase. 
Some benchmark results showing how classical solvers perform on the more challenging MOT20 dataset is given in \cite{motchallengeChallengeResults}.

\section{Cutting stock problem}\label{app:cuttingstock}

In Sec.~\ref{sec:MOT} and Appendix \ref{app:MOT8qubit}, we discussed the Multiple Object Tracking problem, for which we discussed the QUBO formulation and the construction of minimal instances suitable as test problems. In this Appendix, we give a brief analogous overview over the industrially relevant cutting stock problem \cite{kantorovich1960mathematical}. The one-dimensional (1D) cutting stock problem asks to minimize the waste when cutting a number $d_i$ of bar stock with required lengths $l_i$ from a number $N$ of larger bars with a fixed length $L$. We want to optimize the assignment $x_{i,k}\in \mathbb{Z} $ of the pieces to be cut out for the demand type $i\in \{1,...,m\} $ to the different larger bar stock $\{1,...,Q\}$ as

\begin{equation}
\argmax_{x\in \mathbb{Z}^{ N}}  \sum_{i=1}^m \sum_{k=1}^Q l_i x_{i,k} ,\label{eq:CuttingStock}
\end{equation}

with the constraints 

\begin{align}
\sum_{i=1}^m  l_i x_{i,k}&\leq L \quad \forall k \in \{1,...,Q\}, \label{eq:CuttingStockSize}\\
\sum_{k=1}^Q   x_{i,k}&\leq d_i  \quad \forall i \in \{1,...,m\}.\label{eq:CuttingStockDemand}
\end{align}

In this description, all feasible assignments have the same objective energy.
In order to have binary variables for the assignment $x$, we ignore that pieces have a common length and instead model every required piece with an entry $i \in \{1,...,\tilde{m}:= \sum_j d_j\}$ so that every piece is demanded only once and $\tilde{d}:=1$.
The problem is closely related to the binary multiple knapsack problem, for which QUBO formulations have already been proposed and studied \cite{bontekoe2023translating,ohno2024toward,guney2025qubo}.

\textit{Reformulation of inequalities ---} 
An important step for the reformulation of Eq.~\eqref{eq:CuttingStock} into a QUBO problem is to turn all inequality
constraints into equality constraints, which can be achieved by using slack variables $s$:

\begin{equation}
    Ax-b \leq 0 \Leftrightarrow Ax-b=s.
\end{equation}

The slack variables are non-positive and are restricted to values $s_i \in \mathcal{S}_i:= \{ (Ax-b)_i | x \in \mathcal{F}  \}$, where $\mathcal{F}$ is the feasible set for $x$. One usually has to implement integer $s_i$ variables via a binary representation 
\begin{equation}
    s_i= -\sum_{k=0}^{B_{\text{max}}} 2^k s_i^{(k)},
\end{equation}
where $s_i^{(k)}\in\{0,1\}$ and $B_{\text{max}}$ is the minimal number of binary bits needed to represent the possible range of slack variables $s_i$. Note that it does not invalidate the method if the binary decomposition can attain values that are even smaller than the values present in $\mathcal{S}_i$.

After this process the inequality is effectively reformulated as an equality, which can be integrated in the QUBO problem via penalty terms as in Eq.~\ref{eq:MOT_qubo}. Lower bounds for the penalty parameters of the multi-knapsack problem have already been computed, which can be directly applied to our setting \cite{quintero2021characterizing}. 
Alternatives to implement inequalities without additional slack variables are, e.g, the unbalanced penalization formulations \cite{montanez2024unbalanced} or by the use of an ancillary qudit in discretized quantum-optimization schemes~\cite{Bottarelli2025}. 

 The slack variables contribute the most to the scaling in dimension. 
For each inequality, one estimates the largest value the slack variable can attain and then chooses the corresponding number of bits in the binary representation. Due to the inequality given by Eq.~\eqref{eq:CuttingStockSize}, we would need $1+\lfloor \log_2(L) \rfloor$ slack variables for every bar stock and because of the inequality given by Eq.~\eqref{eq:CuttingStockDemand}, $1+\lfloor \log_2(d_i) \rfloor$ slack variables for every demanded piece $i$. However, the later inequalities can also be incorporated without slack variables as we will show now.

 For problem instances with multiple larger bars, the inequalities in Eq.~\eqref{eq:CuttingStockDemand} become nontrivial. In the case with two bars, they read 
\begin{align}
 x_{1,1} + x_{1,2}&\leq 1 ,\\
 x_{2,1} + x_{2,2}&\leq 1 .
\end{align}

Fortunately, one does not need to use slack variables. Instead, one can just add terms $\lambda (x_{1,1}  x_{1,2} + x_{2,1}  x_{2,2} )$ to the QUBO. This is also described as Transformation \#2 in \cite{glover2018tutorial}. 
The same procedure can also be generalized for larger numbers of bars by adding 
\begin{equation}
 \sum_{i=1}^m \left( \left( \sum_{k=1}^Q x_{i,k} \right) \left( \sum_{k=1}^Q x_{i,k}\right) - \sum_{k=1}^Q x_{i,k}^2\right) . 
\end{equation}

\begin{figure}
    \centering
    \includegraphics[width=\linewidth]{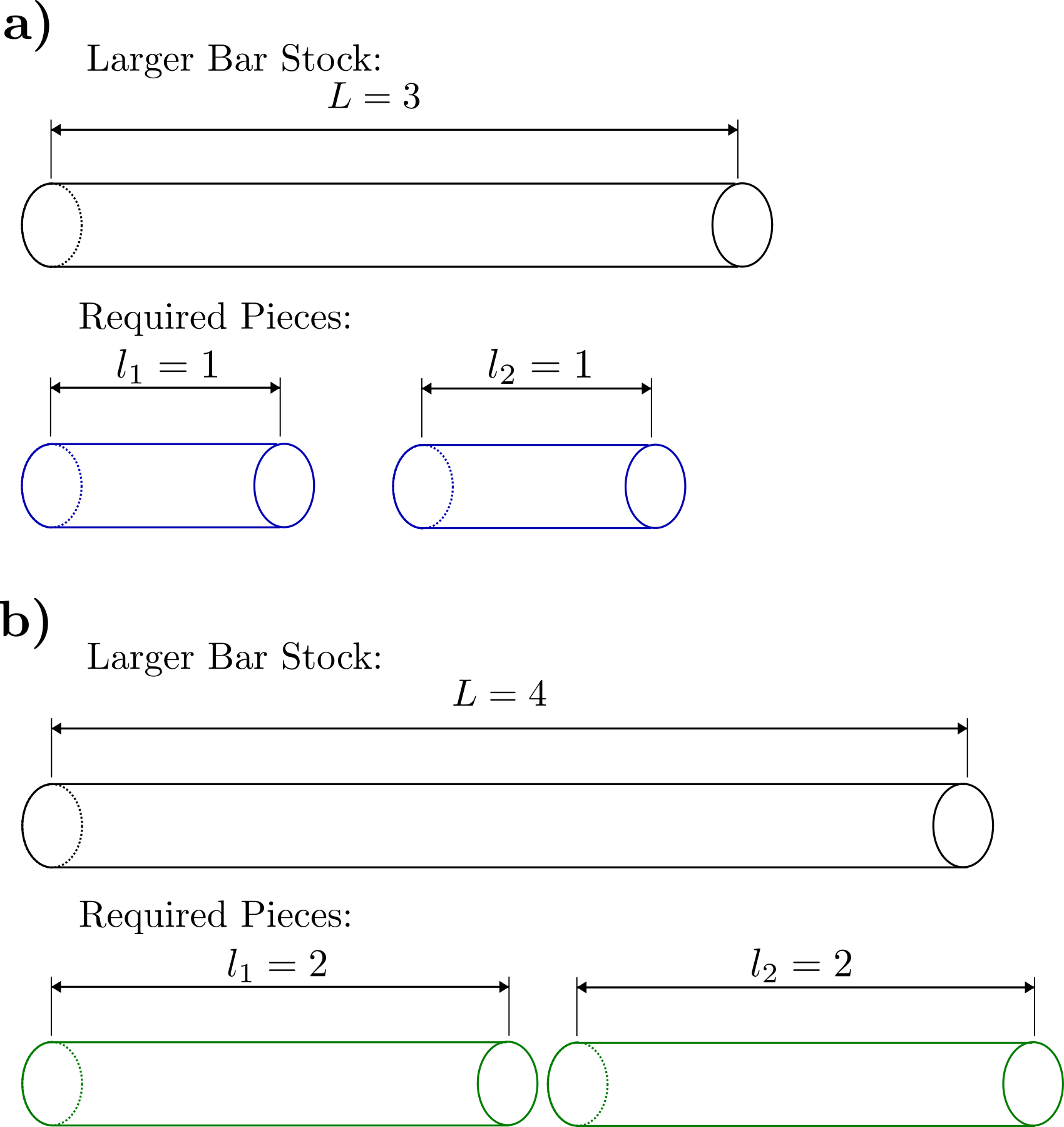}
    \caption{Illustration of the small cutting stock toy problems that we consider for our quantum annealing simulations (see Fig.~\ref{fig:annealing_cutting_stock}). Subfigures a) and b) show problem instances which require four and five qubits, respectively. When encoding the problem into two-body interactions only, an additional ancilla qubit is required (Eq.~\eqref{eq:ancilla_encoding}). In both cases, the trivial optimal solution is that every required piece will be cut out from the larger bar.
    }
    \label{fig:CuttingStockIllustration}
\end{figure}

\textit{Minimal cutting stock problems ---} 
For benchmarking purposes, it is useful to construct 1D cutting stock problems that have a low qubit requirement.
We explicitly construct two trivial cutting stock problems requiring four and five binary variables, respectively, to test our dynamical decoupling protocol. The problems are illustrated in Fig.~\ref{fig:CuttingStockIllustration}. The smaller QUBO problem is constructed to answer the following simple question: Given a single bar of length $L=3$ and required pieces of lengths $l_1=1$ and $l_2=1 $ with demand $d_1= d_2=1$, what should be cut out to maximize the length of cut out pieces? Now the binary variable $x_1:=x_{1,1}$ indicates if the first piece will be cut out and $x_2:=x_{2,1}$ indicates if the second piece will be cut out. The inequalities defined in Eq.~\eqref{eq:CuttingStockDemand} are trivially fulfilled, since there is only one larger bar stock. 
There is only one length constraint (Eq.~\eqref{eq:CuttingStockSize}). It reads

\begin{equation}
    1x_{1}+ 1 x_{2}\leq 3. 
\end{equation}

At the next step, we look at the smallest value that $1x_{1}+ 1 x_{2}- 3$ can attain. This is $-3$ if both binary variables are zero.
Therefore, we will need $1+\lfloor\log_2( 3) \rfloor =2$ slack variables. Using binary representation, we obtain the linear equality

\begin{equation}
\begin{pmatrix}
    1 & 1 &1&2\\
\end{pmatrix}
\begin{pmatrix}
    x_1\\
    x_2\\
    s_1\\
    s_2  
\end{pmatrix}
    = 
    \begin{pmatrix}
    3 
\end{pmatrix}.\label{eq:SingleSizeConst}
\end{equation}

The matrix incorporating the constraints in the QUBO problem can now be obtained as $ A^T A - \textnormal{diag}(A^Tb+ b^T A) $, if the equality constraint in Eq.~\eqref{eq:SingleSizeConst} is denoted as $Ax=b$. This calculation results in the matrix

\begin{equation}
    Q_{\text{const.}}= \lambda
\begin{pmatrix}
   -5&  1&  1&  2\\
 1& -5&  1&  2\\
 1&  1& -5&  2\\
 2&  2&  2& -8
\end{pmatrix}.\label{eq:Qconst}
\end{equation}

The objective part has to be included as a matrix with diagonal $W= (-1,-1,0,0)$ in the QUBO matrix. Converting the QUBO with $\lambda= 1$ (chosen for good annealing performance) to an Ising problem and including an ancilla qubit to avoid bias terms yields the coupling matrix 

\begin{equation}\label{eq:5qubit_CS}
H_{\mathrm{cost}} / J  = 
        \begin{pmatrix}
         0 & -1 & -1 & -0.5 & -1\\
         0 & 0 & 0.5 & 0.5 & 1\\
         0 & 0 & 0 & 0.5 & 1\\
         0 & 0 & 0 & 0 & 1\\
         0 & 0 & 0 & 0 & 0
        \end{pmatrix}.
\end{equation}

The slightly larger problem constructed using five binary variables answers the following question: Given a bar of length $L=4$ and required pieces of lengths $l_1=2$ and $l_2=2 $ with demand $d_1= d_2=1$, what should be cut out to maximize the length of cut out pieces? 
We still have two binary variables $x_i$, which determine if the piece $i$ is cut out or not.
The matrix describing the size constraints can be constructed in a similar fashion as above.
Since we have changed the length to $L=4$, we need $1+\lfloor\log_2( 4) \rfloor =3$ slack variables, resulting in the equality 

\begin{equation}
\begin{pmatrix}
    2 & 2 &1&2&4\\
\end{pmatrix}
\begin{pmatrix}
    x_1\\
    x_2\\
    s_1\\
    s_2 \\
    s_3
\end{pmatrix}
    = 
    \begin{pmatrix}
    4 
\end{pmatrix}.\label{eq:SingleSizeConst2}
\end{equation}

In the same way as in Eq.~\eqref{eq:Qconst}, this equality yields the contribution 

\begin{equation}
    Q_{\text{const.}}= \lambda\begin{pmatrix}-12&   4&   2&   4&   8\\
          4& -12&   2&   4&   8\\
          2&   2&  -7&   2&   4\\
          4&   4&   2& -12&   8\\
          8&   8&   4&   8& -16
    \end{pmatrix}
\end{equation}

to the QUBO matrix.
The objective part is a diagonal matrix with entries $W=(-2,-2,0,0,0)$.
Finally, choosing the penalty strength $\lambda = 1$, converting to an Ising Problem and adding an ancilla variable yields the cost Hamiltonian 

\begin{equation}\label{eq:6qubit_CS}
H_{\mathrm{cost}} / J  = 
        \begin{pmatrix}
         0 & 0.33 & 0.33 & 0.25 & 0.5 & 1\\
         0 & 0 & 0.33 & 1.67 & 0.33 & 0.67\\
         0 & 0 & 0 & 1.67 & 0.33 & 0.67\\
         0 & 0 & 0 & 0 & 1.67 & 0.33\\
         0 & 0 & 0 & 0 & 0 & 0.67 \\
         0 &  0& 0& 0 & 0 & 0
        \end{pmatrix}.
\end{equation}

\begin{figure}
    \centering
    \includegraphics[width=1\linewidth]{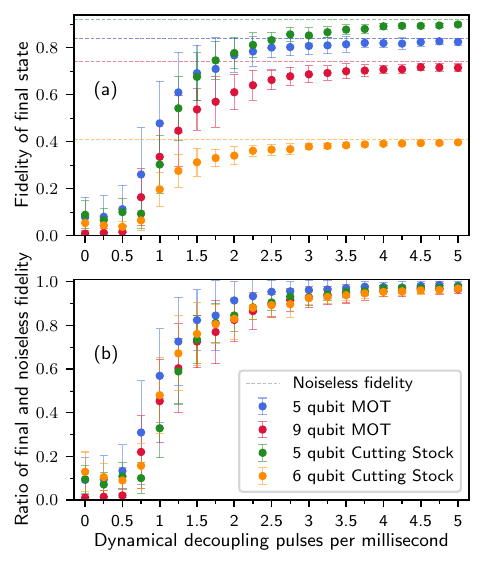}
    \caption{Annealing results for MOT instances with five and nine qubits (see Sec.~\ref{sec:MOT} and Appendix \ref{app:MOT8qubit}), and for cutting stock instances with five and six qubits (see Fig.~\ref{fig:CuttingStockIllustration}), respectively. The annealing time is set to $2.6/J$ for the MOT problems and $26/J$ for the cutting stock problems. A physical sweep time of $100$ ms is assumed for all annealing runs, corresponding to an energy scale of $J=26$ Hz for the MOT problems and $J=260$ Hz for the cutting stock problems. Each data point represents the mean over 50 independent correlated-noise realizations, all sampled from the same noise spectrum as in Figs.~\ref{fig:2peaks_all_corr} and \ref{fig:8qubits_2peaks_all_corr}, with a fixed amplitude of 750 Hz. Error bars denote the standard deviation. (a) Final-state fidelity as a function of the rate of applied dynamical decoupling pulses. Dashed lines indicate the corresponding noiseless final fidelities. (b) Fidelities normalized to their respective noiseless values. Within error bars, all data collapse onto a single curve, showing the generality of the dynamical-decoupling procedure.}
    \label{fig:annealing_cutting_stock}
\end{figure}

\textit{Dynamical decoupling ---}
Figure~\ref{fig:annealing_cutting_stock} presents numerical quantum annealing results for cutting stock instances with 5 and 6 qubits, and MOT instances with 5 and 9 qubits, all subject to identical correlated local-field noise. The time-dependent noise is sampled from a spectrum characterized by two dominant Lorentzian peaks at 50 Hz and 150 Hz, each with a fixed amplitude of 750 Hz. The MOT data correspond to the mean values of the results shown in Figs.~\ref{fig:2peaks_all_corr} and \ref{fig:8qubits_2peaks_all_corr}.

For the cutting stock problems, we generally observe much smaller minimum energy gaps during the annealing sweep than for the MOT instances of similar size (this might be due to the presence of the slack variables; a strong improvement in performance was observed, e.g., in Ref.~\cite{Bottarelli2025} when using a slack-variable free implementation). To still achieve reasonable final fidelities and clearly demonstrate the efficacy of our dynamical decoupling protocol, we set the dimensionless annealing time to $26/J$ for the cutting stock instances, while retaining $2.6/J$ for the MOT cases. To enable a direct comparison between both problem types, we assume an energy scale of $J=260$ Hz for the cutting stock problems and $J=26$ Hz for the MOT problems. These choices yield the same physical annealing duration of 100 ms in both cases, and therefore identical rates of dynamical decoupling pulses per millisecond.

The results in Fig.~\ref{fig:annealing_cutting_stock}(a) show that the final-state fidelities approach the noiseless limits (dashed curves) as the rate of dynamical decoupling pulses increases. In particular, pulse rates of approximately $2.5$ pulses per millisecond are sufficient to achieve good relative performance, with the final fidelity at the end of the sweep depending on the specific problem instance. 

Figure~\ref{fig:annealing_cutting_stock}(b) presents the same data normalized to the corresponding noiseless fidelities, revealing a qualitatively and quantitatively similar behavior. This collapse demonstrates that the applied dynamical-decoupling protocol is robust and largely insensitive to the specific optimization problem.

\section{Sherrington-Kirkpatrick model}\label{app:sk_model}

While the main text focused on application-motivated optimization problems, here we demonstrate that dynamical decoupling (DD) equally mitigates local-field noise in problems without internal structure. As a representative example, we consider the Sherrington–Kirkpatrick (SK) spin-glass model with local fields \cite{Sherrington1975, Panchenko2013}, and present additional numerical results supporting that the DD pulse rate required to suppress noise is largely independent of the system size.

As the cost function in the annealing protocol, we use the SK Hamiltonian 
\begin{align}
    H_{\mathrm{SK}} = \sum_{i\neq j} J^{\mathrm{SK}}_{ij} \sigma^z_i \sigma^z_j + \sum_i h^{\mathrm{SK}} \sigma^z_i,
\end{align}
where $J^{\mathrm{SK}}_{ij}$ and $h^{\mathrm{SK}}$ are independently drawn from Gaussian distributions with zero mean and variances $\sigma^2 = 1$ and $\sigma^2 = 0.3$ respectively. As in the main text, local fields are replaced by two-body terms through the introduction of a single ancilla qubit, ensuring that all problem information is contained in the interaction terms only (cf. Sec.~\ref{sec:annealing}). This allows local-field noise to be mitigated via DD without altering the encoded cost function.

\begin{figure}
    \centering
    \includegraphics[width=1\linewidth]{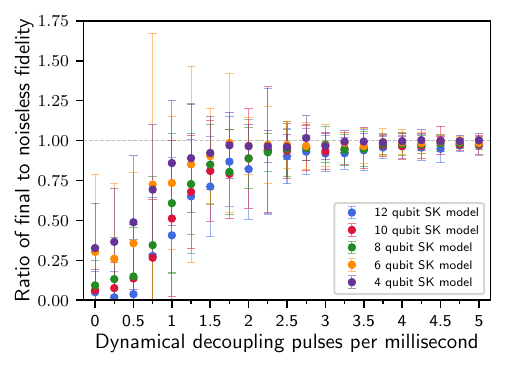}
    \caption{Annealing results for Sherrington-Kirkpatrick (SK) model instances of varying sizes. Annealing time is set to $2.6/J$, corresponding to a physical time of 100 ms when assuming an energy scale of $J=26$ Hz. Each data point corresponds to the mean of 50 random SK instances with different respective correlated-noise realizations. The noise is sampled from the same spectrum as in Figs.~\ref{fig:2peaks_all_corr} and \ref{fig:8qubits_2peaks_all_corr}, defined by Eq.~\eqref{eq:noise_spectrum}, with a fixed noise amplitude of 750 Hz. Error bars denote the standard deviation. The final fidelity achieved after an annealing sweep for each individual problem instance is normalized with respect to the fidelity that instance would achieve without any noise.}
    \label{fig:sk_model}
\end{figure}

Figure \ref{fig:sk_model} summarizes our numerical results. Analogous to the results in the main text, we consider a noise spectrum with two Lorentzian frequency peaks at 50 Hz and 150 Hz (Eq.~\eqref{eq:noise_spectrum}), and a noise amplitude of 750 Hz. We consider the unitless annealing time of $2.6/J$, which corresponds to a physical time of 100 ms, when assuming a characteristic energy scale of $J=26$ Hz. We vary the rate of applied dynamical decoupling pulses and calculate the ratio of the fidelity of the final state after the annealing sweep to the fidelity of the noiseless sweep for each individual problem instance. We consider system sizes ranging from $N=4$ to $N=12$. For each system size and pulse rate, we simulate 50 random SK instances, using a separate noise realization for every run. Data points show the mean ratio of fidelities; error bars denote the standard deviation across problem instances.

At low pulse rates, fidelities are broadly distributed and, for some instances, may even exceed the noiseless reference. This behavior is expected: due to random spectra and varying minimal gaps, certain instances may by chance perform better under the specific noise realization. We also note that smaller system sizes will on average achieve a better final fidelity ratio for small DD pulse rates. This is simply due to the fact that the dimension of the system is small enough for the solution to have a finite amplitude with high likelihood. Nevertheless, for all system sizes we observe a clear convergence to the noiseless fidelity when the DD rate exceeds approximately three pulses per millisecond. The onset of convergence is essentially identical across all system sizes, indicating no significant increase in required DD pulse rate with system size, consistent with our results on the Multiple Object Tracking and cutting stock problems (see Sec.~\ref{sec:dd} and Appendices \ref{app:MOT8qubit} and \ref{app:cuttingstock}). 

This observation also matches our analytical treatment in Sec.~\ref{sec:dd}, which predicts that to leading order the noise enters only through the product of noise amplitude and time interval between consecutive DD pulses, independent of the number of qubits. Our results thus establish that dynamical decoupling is robust with respect to the choice of cost Hamiltonian and system size and continues to suppress local-field noise effectively even for unstructured spin glass problems such as the SK model.

\section{Modulating coupling strengths via dynamical decoupling}\label{app:dd}

In Sec.~\ref{sec:dd}, we discussed how dynamical decoupling control pulses in the form of global spin flips can effectively cancel local field disorder terms in the quantum annealing Hamiltonian (Eq. \ref{eq:annealing_noise}). In this Appendix we explain how a similar scheme of periodically applied global spin flips can also be used to modulate two-body coupling strengths during an annealing protocol effectively \cite{Huber2021, Patrick2024}. This scheme can be potentially utilized to encode QUBO problems into the native all-to-all couplings of the MAGIC setup given by Eq.~\eqref{eq:Hmagic}.

We again consider the time evolution operator of the annealing Hamiltonian during a small time interval $ \Delta t = t_+ - t_- $ with two applied dynamical decoupling control pulses, the first at a time $t_0$ ($t_- < t_0 < t_+$) and the second at time $t_+$ at the end of the interval. In contrast to the calculation in Sec.~\ref{sec:dd}, we now consider time intervals before and after the first pulse that can be of different lengths, i.e., $\Delta t_- = t_0 - t_- \neq t_+ - t_0 = \Delta t_+$. Before and after the control pulses, the system evolves with the usual annealing Hamiltonian composed of driving and cost Hamiltonian, where we again assume that the problem is fully encoded into the two-body interaction terms. For now, we consider the ideal case without any noise. The time evolution operator during the first part of the small time interval (from time $t_-$ to $t_0$) can then be written as

\begin{align}\label{eq:app_first_int}
    U_{t_- \to t_0} &\approx \exp \big(-i \sum_{ij}J_{ij} \sigma_i^z \sigma_j^z \Delta t_{-} -i C(\bar{t}_-) H_{\text{driving}} \Delta t_{-}\big),
\end{align}

where we only keep terms of first order in $\Delta t_-$ and define the intermediate time $\bar{t}_- = (t_- + t_0)/2$.

To modulate the coupling strengths between specific sets of ions, the dynamical decoupling control pulses are no longer global spin flips, but only flip a subset of ions. For simplicity and without loss of generality we assume that each pulse only flips a single qubit ($\sigma_k^x$). The time evolution in between the two control pulses ($U_{t_0 \to t_+}$) has the analogous form of Eq.~\eqref{eq:app_first_int}, and the evolution over the entire time interval can be calculated to first order as

\begin{align}\label{eq:app_J_mod}
      \sigma_k^x U_{t_0 \to t_+} \sigma_k^x U_{t_- \to t_0} &\approx \exp\bigg(-i \sum_{i,j \neq k}J_{ij} \sigma_i^z \sigma_j^z \Delta t \nonumber \\
     &-i \sum_{i \neq k}J_{ik} \sigma_i^z \sigma_j^z (\Delta t_- - \Delta t_{+}) \nonumber \\
     &-i C(\Bar{t}) H_{\text{drive}} \Delta t \bigg),
\end{align}
with $\Bar{t} = (t_+ + t_-)/2$.
All two-body interactions with qubit $k$ can in this way be uniformly modified by changing the relative lengths of time intervals $\Delta t_{-}$ and $\Delta t_{+}$ within the overall time interval $\Delta t = \Delta t_{-} + \Delta t_{+}$.

Generalizing this setting to include local terms $\delta h_i^z$, we would simultaneously modify the strength of the local field experienced by ion $k$: $\delta h_k^z \to \delta h_k^z (\Delta t_- - \Delta t_+)/\Delta t$. If these local fields encode the optimization problem to be solved (as in the protocol discussed in Appendix \ref{app:alt_protocol}), this modification has to be taken into account and the field strength adjusted accordingly.
If, instead, the local fields are undesired disorder terms, they can be effectively mitigated in the same manner as discussed in Sec.~\ref{sec:dd}. To that end, one can interleave both the global spin flips and the spin flips acting on a subset of qubits. For example, defining $U_{\Delta t_1} =  \sigma_k^x U_{{t_{1}} \to t_{1,+}} \sigma_k^x U_{t_{1,-} \to t_{1}}$ ($\Delta t_{1} = t_{1,+} - t_{1,-}$) and similarly $U_{\Delta t_2}$ as two time evolution operators following Eq.~\eqref{eq:app_J_mod}, one can consider the evolution $ \sigma^x U_{\Delta t_2} \sigma^x U_{\Delta t_1}$ over the total time interval $\Delta t_1 + \Delta t_2$, where $\sigma^x = \bigotimes_i \sigma^x_i$ denotes a global spin flip. By setting $\Delta t_1 = \Delta t_2$ and choosing the relative time intervals between the spin flips appropriately, one can modulate the coupling strengths while simultaneously suppressing local field disorder terms.

\section{Numerical methods}\label{app:methods}

In this Appendix, we explain how the noise spectra that we use in the numerical simulations in the main text (Sec.~\ref{sec:dd}) are defined and local field fluctuation samples are generated. Further, we discuss how the dynamical decoupling pulses are distributed throughout the numerically discretized annealing sweep.

\textit{Generating the noise samples ---} As discussed in Sec.~\ref{subsec:exp_parameters}, realistic trapped-ion experiments typically exhibit local field noise with dominant frequencies at 50~Hz (corresponding to the electrical grid frequency) and its odd harmonics. We model this characteristic noise spectrum numerically using two Lorentzian peaks: 

\begin{align}\label{eq:noise_spectrum}
    S(f) = \frac{1}{2 \pi} \left( \frac{\gamma}{(f - f_1)^2 + \gamma^2} + \frac{\beta}{(f - f_3)^2 + \beta^2} \right),
\end{align}

where $f_1 = 50$~Hz and $f_3 = 150$~Hz represent the fundamental and third harmonic frequency, respectively. The parameters $\gamma$ and $\beta$ denote the corresponding half-widths at half-maximum. To reflect typical experimental conditions where higher harmonics have reduced amplitude, we set the second peak to one-third of the intensity of the first by choosing $\beta = 3\gamma$.

We generate noise samples using the approximate frequency domain method of Ref.~\cite{percival1992}. The sampling covers $N = 50\,000$ points over 100~ms, corresponding to our in the main text assumed energy scale of $J = 26$~Hz and chosen annealing duration $2.6 J^{-1}$. Following the sampling method's requirements, we evaluate the spectrum at $M = 3N$ equidistant frequencies to ensure an accurate representation of the noise spectrum.

\begin{figure}
    \centering
    \includegraphics[width=1\linewidth]{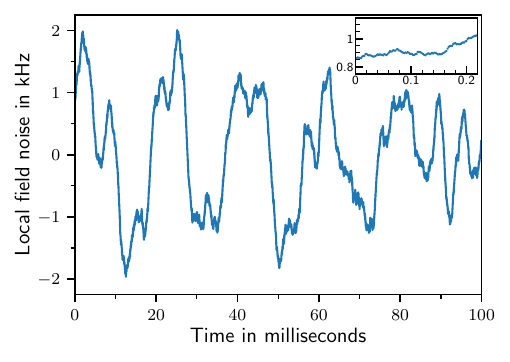}
    \caption{Example of a time trace of local field noise with 50000 sampling points over 100 milliseconds, generated from a noise spectrum with two Lorentzian peaks at 50~Hz and 150~Hz (Eq.~\ref{eq:noise_spectrum}) with a noise amplitude of 750~Hz. The inset shows a zoom onto the first 0.2 milliseconds (100 samples), which corresponds to the time interval between two dynamical decoupling pulses for the highest pulse rate of 5~kHz we consider in this work (see Sec.~\ref{sec:dd}).}
    \label{fig:noise_samples}
\end{figure}

This discretization of the spectrum necessitates sufficiently broad Lorentzian peaks, leading us to choose $\gamma = 3$~Hz. While experimental noise peaks are typically much narrower, our analysis in Fig.~\ref{fig:1peak_all_corr} demonstrates that the dynamical decoupling protocol exhibits robust performance across different noise frequencies. We therefore expect this approximation to have a negligible impact on our numerical results for the annealing sweeps.
The samples for the noise spectra with a single Lorentzian peak (again setting $\gamma = 3$~Hz), utilized for the simulation results in Fig.~\ref{fig:1peak_all_corr}, are generated analogously to the method outlined above.

In Fig.~\ref{fig:noise_samples}, we show an example of the time evolution of a noisy local field with $50\,000$ sampling points, generated for the spectrum with two Lorentzian peaks at 50~Hz and 150~Hz and a noise amplitude of 750 Hz. The main figure shows the local field noise over the entire time evolution of 100 milliseconds (corresponding to the time of an annealing sweep in the main text), while the inset zooms in on the first 0.2 milliseconds. This smaller time interval  corresponds to the time between two dynamical decoupling pulses for the highest pulse rate of 5~kHz we consider in the main text (see Sec.~\ref{sec:dd}) and is resolved into 100 distinct noise sampling points. 

\textit{Numerical simulation time scales ---}
Our numerical implementation involves two distinct timescales that require careful treatment. The annealing sweep is discretized into $50\,000$ time steps, each lasting $100\,\text{ms} / 50\,000 = 0.002\,\text{ms}$ for our chosen parameters. This fine discretization ensures a sufficient number of noise samples between dynamical decoupling pulses while approximating the continuous time-dependent fluctuations. For our highest considered pulse rate of 500 dynamical decoupling pulses per sweep, this provides 100 distinct noise samples between consecutive pulses.

Since we maintain a fixed number of noise samples (50\,000) while varying the number of dynamical decoupling pulses, achieving perfectly uniform pulse spacing is not always possible. 
For example, with 300 pulses over the total sweep duration, the ideal spacing would be on average $50\,000/300 \approx 166.67$ steps per interval. We implement this by using 166 steps between pulses for the first portion of the sweep and 167 steps for the remainder, ensuring the total remains exactly 50\,000 steps.
Our numerical tests demonstrate that small, random variations in pulse spacing have negligible impact on final fidelity, provided the pulses remain approximately equidistant (see discussion in Sec.~\ref{subsec:dd_time_evolution}). Alternative periodic spacing patterns such as (166, 167, 167, 166, 167, 167, ...) also yield essentially identical results to the distribution described above.

However, other systematic alternating patterns can cause significant performance degradation. For instance, with 400 pulses over 50\,000 steps, uniform spacing of 125 steps between pulses recovers the ideal noiseless fidelity at the end of the sweep. In contrast, an alternating pattern of (124, 126, 124, 126, ...) results in a vanishing final fidelity. This failure occurs because the systematic alternation allows the small timing errors to accumulate coherently throughout the annealing process, ultimately destroying the effectiveness of the dynamical decoupling scheme.

%% file: main.bbl
\begin{thebibliography}{100}%
\makeatletter
\providecommand \@ifxundefined [1]{%
 \@ifx{#1\undefined}
}%
\providecommand \@ifnum [1]{%
 \ifnum #1\expandafter \@firstoftwo
 \else \expandafter \@secondoftwo
 \fi
}%
\providecommand \@ifx [1]{%
 \ifx #1\expandafter \@firstoftwo
 \else \expandafter \@secondoftwo
 \fi
}%
\providecommand \natexlab [1]{#1}%
\providecommand \enquote  [1]{``#1''}%
\providecommand \bibnamefont  [1]{#1}%
\providecommand \bibfnamefont [1]{#1}%
\providecommand \citenamefont [1]{#1}%
\providecommand \href@noop [0]{\@secondoftwo}%
\providecommand \href [0]{\begingroup \@sanitize@url \@href}%
\providecommand \@href[1]{\@@startlink{#1}\@@href}%
\providecommand \@@href[1]{\endgroup#1\@@endlink}%
\providecommand \@sanitize@url [0]{\catcode `\\12\catcode `\$12\catcode `\&12\catcode `\#12\catcode `\^12\catcode `\_12\catcode `\%12\relax}%
\providecommand \@@startlink[1]{}%
\providecommand \@@endlink[0]{}%
\providecommand \url  [0]{\begingroup\@sanitize@url \@url }%
\providecommand \@url [1]{\endgroup\@href {#1}{\urlprefix }}%
\providecommand \urlprefix  [0]{URL }%
\providecommand \Eprint [0]{\href }%
\providecommand \doibase [0]{https://doi.org/}%
\providecommand \selectlanguage [0]{\@gobble}%
\providecommand \bibinfo  [0]{\@secondoftwo}%
\providecommand \bibfield  [0]{\@secondoftwo}%
\providecommand \translation [1]{[#1]}%
\providecommand \BibitemOpen [0]{}%
\providecommand \bibitemStop [0]{}%
\providecommand \bibitemNoStop [0]{.\EOS\space}%
\providecommand \EOS [0]{\spacefactor3000\relax}%
\providecommand \BibitemShut  [1]{\csname bibitem#1\endcsname}%
\let\auto@bib@innerbib\@empty
\bibitem [{\citenamefont {Ladd}\ \emph {et~al.}(2010)\citenamefont {Ladd}, \citenamefont {Jelezko}, \citenamefont {Laflamme}, \citenamefont {Nakamura}, \citenamefont {Monroe},\ and\ \citenamefont {O’Brien}}]{Ladd2010}%
  \BibitemOpen
  \bibfield  {author} {\bibinfo {author} {\bibfnamefont {T.~D.}\ \bibnamefont {Ladd}}, \bibinfo {author} {\bibfnamefont {F.}~\bibnamefont {Jelezko}}, \bibinfo {author} {\bibfnamefont {R.}~\bibnamefont {Laflamme}}, \bibinfo {author} {\bibfnamefont {Y.}~\bibnamefont {Nakamura}}, \bibinfo {author} {\bibfnamefont {C.}~\bibnamefont {Monroe}},\ and\ \bibinfo {author} {\bibfnamefont {J.~L.}\ \bibnamefont {O’Brien}},\ }\href {https://doi.org/10.1038/nature08812} {\bibfield  {journal} {\bibinfo  {journal} {Nature}\ }\textbf {\bibinfo {volume} {464}},\ \bibinfo {pages} {45} (\bibinfo {year} {2010})}\BibitemShut {NoStop}%
\bibitem [{\citenamefont {Fedorov}\ \emph {et~al.}(2022)\citenamefont {Fedorov}, \citenamefont {Gisin}, \citenamefont {Beloussov},\ and\ \citenamefont {Lvovsky}}]{Fedorov2022}%
  \BibitemOpen
  \bibfield  {author} {\bibinfo {author} {\bibfnamefont {A.~K.}\ \bibnamefont {Fedorov}}, \bibinfo {author} {\bibfnamefont {N.}~\bibnamefont {Gisin}}, \bibinfo {author} {\bibfnamefont {S.~M.}\ \bibnamefont {Beloussov}},\ and\ \bibinfo {author} {\bibfnamefont {A.~I.}\ \bibnamefont {Lvovsky}},\ }\href {https://doi.org/10.48550/ARXIV.2203.17181} {\bibinfo {title} {Quantum computing at the quantum advantage threshold: a down-to-business review}} (\bibinfo {year} {2022})\BibitemShut {NoStop}%
\bibitem [{\citenamefont {Beck}\ \emph {et~al.}(2023)\citenamefont {Beck}, \citenamefont {Carlson}, \citenamefont {Davoudi}, \citenamefont {Formaggio}, \citenamefont {Quaglioni}, \citenamefont {Savage}, \citenamefont {Barata}, \citenamefont {Bhattacharya}, \citenamefont {Bishof}, \citenamefont {Cloet}, \citenamefont {Delgado}, \citenamefont {DeMarco}, \citenamefont {Fink}, \citenamefont {Florio}, \citenamefont {Francois}, \citenamefont {Grabowska}, \citenamefont {Hoogerheide}, \citenamefont {Huang}, \citenamefont {Ikeda}, \citenamefont {Illa}, \citenamefont {Joo}, \citenamefont {Kharzeev}, \citenamefont {Kowalski}, \citenamefont {Lai}, \citenamefont {Leach}, \citenamefont {Loer}, \citenamefont {Low}, \citenamefont {Martin}, \citenamefont {Moore}, \citenamefont {Mehen}, \citenamefont {Mueller}, \citenamefont {Mulligan}, \citenamefont {Mumm}, \citenamefont {Pederiva}, \citenamefont {Pisarski}, \citenamefont {Ploskon}, \citenamefont {Reddy}, \citenamefont {Rupak}, \citenamefont {Singh}, \citenamefont {Singh},
  \citenamefont {Stetcu}, \citenamefont {Stryker}, \citenamefont {Szypryt}, \citenamefont {Valgushev}, \citenamefont {VanDevender}, \citenamefont {Watkins}, \citenamefont {Wilson}, \citenamefont {Yao}, \citenamefont {Afanasev}, \citenamefont {Balantekin}, \citenamefont {Baroni}, \citenamefont {Bunker}, \citenamefont {Chakraborty}, \citenamefont {Chernyshev}, \citenamefont {Cirigliano}, \citenamefont {Clark}, \citenamefont {Dhiman}, \citenamefont {Du}, \citenamefont {Dutta}, \citenamefont {Edwards}, \citenamefont {Flores}, \citenamefont {Galindo-Uribarri}, \citenamefont {Ruiz}, \citenamefont {Gueorguiev}, \citenamefont {Guo}, \citenamefont {Hansen}, \citenamefont {Hernandez}, \citenamefont {Hattori}, \citenamefont {Hauke}, \citenamefont {Hjorth-Jensen}, \citenamefont {Jankowski}, \citenamefont {Johnson}, \citenamefont {Lacroix}, \citenamefont {Lee}, \citenamefont {Lin}, \citenamefont {Liu}, \citenamefont {Llanes-Estrada}, \citenamefont {Looney}, \citenamefont {Lukin}, \citenamefont {Mercenne}, \citenamefont
  {Miller}, \citenamefont {Mottola}, \citenamefont {Mueller}, \citenamefont {Nachman}, \citenamefont {Negele}, \citenamefont {Orrell}, \citenamefont {Patwardhan}, \citenamefont {Phillips}, \citenamefont {Poole}, \citenamefont {Qualters}, \citenamefont {Rumore}, \citenamefont {Schaefer}, \citenamefont {Scott}, \citenamefont {Singh}, \citenamefont {Vary}, \citenamefont {Galvez-Viruet}, \citenamefont {Wendt}, \citenamefont {Xing}, \citenamefont {Yang}, \citenamefont {Young},\ and\ \citenamefont {Zhao}}]{Beck2023}%
  \BibitemOpen
  \bibfield  {author} {\bibinfo {author} {\bibfnamefont {D.}~\bibnamefont {Beck}}, \bibinfo {author} {\bibfnamefont {J.}~\bibnamefont {Carlson}}, \bibinfo {author} {\bibfnamefont {Z.}~\bibnamefont {Davoudi}}, \bibinfo {author} {\bibfnamefont {J.}~\bibnamefont {Formaggio}}, \bibinfo {author} {\bibfnamefont {S.}~\bibnamefont {Quaglioni}}, \bibinfo {author} {\bibfnamefont {M.}~\bibnamefont {Savage}}, \bibinfo {author} {\bibfnamefont {J.}~\bibnamefont {Barata}}, \bibinfo {author} {\bibfnamefont {T.}~\bibnamefont {Bhattacharya}}, \bibinfo {author} {\bibfnamefont {M.}~\bibnamefont {Bishof}}, \bibinfo {author} {\bibfnamefont {I.}~\bibnamefont {Cloet}}, \bibinfo {author} {\bibfnamefont {A.}~\bibnamefont {Delgado}}, \bibinfo {author} {\bibfnamefont {M.}~\bibnamefont {DeMarco}}, \bibinfo {author} {\bibfnamefont {C.}~\bibnamefont {Fink}}, \bibinfo {author} {\bibfnamefont {A.}~\bibnamefont {Florio}}, \bibinfo {author} {\bibfnamefont {M.}~\bibnamefont {Francois}}, \bibinfo {author} {\bibfnamefont {D.}~\bibnamefont
  {Grabowska}}, \bibinfo {author} {\bibfnamefont {S.}~\bibnamefont {Hoogerheide}}, \bibinfo {author} {\bibfnamefont {M.}~\bibnamefont {Huang}}, \bibinfo {author} {\bibfnamefont {K.}~\bibnamefont {Ikeda}}, \bibinfo {author} {\bibfnamefont {M.}~\bibnamefont {Illa}}, \bibinfo {author} {\bibfnamefont {K.}~\bibnamefont {Joo}}, \bibinfo {author} {\bibfnamefont {D.}~\bibnamefont {Kharzeev}}, \bibinfo {author} {\bibfnamefont {K.}~\bibnamefont {Kowalski}}, \bibinfo {author} {\bibfnamefont {W.~K.}\ \bibnamefont {Lai}}, \bibinfo {author} {\bibfnamefont {K.}~\bibnamefont {Leach}}, \bibinfo {author} {\bibfnamefont {B.}~\bibnamefont {Loer}}, \bibinfo {author} {\bibfnamefont {I.}~\bibnamefont {Low}}, \bibinfo {author} {\bibfnamefont {J.}~\bibnamefont {Martin}}, \bibinfo {author} {\bibfnamefont {D.}~\bibnamefont {Moore}}, \bibinfo {author} {\bibfnamefont {T.}~\bibnamefont {Mehen}}, \bibinfo {author} {\bibfnamefont {N.}~\bibnamefont {Mueller}}, \bibinfo {author} {\bibfnamefont {J.}~\bibnamefont {Mulligan}}, \bibinfo {author}
  {\bibfnamefont {P.}~\bibnamefont {Mumm}}, \bibinfo {author} {\bibfnamefont {F.}~\bibnamefont {Pederiva}}, \bibinfo {author} {\bibfnamefont {R.}~\bibnamefont {Pisarski}}, \bibinfo {author} {\bibfnamefont {M.}~\bibnamefont {Ploskon}}, \bibinfo {author} {\bibfnamefont {S.}~\bibnamefont {Reddy}}, \bibinfo {author} {\bibfnamefont {G.}~\bibnamefont {Rupak}}, \bibinfo {author} {\bibfnamefont {H.}~\bibnamefont {Singh}}, \bibinfo {author} {\bibfnamefont {M.}~\bibnamefont {Singh}}, \bibinfo {author} {\bibfnamefont {I.}~\bibnamefont {Stetcu}}, \bibinfo {author} {\bibfnamefont {J.}~\bibnamefont {Stryker}}, \bibinfo {author} {\bibfnamefont {P.}~\bibnamefont {Szypryt}}, \bibinfo {author} {\bibfnamefont {S.}~\bibnamefont {Valgushev}}, \bibinfo {author} {\bibfnamefont {B.}~\bibnamefont {VanDevender}}, \bibinfo {author} {\bibfnamefont {S.}~\bibnamefont {Watkins}}, \bibinfo {author} {\bibfnamefont {C.}~\bibnamefont {Wilson}}, \bibinfo {author} {\bibfnamefont {X.}~\bibnamefont {Yao}}, \bibinfo {author} {\bibfnamefont
  {A.}~\bibnamefont {Afanasev}}, \bibinfo {author} {\bibfnamefont {A.~B.}\ \bibnamefont {Balantekin}}, \bibinfo {author} {\bibfnamefont {A.}~\bibnamefont {Baroni}}, \bibinfo {author} {\bibfnamefont {R.}~\bibnamefont {Bunker}}, \bibinfo {author} {\bibfnamefont {B.}~\bibnamefont {Chakraborty}}, \bibinfo {author} {\bibfnamefont {I.}~\bibnamefont {Chernyshev}}, \bibinfo {author} {\bibfnamefont {V.}~\bibnamefont {Cirigliano}}, \bibinfo {author} {\bibfnamefont {B.}~\bibnamefont {Clark}}, \bibinfo {author} {\bibfnamefont {S.~K.}\ \bibnamefont {Dhiman}}, \bibinfo {author} {\bibfnamefont {W.}~\bibnamefont {Du}}, \bibinfo {author} {\bibfnamefont {D.}~\bibnamefont {Dutta}}, \bibinfo {author} {\bibfnamefont {R.}~\bibnamefont {Edwards}}, \bibinfo {author} {\bibfnamefont {A.}~\bibnamefont {Flores}}, \bibinfo {author} {\bibfnamefont {A.}~\bibnamefont {Galindo-Uribarri}}, \bibinfo {author} {\bibfnamefont {R.~F.~G.}\ \bibnamefont {Ruiz}}, \bibinfo {author} {\bibfnamefont {V.}~\bibnamefont {Gueorguiev}}, \bibinfo {author}
  {\bibfnamefont {F.}~\bibnamefont {Guo}}, \bibinfo {author} {\bibfnamefont {E.}~\bibnamefont {Hansen}}, \bibinfo {author} {\bibfnamefont {H.}~\bibnamefont {Hernandez}}, \bibinfo {author} {\bibfnamefont {K.}~\bibnamefont {Hattori}}, \bibinfo {author} {\bibfnamefont {P.}~\bibnamefont {Hauke}}, \bibinfo {author} {\bibfnamefont {M.}~\bibnamefont {Hjorth-Jensen}}, \bibinfo {author} {\bibfnamefont {K.}~\bibnamefont {Jankowski}}, \bibinfo {author} {\bibfnamefont {C.}~\bibnamefont {Johnson}}, \bibinfo {author} {\bibfnamefont {D.}~\bibnamefont {Lacroix}}, \bibinfo {author} {\bibfnamefont {D.}~\bibnamefont {Lee}}, \bibinfo {author} {\bibfnamefont {H.-W.}\ \bibnamefont {Lin}}, \bibinfo {author} {\bibfnamefont {X.}~\bibnamefont {Liu}}, \bibinfo {author} {\bibfnamefont {F.~J.}\ \bibnamefont {Llanes-Estrada}}, \bibinfo {author} {\bibfnamefont {J.}~\bibnamefont {Looney}}, \bibinfo {author} {\bibfnamefont {M.}~\bibnamefont {Lukin}}, \bibinfo {author} {\bibfnamefont {A.}~\bibnamefont {Mercenne}}, \bibinfo {author}
  {\bibfnamefont {J.}~\bibnamefont {Miller}}, \bibinfo {author} {\bibfnamefont {E.}~\bibnamefont {Mottola}}, \bibinfo {author} {\bibfnamefont {B.}~\bibnamefont {Mueller}}, \bibinfo {author} {\bibfnamefont {B.}~\bibnamefont {Nachman}}, \bibinfo {author} {\bibfnamefont {J.}~\bibnamefont {Negele}}, \bibinfo {author} {\bibfnamefont {J.}~\bibnamefont {Orrell}}, \bibinfo {author} {\bibfnamefont {A.}~\bibnamefont {Patwardhan}}, \bibinfo {author} {\bibfnamefont {D.}~\bibnamefont {Phillips}}, \bibinfo {author} {\bibfnamefont {S.}~\bibnamefont {Poole}}, \bibinfo {author} {\bibfnamefont {I.}~\bibnamefont {Qualters}}, \bibinfo {author} {\bibfnamefont {M.}~\bibnamefont {Rumore}}, \bibinfo {author} {\bibfnamefont {T.}~\bibnamefont {Schaefer}}, \bibinfo {author} {\bibfnamefont {J.}~\bibnamefont {Scott}}, \bibinfo {author} {\bibfnamefont {R.}~\bibnamefont {Singh}}, \bibinfo {author} {\bibfnamefont {J.}~\bibnamefont {Vary}}, \bibinfo {author} {\bibfnamefont {J.-J.}\ \bibnamefont {Galvez-Viruet}}, \bibinfo {author}
  {\bibfnamefont {K.}~\bibnamefont {Wendt}}, \bibinfo {author} {\bibfnamefont {H.}~\bibnamefont {Xing}}, \bibinfo {author} {\bibfnamefont {L.}~\bibnamefont {Yang}}, \bibinfo {author} {\bibfnamefont {G.}~\bibnamefont {Young}},\ and\ \bibinfo {author} {\bibfnamefont {F.}~\bibnamefont {Zhao}},\ }\href {https://doi.org/10.48550/ARXIV.2303.00113} {\bibinfo {title} {Quantum information science and technology for nuclear physics. input into u.s. long-range planning, 2023}} (\bibinfo {year} {2023})\BibitemShut {NoStop}%
\bibitem [{\citenamefont {Hoefler}\ \emph {et~al.}(2023)\citenamefont {Hoefler}, \citenamefont {Häner},\ and\ \citenamefont {Troyer}}]{Hoefler2023}%
  \BibitemOpen
  \bibfield  {author} {\bibinfo {author} {\bibfnamefont {T.}~\bibnamefont {Hoefler}}, \bibinfo {author} {\bibfnamefont {T.}~\bibnamefont {Häner}},\ and\ \bibinfo {author} {\bibfnamefont {M.}~\bibnamefont {Troyer}},\ }\href {https://doi.org/10.1145/3571725} {\bibfield  {journal} {\bibinfo  {journal} {Communications of the ACM}\ }\textbf {\bibinfo {volume} {66}},\ \bibinfo {pages} {82} (\bibinfo {year} {2023})}\BibitemShut {NoStop}%
\bibitem [{\citenamefont {Scholten}\ \emph {et~al.}(2024)\citenamefont {Scholten}, \citenamefont {Williams}, \citenamefont {Moody}, \citenamefont {Mosca}, \citenamefont {Hurley}, \citenamefont {Zeng}, \citenamefont {Troyer},\ and\ \citenamefont {Gambetta}}]{Scholten2024}%
  \BibitemOpen
  \bibfield  {author} {\bibinfo {author} {\bibfnamefont {T.~L.}\ \bibnamefont {Scholten}}, \bibinfo {author} {\bibfnamefont {C.~J.}\ \bibnamefont {Williams}}, \bibinfo {author} {\bibfnamefont {D.}~\bibnamefont {Moody}}, \bibinfo {author} {\bibfnamefont {M.}~\bibnamefont {Mosca}}, \bibinfo {author} {\bibfnamefont {W.}~\bibnamefont {Hurley}}, \bibinfo {author} {\bibfnamefont {W.~J.}\ \bibnamefont {Zeng}}, \bibinfo {author} {\bibfnamefont {M.}~\bibnamefont {Troyer}},\ and\ \bibinfo {author} {\bibfnamefont {J.~M.}\ \bibnamefont {Gambetta}},\ }\href {https://doi.org/10.48550/ARXIV.2401.16317} {\bibinfo {title} {Assessing the benefits and risks of quantum computers}} (\bibinfo {year} {2024})\BibitemShut {NoStop}%
\bibitem [{\citenamefont {Troyer}\ \emph {et~al.}(2024)\citenamefont {Troyer}, \citenamefont {Benjamin},\ and\ \citenamefont {Gevorkian}}]{Troyer2024}%
  \BibitemOpen
  \bibfield  {author} {\bibinfo {author} {\bibfnamefont {M.}~\bibnamefont {Troyer}}, \bibinfo {author} {\bibfnamefont {E.~V.}\ \bibnamefont {Benjamin}},\ and\ \bibinfo {author} {\bibfnamefont {A.}~\bibnamefont {Gevorkian}},\ }\href {https://doi.org/10.48550/ARXIV.2403.02921} {\bibinfo {title} {Quantum for good and the societal impact of quantum computing}} (\bibinfo {year} {2024})\BibitemShut {NoStop}%
\bibitem [{\citenamefont {Santoro}\ \emph {et~al.}(2002)\citenamefont {Santoro}, \citenamefont {Marto{\v{n}}\'ak}, \citenamefont {Tosatti},\ and\ \citenamefont {Car}}]{Santoro2002}%
  \BibitemOpen
  \bibfield  {author} {\bibinfo {author} {\bibfnamefont {G.~E.}\ \bibnamefont {Santoro}}, \bibinfo {author} {\bibfnamefont {R.}~\bibnamefont {Marto{\v{n}}\'ak}}, \bibinfo {author} {\bibfnamefont {E.}~\bibnamefont {Tosatti}},\ and\ \bibinfo {author} {\bibfnamefont {R.}~\bibnamefont {Car}},\ }\href {https://doi.org/10.1126/science.1068774} {\bibfield  {journal} {\bibinfo  {journal} {Science}\ }\textbf {\bibinfo {volume} {295}},\ \bibinfo {pages} {2427} (\bibinfo {year} {2002})}\BibitemShut {NoStop}%
\bibitem [{\citenamefont {Marto{\v{n}}\'ak}\ \emph {et~al.}(2004)\citenamefont {Marto{\v{n}}\'ak}, \citenamefont {Santoro},\ and\ \citenamefont {Tosatti}}]{Martonak2004}%
  \BibitemOpen
  \bibfield  {author} {\bibinfo {author} {\bibfnamefont {R.}~\bibnamefont {Marto{\v{n}}\'ak}}, \bibinfo {author} {\bibfnamefont {G.~E.}\ \bibnamefont {Santoro}},\ and\ \bibinfo {author} {\bibfnamefont {E.}~\bibnamefont {Tosatti}},\ }\href {https://doi.org/10.1103/physreve.70.057701} {\bibfield  {journal} {\bibinfo  {journal} {Physical Review E}\ }\textbf {\bibinfo {volume} {70}},\ \bibinfo {pages} {057701} (\bibinfo {year} {2004})}\BibitemShut {NoStop}%
\bibitem [{\citenamefont {Hauke}\ \emph {et~al.}(2020)\citenamefont {Hauke}, \citenamefont {Katzgraber}, \citenamefont {Lechner}, \citenamefont {Nishimori},\ and\ \citenamefont {Oliver}}]{Hauke2020}%
  \BibitemOpen
  \bibfield  {author} {\bibinfo {author} {\bibfnamefont {P.}~\bibnamefont {Hauke}}, \bibinfo {author} {\bibfnamefont {H.~G.}\ \bibnamefont {Katzgraber}}, \bibinfo {author} {\bibfnamefont {W.}~\bibnamefont {Lechner}}, \bibinfo {author} {\bibfnamefont {H.}~\bibnamefont {Nishimori}},\ and\ \bibinfo {author} {\bibfnamefont {W.~D.}\ \bibnamefont {Oliver}},\ }\href {https://doi.org/10.1088/1361-6633/ab85b8} {\bibfield  {journal} {\bibinfo  {journal} {Reports on Progress in Physics}\ }\textbf {\bibinfo {volume} {83}},\ \bibinfo {pages} {054401} (\bibinfo {year} {2020})}\BibitemShut {NoStop}%
\bibitem [{\citenamefont {Rajak}\ \emph {et~al.}(2022)\citenamefont {Rajak}, \citenamefont {Suzuki}, \citenamefont {Dutta},\ and\ \citenamefont {Chakrabarti}}]{Rajak2022}%
  \BibitemOpen
  \bibfield  {author} {\bibinfo {author} {\bibfnamefont {A.}~\bibnamefont {Rajak}}, \bibinfo {author} {\bibfnamefont {S.}~\bibnamefont {Suzuki}}, \bibinfo {author} {\bibfnamefont {A.}~\bibnamefont {Dutta}},\ and\ \bibinfo {author} {\bibfnamefont {B.~K.}\ \bibnamefont {Chakrabarti}},\ }\bibfield  {journal} {\bibinfo  {journal} {Philosophical Transactions of the Royal Society A: Mathematical, Physical and Engineering Sciences}\ }\textbf {\bibinfo {volume} {381}},\ \href {https://doi.org/10.1098/rsta.2021.0417} {10.1098/rsta.2021.0417} (\bibinfo {year} {2022})\BibitemShut {NoStop}%
\bibitem [{\citenamefont {Yarkoni}\ \emph {et~al.}(2022)\citenamefont {Yarkoni}, \citenamefont {Raponi}, \citenamefont {Bäck},\ and\ \citenamefont {Schmitt}}]{Yarkoni2022}%
  \BibitemOpen
  \bibfield  {author} {\bibinfo {author} {\bibfnamefont {S.}~\bibnamefont {Yarkoni}}, \bibinfo {author} {\bibfnamefont {E.}~\bibnamefont {Raponi}}, \bibinfo {author} {\bibfnamefont {T.}~\bibnamefont {Bäck}},\ and\ \bibinfo {author} {\bibfnamefont {S.}~\bibnamefont {Schmitt}},\ }\href {https://doi.org/10.1088/1361-6633/ac8c54} {\bibfield  {journal} {\bibinfo  {journal} {Reports on Progress in Physics}\ }\textbf {\bibinfo {volume} {85}},\ \bibinfo {pages} {104001} (\bibinfo {year} {2022})}\BibitemShut {NoStop}%
\bibitem [{\citenamefont {Arute}\ \emph {et~al.}(2019)\citenamefont {Arute}, \citenamefont {Arya}, \citenamefont {Babbush}, \citenamefont {Bacon}, \citenamefont {Bardin}, \citenamefont {Barends}, \citenamefont {Biswas}, \citenamefont {Boixo}, \citenamefont {Brandao}, \citenamefont {Buell}, \citenamefont {Burkett}, \citenamefont {Chen}, \citenamefont {Chen}, \citenamefont {Chiaro}, \citenamefont {Collins}, \citenamefont {Courtney}, \citenamefont {Dunsworth}, \citenamefont {Farhi}, \citenamefont {Foxen}, \citenamefont {Fowler}, \citenamefont {Gidney}, \citenamefont {Giustina}, \citenamefont {Graff}, \citenamefont {Guerin}, \citenamefont {Habegger}, \citenamefont {Harrigan}, \citenamefont {Hartmann}, \citenamefont {Ho}, \citenamefont {Hoffmann}, \citenamefont {Huang}, \citenamefont {Humble}, \citenamefont {Isakov}, \citenamefont {Jeffrey}, \citenamefont {Jiang}, \citenamefont {Kafri}, \citenamefont {Kechedzhi}, \citenamefont {Kelly}, \citenamefont {Klimov}, \citenamefont {Knysh}, \citenamefont {Korotkov},
  \citenamefont {Kostritsa}, \citenamefont {Landhuis}, \citenamefont {Lindmark}, \citenamefont {Lucero}, \citenamefont {Lyakh}, \citenamefont {Mandrà}, \citenamefont {McClean}, \citenamefont {McEwen}, \citenamefont {Megrant}, \citenamefont {Mi}, \citenamefont {Michielsen}, \citenamefont {Mohseni}, \citenamefont {Mutus}, \citenamefont {Naaman}, \citenamefont {Neeley}, \citenamefont {Neill}, \citenamefont {Niu}, \citenamefont {Ostby}, \citenamefont {Petukhov}, \citenamefont {Platt}, \citenamefont {Quintana}, \citenamefont {Rieffel}, \citenamefont {Roushan}, \citenamefont {Rubin}, \citenamefont {Sank}, \citenamefont {Satzinger}, \citenamefont {Smelyanskiy}, \citenamefont {Sung}, \citenamefont {Trevithick}, \citenamefont {Vainsencher}, \citenamefont {Villalonga}, \citenamefont {White}, \citenamefont {Yao}, \citenamefont {Yeh}, \citenamefont {Zalcman}, \citenamefont {Neven},\ and\ \citenamefont {Martinis}}]{Arute2019}%
  \BibitemOpen
  \bibfield  {author} {\bibinfo {author} {\bibfnamefont {F.}~\bibnamefont {Arute}}, \bibinfo {author} {\bibfnamefont {K.}~\bibnamefont {Arya}}, \bibinfo {author} {\bibfnamefont {R.}~\bibnamefont {Babbush}}, \bibinfo {author} {\bibfnamefont {D.}~\bibnamefont {Bacon}}, \bibinfo {author} {\bibfnamefont {J.~C.}\ \bibnamefont {Bardin}}, \bibinfo {author} {\bibfnamefont {R.}~\bibnamefont {Barends}}, \bibinfo {author} {\bibfnamefont {R.}~\bibnamefont {Biswas}}, \bibinfo {author} {\bibfnamefont {S.}~\bibnamefont {Boixo}}, \bibinfo {author} {\bibfnamefont {F.~G. S.~L.}\ \bibnamefont {Brandao}}, \bibinfo {author} {\bibfnamefont {D.~A.}\ \bibnamefont {Buell}}, \bibinfo {author} {\bibfnamefont {B.}~\bibnamefont {Burkett}}, \bibinfo {author} {\bibfnamefont {Y.}~\bibnamefont {Chen}}, \bibinfo {author} {\bibfnamefont {Z.}~\bibnamefont {Chen}}, \bibinfo {author} {\bibfnamefont {B.}~\bibnamefont {Chiaro}}, \bibinfo {author} {\bibfnamefont {R.}~\bibnamefont {Collins}}, \bibinfo {author} {\bibfnamefont {W.}~\bibnamefont
  {Courtney}}, \bibinfo {author} {\bibfnamefont {A.}~\bibnamefont {Dunsworth}}, \bibinfo {author} {\bibfnamefont {E.}~\bibnamefont {Farhi}}, \bibinfo {author} {\bibfnamefont {B.}~\bibnamefont {Foxen}}, \bibinfo {author} {\bibfnamefont {A.}~\bibnamefont {Fowler}}, \bibinfo {author} {\bibfnamefont {C.}~\bibnamefont {Gidney}}, \bibinfo {author} {\bibfnamefont {M.}~\bibnamefont {Giustina}}, \bibinfo {author} {\bibfnamefont {R.}~\bibnamefont {Graff}}, \bibinfo {author} {\bibfnamefont {K.}~\bibnamefont {Guerin}}, \bibinfo {author} {\bibfnamefont {S.}~\bibnamefont {Habegger}}, \bibinfo {author} {\bibfnamefont {M.~P.}\ \bibnamefont {Harrigan}}, \bibinfo {author} {\bibfnamefont {M.~J.}\ \bibnamefont {Hartmann}}, \bibinfo {author} {\bibfnamefont {A.}~\bibnamefont {Ho}}, \bibinfo {author} {\bibfnamefont {M.}~\bibnamefont {Hoffmann}}, \bibinfo {author} {\bibfnamefont {T.}~\bibnamefont {Huang}}, \bibinfo {author} {\bibfnamefont {T.~S.}\ \bibnamefont {Humble}}, \bibinfo {author} {\bibfnamefont {S.~V.}\ \bibnamefont
  {Isakov}}, \bibinfo {author} {\bibfnamefont {E.}~\bibnamefont {Jeffrey}}, \bibinfo {author} {\bibfnamefont {Z.}~\bibnamefont {Jiang}}, \bibinfo {author} {\bibfnamefont {D.}~\bibnamefont {Kafri}}, \bibinfo {author} {\bibfnamefont {K.}~\bibnamefont {Kechedzhi}}, \bibinfo {author} {\bibfnamefont {J.}~\bibnamefont {Kelly}}, \bibinfo {author} {\bibfnamefont {P.~V.}\ \bibnamefont {Klimov}}, \bibinfo {author} {\bibfnamefont {S.}~\bibnamefont {Knysh}}, \bibinfo {author} {\bibfnamefont {A.}~\bibnamefont {Korotkov}}, \bibinfo {author} {\bibfnamefont {F.}~\bibnamefont {Kostritsa}}, \bibinfo {author} {\bibfnamefont {D.}~\bibnamefont {Landhuis}}, \bibinfo {author} {\bibfnamefont {M.}~\bibnamefont {Lindmark}}, \bibinfo {author} {\bibfnamefont {E.}~\bibnamefont {Lucero}}, \bibinfo {author} {\bibfnamefont {D.}~\bibnamefont {Lyakh}}, \bibinfo {author} {\bibfnamefont {S.}~\bibnamefont {Mandrà}}, \bibinfo {author} {\bibfnamefont {J.~R.}\ \bibnamefont {McClean}}, \bibinfo {author} {\bibfnamefont {M.}~\bibnamefont {McEwen}},
  \bibinfo {author} {\bibfnamefont {A.}~\bibnamefont {Megrant}}, \bibinfo {author} {\bibfnamefont {X.}~\bibnamefont {Mi}}, \bibinfo {author} {\bibfnamefont {K.}~\bibnamefont {Michielsen}}, \bibinfo {author} {\bibfnamefont {M.}~\bibnamefont {Mohseni}}, \bibinfo {author} {\bibfnamefont {J.}~\bibnamefont {Mutus}}, \bibinfo {author} {\bibfnamefont {O.}~\bibnamefont {Naaman}}, \bibinfo {author} {\bibfnamefont {M.}~\bibnamefont {Neeley}}, \bibinfo {author} {\bibfnamefont {C.}~\bibnamefont {Neill}}, \bibinfo {author} {\bibfnamefont {M.~Y.}\ \bibnamefont {Niu}}, \bibinfo {author} {\bibfnamefont {E.}~\bibnamefont {Ostby}}, \bibinfo {author} {\bibfnamefont {A.}~\bibnamefont {Petukhov}}, \bibinfo {author} {\bibfnamefont {J.~C.}\ \bibnamefont {Platt}}, \bibinfo {author} {\bibfnamefont {C.}~\bibnamefont {Quintana}}, \bibinfo {author} {\bibfnamefont {E.~G.}\ \bibnamefont {Rieffel}}, \bibinfo {author} {\bibfnamefont {P.}~\bibnamefont {Roushan}}, \bibinfo {author} {\bibfnamefont {N.~C.}\ \bibnamefont {Rubin}}, \bibinfo
  {author} {\bibfnamefont {D.}~\bibnamefont {Sank}}, \bibinfo {author} {\bibfnamefont {K.~J.}\ \bibnamefont {Satzinger}}, \bibinfo {author} {\bibfnamefont {V.}~\bibnamefont {Smelyanskiy}}, \bibinfo {author} {\bibfnamefont {K.~J.}\ \bibnamefont {Sung}}, \bibinfo {author} {\bibfnamefont {M.~D.}\ \bibnamefont {Trevithick}}, \bibinfo {author} {\bibfnamefont {A.}~\bibnamefont {Vainsencher}}, \bibinfo {author} {\bibfnamefont {B.}~\bibnamefont {Villalonga}}, \bibinfo {author} {\bibfnamefont {T.}~\bibnamefont {White}}, \bibinfo {author} {\bibfnamefont {Z.~J.}\ \bibnamefont {Yao}}, \bibinfo {author} {\bibfnamefont {P.}~\bibnamefont {Yeh}}, \bibinfo {author} {\bibfnamefont {A.}~\bibnamefont {Zalcman}}, \bibinfo {author} {\bibfnamefont {H.}~\bibnamefont {Neven}},\ and\ \bibinfo {author} {\bibfnamefont {J.~M.}\ \bibnamefont {Martinis}},\ }\href {https://doi.org/10.1038/s41586-019-1666-5} {\bibfield  {journal} {\bibinfo  {journal} {Nature}\ }\textbf {\bibinfo {volume} {574}},\ \bibinfo {pages} {505} (\bibinfo {year}
  {2019})}\BibitemShut {NoStop}%
\bibitem [{\citenamefont {Kjaergaard}\ \emph {et~al.}(2020)\citenamefont {Kjaergaard}, \citenamefont {Schwartz}, \citenamefont {Braum\"uller}, \citenamefont {Krantz}, \citenamefont {Wang}, \citenamefont {Gustavsson},\ and\ \citenamefont {Oliver}}]{Kjaergaard2020}%
  \BibitemOpen
  \bibfield  {author} {\bibinfo {author} {\bibfnamefont {M.}~\bibnamefont {Kjaergaard}}, \bibinfo {author} {\bibfnamefont {M.~E.}\ \bibnamefont {Schwartz}}, \bibinfo {author} {\bibfnamefont {J.}~\bibnamefont {Braum\"uller}}, \bibinfo {author} {\bibfnamefont {P.}~\bibnamefont {Krantz}}, \bibinfo {author} {\bibfnamefont {J.~I.}\ \bibnamefont {Wang}}, \bibinfo {author} {\bibfnamefont {S.}~\bibnamefont {Gustavsson}},\ and\ \bibinfo {author} {\bibfnamefont {W.~D.}\ \bibnamefont {Oliver}},\ }\href {https://doi.org/10.1140/epjb/s10051-020-00634-2} {\bibfield  {journal} {\bibinfo  {journal} {Annual Review of Condensed Matter Physics}\ }\textbf {\bibinfo {volume} {11}},\ \bibinfo {pages} {369–395} (\bibinfo {year} {2020})}\BibitemShut {NoStop}%
\bibitem [{\citenamefont {Bravyi}\ \emph {et~al.}(2024)\citenamefont {Bravyi}, \citenamefont {Cross}, \citenamefont {Gambetta}, \citenamefont {Maslov}, \citenamefont {Rall},\ and\ \citenamefont {Yoder}}]{Bravyi2024}%
  \BibitemOpen
  \bibfield  {author} {\bibinfo {author} {\bibfnamefont {S.}~\bibnamefont {Bravyi}}, \bibinfo {author} {\bibfnamefont {A.~W.}\ \bibnamefont {Cross}}, \bibinfo {author} {\bibfnamefont {J.~M.}\ \bibnamefont {Gambetta}}, \bibinfo {author} {\bibfnamefont {D.}~\bibnamefont {Maslov}}, \bibinfo {author} {\bibfnamefont {P.}~\bibnamefont {Rall}},\ and\ \bibinfo {author} {\bibfnamefont {T.~J.}\ \bibnamefont {Yoder}},\ }\href {https://doi.org/10.1038/s41586-024-07107-7} {\bibfield  {journal} {\bibinfo  {journal} {Nature}\ }\textbf {\bibinfo {volume} {627}},\ \bibinfo {pages} {778} (\bibinfo {year} {2024})}\BibitemShut {NoStop}%
\bibitem [{\citenamefont {Boixo}\ \emph {et~al.}(2014)\citenamefont {Boixo}, \citenamefont {R{\o}nnow}, \citenamefont {Isakov}, \citenamefont {Wang}, \citenamefont {Wecker}, \citenamefont {Lidar}, \citenamefont {Martinis},\ and\ \citenamefont {Troyer}}]{Boixo2014}%
  \BibitemOpen
  \bibfield  {author} {\bibinfo {author} {\bibfnamefont {S.}~\bibnamefont {Boixo}}, \bibinfo {author} {\bibfnamefont {T.~F.}\ \bibnamefont {R{\o}nnow}}, \bibinfo {author} {\bibfnamefont {S.~V.}\ \bibnamefont {Isakov}}, \bibinfo {author} {\bibfnamefont {Z.}~\bibnamefont {Wang}}, \bibinfo {author} {\bibfnamefont {D.}~\bibnamefont {Wecker}}, \bibinfo {author} {\bibfnamefont {D.~A.}\ \bibnamefont {Lidar}}, \bibinfo {author} {\bibfnamefont {J.~M.}\ \bibnamefont {Martinis}},\ and\ \bibinfo {author} {\bibfnamefont {M.}~\bibnamefont {Troyer}},\ }\href {https://doi.org/https://doi.org/10.1038/nphys2900} {\bibfield  {journal} {\bibinfo  {journal} {Nature physics}\ }\textbf {\bibinfo {volume} {10}},\ \bibinfo {pages} {218} (\bibinfo {year} {2014})}\BibitemShut {NoStop}%
\bibitem [{\citenamefont {Werninghaus}\ \emph {et~al.}(2021)\citenamefont {Werninghaus}, \citenamefont {Egger}, \citenamefont {Roy}, \citenamefont {Machnes}, \citenamefont {Wilhelm},\ and\ \citenamefont {Filipp}}]{Werninghaus2021}%
  \BibitemOpen
  \bibfield  {author} {\bibinfo {author} {\bibfnamefont {M.}~\bibnamefont {Werninghaus}}, \bibinfo {author} {\bibfnamefont {D.~J.}\ \bibnamefont {Egger}}, \bibinfo {author} {\bibfnamefont {F.}~\bibnamefont {Roy}}, \bibinfo {author} {\bibfnamefont {S.}~\bibnamefont {Machnes}}, \bibinfo {author} {\bibfnamefont {F.~K.}\ \bibnamefont {Wilhelm}},\ and\ \bibinfo {author} {\bibfnamefont {S.}~\bibnamefont {Filipp}},\ }\bibfield  {journal} {\bibinfo  {journal} {npj Quantum Information}\ }\textbf {\bibinfo {volume} {7}},\ \href {https://doi.org/10.1038/s41534-020-00346-2} {10.1038/s41534-020-00346-2} (\bibinfo {year} {2021})\BibitemShut {NoStop}%
\bibitem [{\citenamefont {Brodoloni}\ \emph {et~al.}(2025)\citenamefont {Brodoloni}, \citenamefont {Vovrosh}, \citenamefont {Julià-Farr\'e}, \citenamefont {Dauphin},\ and\ \citenamefont {Pilati}}]{Brodoloni2025}%
  \BibitemOpen
  \bibfield  {author} {\bibinfo {author} {\bibfnamefont {L.}~\bibnamefont {Brodoloni}}, \bibinfo {author} {\bibfnamefont {J.}~\bibnamefont {Vovrosh}}, \bibinfo {author} {\bibfnamefont {S.}~\bibnamefont {Julià-Farr\'e}}, \bibinfo {author} {\bibfnamefont {A.}~\bibnamefont {Dauphin}},\ and\ \bibinfo {author} {\bibfnamefont {S.}~\bibnamefont {Pilati}},\ }\href {https://doi.org/10.48550/ARXIV.2505.05117} {\bibinfo {title} {Spin-glass quantum phase transition in amorphous arrays of rydberg atoms}} (\bibinfo {year} {2025})\BibitemShut {NoStop}%
\bibitem [{\citenamefont {Gonz\'alez-Cuadra}\ \emph {et~al.}(2025)\citenamefont {Gonz\'alez-Cuadra}, \citenamefont {Hamdan}, \citenamefont {Zache}, \citenamefont {Braverman}, \citenamefont {Kornja{\v{c}}a}, \citenamefont {Lukin}, \citenamefont {Cant\'u}, \citenamefont {Liu}, \citenamefont {Wang}, \citenamefont {Keesling}, \citenamefont {Lukin}, \citenamefont {Zoller},\ and\ \citenamefont {Bylinskii}}]{GonzalezCuadra2025}%
  \BibitemOpen
  \bibfield  {author} {\bibinfo {author} {\bibfnamefont {D.}~\bibnamefont {Gonz\'alez-Cuadra}}, \bibinfo {author} {\bibfnamefont {M.}~\bibnamefont {Hamdan}}, \bibinfo {author} {\bibfnamefont {T.~V.}\ \bibnamefont {Zache}}, \bibinfo {author} {\bibfnamefont {B.}~\bibnamefont {Braverman}}, \bibinfo {author} {\bibfnamefont {M.}~\bibnamefont {Kornja{\v{c}}a}}, \bibinfo {author} {\bibfnamefont {A.}~\bibnamefont {Lukin}}, \bibinfo {author} {\bibfnamefont {S.~H.}\ \bibnamefont {Cant\'u}}, \bibinfo {author} {\bibfnamefont {F.}~\bibnamefont {Liu}}, \bibinfo {author} {\bibfnamefont {S.-T.}\ \bibnamefont {Wang}}, \bibinfo {author} {\bibfnamefont {A.}~\bibnamefont {Keesling}}, \bibinfo {author} {\bibfnamefont {M.~D.}\ \bibnamefont {Lukin}}, \bibinfo {author} {\bibfnamefont {P.}~\bibnamefont {Zoller}},\ and\ \bibinfo {author} {\bibfnamefont {A.}~\bibnamefont {Bylinskii}},\ }\href {https://doi.org/10.1038/s41586-025-09051-6} {\bibfield  {journal} {\bibinfo  {journal} {Nature}\ }\textbf {\bibinfo {volume} {642}},\ \bibinfo
  {pages} {321} (\bibinfo {year} {2025})}\BibitemShut {NoStop}%
\bibitem [{\citenamefont {Wineland}\ \emph {et~al.}(2003)\citenamefont {Wineland}, \citenamefont {Barrett}, \citenamefont {Britton}, \citenamefont {Chiaverini}, \citenamefont {DeMarco}, \citenamefont {Itano}, \citenamefont {Jelenković}, \citenamefont {Langer}, \citenamefont {Leibfried}, \citenamefont {Meyer}, \citenamefont {Rosenband},\ and\ \citenamefont {Schätz}}]{Wineland2003}%
  \BibitemOpen
  \bibfield  {author} {\bibinfo {author} {\bibfnamefont {D.~J.}\ \bibnamefont {Wineland}}, \bibinfo {author} {\bibfnamefont {M.}~\bibnamefont {Barrett}}, \bibinfo {author} {\bibfnamefont {J.}~\bibnamefont {Britton}}, \bibinfo {author} {\bibfnamefont {J.}~\bibnamefont {Chiaverini}}, \bibinfo {author} {\bibfnamefont {B.}~\bibnamefont {DeMarco}}, \bibinfo {author} {\bibfnamefont {W.~M.}\ \bibnamefont {Itano}}, \bibinfo {author} {\bibfnamefont {B.}~\bibnamefont {Jelenković}}, \bibinfo {author} {\bibfnamefont {C.}~\bibnamefont {Langer}}, \bibinfo {author} {\bibfnamefont {D.}~\bibnamefont {Leibfried}}, \bibinfo {author} {\bibfnamefont {V.}~\bibnamefont {Meyer}}, \bibinfo {author} {\bibfnamefont {T.}~\bibnamefont {Rosenband}},\ and\ \bibinfo {author} {\bibfnamefont {T.}~\bibnamefont {Schätz}},\ }\href {https://doi.org/10.1098/rsta.2003.1205} {\bibfield  {journal} {\bibinfo  {journal} {Philosophical Transactions of the Royal Society of London. Series A: Mathematical, Physical and Engineering Sciences}\ }\textbf
  {\bibinfo {volume} {361}},\ \bibinfo {pages} {1349} (\bibinfo {year} {2003})}\BibitemShut {NoStop}%
\bibitem [{\citenamefont {Leibfried}\ \emph {et~al.}(2003)\citenamefont {Leibfried}, \citenamefont {Blatt}, \citenamefont {Monroe},\ and\ \citenamefont {Wineland}}]{Leibfried2003a}%
  \BibitemOpen
  \bibfield  {author} {\bibinfo {author} {\bibfnamefont {D.}~\bibnamefont {Leibfried}}, \bibinfo {author} {\bibfnamefont {R.}~\bibnamefont {Blatt}}, \bibinfo {author} {\bibfnamefont {C.}~\bibnamefont {Monroe}},\ and\ \bibinfo {author} {\bibfnamefont {D.}~\bibnamefont {Wineland}},\ }\href {https://doi.org/10.1103/revmodphys.75.281} {\bibfield  {journal} {\bibinfo  {journal} {Reviews of Modern Physics}\ }\textbf {\bibinfo {volume} {75}},\ \bibinfo {pages} {281} (\bibinfo {year} {2003})}\BibitemShut {NoStop}%
\bibitem [{\citenamefont {Häffner}\ \emph {et~al.}(2008)\citenamefont {Häffner}, \citenamefont {Roos},\ and\ \citenamefont {Blatt}}]{HAFFNER2008}%
  \BibitemOpen
  \bibfield  {author} {\bibinfo {author} {\bibfnamefont {H.}~\bibnamefont {Häffner}}, \bibinfo {author} {\bibfnamefont {C.}~\bibnamefont {Roos}},\ and\ \bibinfo {author} {\bibfnamefont {R.}~\bibnamefont {Blatt}},\ }\href {https://doi.org/10.1016/j.physrep.2008.09.003} {\bibfield  {journal} {\bibinfo  {journal} {Physics Reports}\ }\textbf {\bibinfo {volume} {469}},\ \bibinfo {pages} {155} (\bibinfo {year} {2008})}\BibitemShut {NoStop}%
\bibitem [{\citenamefont {Bruzewicz}\ \emph {et~al.}(2019)\citenamefont {Bruzewicz}, \citenamefont {Chiaverini}, \citenamefont {McConnell},\ and\ \citenamefont {Sage}}]{Bruzewicz2019}%
  \BibitemOpen
  \bibfield  {author} {\bibinfo {author} {\bibfnamefont {C.~D.}\ \bibnamefont {Bruzewicz}}, \bibinfo {author} {\bibfnamefont {J.}~\bibnamefont {Chiaverini}}, \bibinfo {author} {\bibfnamefont {R.}~\bibnamefont {McConnell}},\ and\ \bibinfo {author} {\bibfnamefont {J.~M.}\ \bibnamefont {Sage}},\ }\bibfield  {journal} {\bibinfo  {journal} {Applied Physics Reviews}\ }\textbf {\bibinfo {volume} {6}},\ \href {https://doi.org/10.1063/1.5088164} {10.1063/1.5088164} (\bibinfo {year} {2019})\BibitemShut {NoStop}%
\bibitem [{\citenamefont {Srinivas}\ \emph {et~al.}(2021)\citenamefont {Srinivas}, \citenamefont {Burd}, \citenamefont {Knaack}, \citenamefont {Sutherland}, \citenamefont {Kwiatkowski}, \citenamefont {Glancy}, \citenamefont {Knill}, \citenamefont {Wineland}, \citenamefont {Leibfried}, \citenamefont {Wilson}, \citenamefont {Allcock},\ and\ \citenamefont {Slichter}}]{Srinivas2021}%
  \BibitemOpen
  \bibfield  {author} {\bibinfo {author} {\bibfnamefont {R.}~\bibnamefont {Srinivas}}, \bibinfo {author} {\bibfnamefont {S.~C.}\ \bibnamefont {Burd}}, \bibinfo {author} {\bibfnamefont {H.~M.}\ \bibnamefont {Knaack}}, \bibinfo {author} {\bibfnamefont {R.~T.}\ \bibnamefont {Sutherland}}, \bibinfo {author} {\bibfnamefont {A.}~\bibnamefont {Kwiatkowski}}, \bibinfo {author} {\bibfnamefont {S.}~\bibnamefont {Glancy}}, \bibinfo {author} {\bibfnamefont {E.}~\bibnamefont {Knill}}, \bibinfo {author} {\bibfnamefont {D.~J.}\ \bibnamefont {Wineland}}, \bibinfo {author} {\bibfnamefont {D.}~\bibnamefont {Leibfried}}, \bibinfo {author} {\bibfnamefont {A.~C.}\ \bibnamefont {Wilson}}, \bibinfo {author} {\bibfnamefont {D.~T.~C.}\ \bibnamefont {Allcock}},\ and\ \bibinfo {author} {\bibfnamefont {D.~H.}\ \bibnamefont {Slichter}},\ }\href {https://doi.org/10.1038/s41586-021-03809-4} {\bibfield  {journal} {\bibinfo  {journal} {Nature}\ }\textbf {\bibinfo {volume} {597}},\ \bibinfo {pages} {209} (\bibinfo {year} {2021})}\BibitemShut
  {NoStop}%
\bibitem [{\citenamefont {Chen}\ \emph {et~al.}(2024)\citenamefont {Chen}, \citenamefont {Nielsen}, \citenamefont {Ebert}, \citenamefont {Inlek}, \citenamefont {Wright}, \citenamefont {Chaplin}, \citenamefont {Maksymov}, \citenamefont {P\'aez}, \citenamefont {Poudel}, \citenamefont {Maunz},\ and\ \citenamefont {Gamble}}]{Chen2024}%
  \BibitemOpen
  \bibfield  {author} {\bibinfo {author} {\bibfnamefont {J.-S.}\ \bibnamefont {Chen}}, \bibinfo {author} {\bibfnamefont {E.}~\bibnamefont {Nielsen}}, \bibinfo {author} {\bibfnamefont {M.}~\bibnamefont {Ebert}}, \bibinfo {author} {\bibfnamefont {V.}~\bibnamefont {Inlek}}, \bibinfo {author} {\bibfnamefont {K.}~\bibnamefont {Wright}}, \bibinfo {author} {\bibfnamefont {V.}~\bibnamefont {Chaplin}}, \bibinfo {author} {\bibfnamefont {A.}~\bibnamefont {Maksymov}}, \bibinfo {author} {\bibfnamefont {E.}~\bibnamefont {P\'aez}}, \bibinfo {author} {\bibfnamefont {A.}~\bibnamefont {Poudel}}, \bibinfo {author} {\bibfnamefont {P.}~\bibnamefont {Maunz}},\ and\ \bibinfo {author} {\bibfnamefont {J.}~\bibnamefont {Gamble}},\ }\href {https://doi.org/10.22331/q-2024-11-07-1516} {\bibfield  {journal} {\bibinfo  {journal} {Quantum}\ }\textbf {\bibinfo {volume} {8}},\ \bibinfo {pages} {1516} (\bibinfo {year} {2024})}\BibitemShut {NoStop}%
\bibitem [{\citenamefont {Liu}\ \emph {et~al.}(2025)\citenamefont {Liu}, \citenamefont {Shaydulin}, \citenamefont {Niroula}, \citenamefont {DeCross}, \citenamefont {Hung}, \citenamefont {Kon}, \citenamefont {Cervero-Martín}, \citenamefont {Chakraborty}, \citenamefont {Amer}, \citenamefont {Aaronson}, \citenamefont {Acharya}, \citenamefont {Alexeev}, \citenamefont {Berg}, \citenamefont {Chakrabarti}, \citenamefont {Curchod}, \citenamefont {Dreiling}, \citenamefont {Erickson}, \citenamefont {Foltz}, \citenamefont {Foss-Feig}, \citenamefont {Hayes}, \citenamefont {Humble}, \citenamefont {Kumar}, \citenamefont {Larson}, \citenamefont {Lykov}, \citenamefont {Mills}, \citenamefont {Moses}, \citenamefont {Neyenhuis}, \citenamefont {Eloul}, \citenamefont {Siegfried}, \citenamefont {Walker}, \citenamefont {Lim},\ and\ \citenamefont {Pistoia}}]{Liu2025}%
  \BibitemOpen
  \bibfield  {author} {\bibinfo {author} {\bibfnamefont {M.}~\bibnamefont {Liu}}, \bibinfo {author} {\bibfnamefont {R.}~\bibnamefont {Shaydulin}}, \bibinfo {author} {\bibfnamefont {P.}~\bibnamefont {Niroula}}, \bibinfo {author} {\bibfnamefont {M.}~\bibnamefont {DeCross}}, \bibinfo {author} {\bibfnamefont {S.-H.}\ \bibnamefont {Hung}}, \bibinfo {author} {\bibfnamefont {W.~Y.}\ \bibnamefont {Kon}}, \bibinfo {author} {\bibfnamefont {E.}~\bibnamefont {Cervero-Martín}}, \bibinfo {author} {\bibfnamefont {K.}~\bibnamefont {Chakraborty}}, \bibinfo {author} {\bibfnamefont {O.}~\bibnamefont {Amer}}, \bibinfo {author} {\bibfnamefont {S.}~\bibnamefont {Aaronson}}, \bibinfo {author} {\bibfnamefont {A.}~\bibnamefont {Acharya}}, \bibinfo {author} {\bibfnamefont {Y.}~\bibnamefont {Alexeev}}, \bibinfo {author} {\bibfnamefont {K.~J.}\ \bibnamefont {Berg}}, \bibinfo {author} {\bibfnamefont {S.}~\bibnamefont {Chakrabarti}}, \bibinfo {author} {\bibfnamefont {F.~J.}\ \bibnamefont {Curchod}}, \bibinfo {author} {\bibfnamefont
  {J.~M.}\ \bibnamefont {Dreiling}}, \bibinfo {author} {\bibfnamefont {N.}~\bibnamefont {Erickson}}, \bibinfo {author} {\bibfnamefont {C.}~\bibnamefont {Foltz}}, \bibinfo {author} {\bibfnamefont {M.}~\bibnamefont {Foss-Feig}}, \bibinfo {author} {\bibfnamefont {D.}~\bibnamefont {Hayes}}, \bibinfo {author} {\bibfnamefont {T.~S.}\ \bibnamefont {Humble}}, \bibinfo {author} {\bibfnamefont {N.}~\bibnamefont {Kumar}}, \bibinfo {author} {\bibfnamefont {J.}~\bibnamefont {Larson}}, \bibinfo {author} {\bibfnamefont {D.}~\bibnamefont {Lykov}}, \bibinfo {author} {\bibfnamefont {M.}~\bibnamefont {Mills}}, \bibinfo {author} {\bibfnamefont {S.~A.}\ \bibnamefont {Moses}}, \bibinfo {author} {\bibfnamefont {B.}~\bibnamefont {Neyenhuis}}, \bibinfo {author} {\bibfnamefont {S.}~\bibnamefont {Eloul}}, \bibinfo {author} {\bibfnamefont {P.}~\bibnamefont {Siegfried}}, \bibinfo {author} {\bibfnamefont {J.}~\bibnamefont {Walker}}, \bibinfo {author} {\bibfnamefont {C.}~\bibnamefont {Lim}},\ and\ \bibinfo {author} {\bibfnamefont
  {M.}~\bibnamefont {Pistoia}},\ }\href {https://doi.org/10.1038/s41586-025-08737-1} {\bibfield  {journal} {\bibinfo  {journal} {Nature}\ }\textbf {\bibinfo {volume} {640}},\ \bibinfo {pages} {343} (\bibinfo {year} {2025})}\BibitemShut {NoStop}%
\bibitem [{\citenamefont {Meth}\ \emph {et~al.}(2025)\citenamefont {Meth}, \citenamefont {Zhang}, \citenamefont {Haase}, \citenamefont {Edmunds}, \citenamefont {Postler}, \citenamefont {Jena}, \citenamefont {Steiner}, \citenamefont {Dellantonio}, \citenamefont {Blatt}, \citenamefont {Zoller}, \citenamefont {Monz}, \citenamefont {Schindler}, \citenamefont {Muschik},\ and\ \citenamefont {Ringbauer}}]{Meth2025}%
  \BibitemOpen
  \bibfield  {author} {\bibinfo {author} {\bibfnamefont {M.}~\bibnamefont {Meth}}, \bibinfo {author} {\bibfnamefont {J.}~\bibnamefont {Zhang}}, \bibinfo {author} {\bibfnamefont {J.~F.}\ \bibnamefont {Haase}}, \bibinfo {author} {\bibfnamefont {C.}~\bibnamefont {Edmunds}}, \bibinfo {author} {\bibfnamefont {L.}~\bibnamefont {Postler}}, \bibinfo {author} {\bibfnamefont {A.~J.}\ \bibnamefont {Jena}}, \bibinfo {author} {\bibfnamefont {A.}~\bibnamefont {Steiner}}, \bibinfo {author} {\bibfnamefont {L.}~\bibnamefont {Dellantonio}}, \bibinfo {author} {\bibfnamefont {R.}~\bibnamefont {Blatt}}, \bibinfo {author} {\bibfnamefont {P.}~\bibnamefont {Zoller}}, \bibinfo {author} {\bibfnamefont {T.}~\bibnamefont {Monz}}, \bibinfo {author} {\bibfnamefont {P.}~\bibnamefont {Schindler}}, \bibinfo {author} {\bibfnamefont {C.}~\bibnamefont {Muschik}},\ and\ \bibinfo {author} {\bibfnamefont {M.}~\bibnamefont {Ringbauer}},\ }\href {https://doi.org/10.1038/s41567-025-02797-w} {\bibfield  {journal} {\bibinfo  {journal} {Nature Physics}\
  }\textbf {\bibinfo {volume} {21}},\ \bibinfo {pages} {570} (\bibinfo {year} {2025})}\BibitemShut {NoStop}%
\bibitem [{\citenamefont {Mintert}\ and\ \citenamefont {Wunderlich}(2001)}]{Mintert2001}%
  \BibitemOpen
  \bibfield  {author} {\bibinfo {author} {\bibfnamefont {F.}~\bibnamefont {Mintert}}\ and\ \bibinfo {author} {\bibfnamefont {C.}~\bibnamefont {Wunderlich}},\ }\href {https://doi.org/10.1103/physrevlett.87.257904} {\bibfield  {journal} {\bibinfo  {journal} {Physical Review Letters}\ }\textbf {\bibinfo {volume} {87}},\ \bibinfo {pages} {257904} (\bibinfo {year} {2001})}\BibitemShut {NoStop}%
\bibitem [{\citenamefont {Wunderlich}(2002)}]{Wunderlich2002}%
  \BibitemOpen
  \bibfield  {author} {\bibinfo {author} {\bibfnamefont {C.}~\bibnamefont {Wunderlich}},\ }\bibinfo {title} {Conditional spin resonance with trapped ions},\ in\ \href {https://doi.org/10.1007/978-3-662-04897-9_25} {\emph {\bibinfo {booktitle} {Laser Physics at the Limits}}}\ (\bibinfo  {publisher} {Springer Berlin Heidelberg},\ \bibinfo {year} {2002})\ pp.\ \bibinfo {pages} {261--273}\BibitemShut {NoStop}%
\bibitem [{\citenamefont {Johanning}\ \emph {et~al.}(2009)\citenamefont {Johanning}, \citenamefont {Var{\'{o}}n},\ and\ \citenamefont {Wunderlich}}]{Johanning2009}%
  \BibitemOpen
  \bibfield  {author} {\bibinfo {author} {\bibfnamefont {M.}~\bibnamefont {Johanning}}, \bibinfo {author} {\bibfnamefont {A.~F.}\ \bibnamefont {Var{\'{o}}n}},\ and\ \bibinfo {author} {\bibfnamefont {C.}~\bibnamefont {Wunderlich}},\ }\href {https://doi.org/10.1088/0953-4075/42/15/154009} {\bibfield  {journal} {\bibinfo  {journal} {Journal of Physics B: Atomic, Molecular and Optical Physics}\ }\textbf {\bibinfo {volume} {42}},\ \bibinfo {pages} {154009} (\bibinfo {year} {2009})}\BibitemShut {NoStop}%
\bibitem [{\citenamefont {Zippilli}\ \emph {et~al.}(2014)\citenamefont {Zippilli}, \citenamefont {Johanning}, \citenamefont {Giampaolo}, \citenamefont {Wunderlich},\ and\ \citenamefont {Illuminati}}]{Zippilli2014}%
  \BibitemOpen
  \bibfield  {author} {\bibinfo {author} {\bibfnamefont {S.}~\bibnamefont {Zippilli}}, \bibinfo {author} {\bibfnamefont {M.}~\bibnamefont {Johanning}}, \bibinfo {author} {\bibfnamefont {S.~M.}\ \bibnamefont {Giampaolo}}, \bibinfo {author} {\bibfnamefont {C.}~\bibnamefont {Wunderlich}},\ and\ \bibinfo {author} {\bibfnamefont {F.}~\bibnamefont {Illuminati}},\ }\href {https://doi.org/10.1103/physreva.89.042308} {\bibfield  {journal} {\bibinfo  {journal} {Physical Review A}\ }\textbf {\bibinfo {volume} {89}},\ \bibinfo {pages} {042308} (\bibinfo {year} {2014})}\BibitemShut {NoStop}%
\bibitem [{\citenamefont {Piltz}\ \emph {et~al.}(2016)\citenamefont {Piltz}, \citenamefont {Sriarunothai}, \citenamefont {Ivanov}, \citenamefont {W\"olk},\ and\ \citenamefont {Wunderlich}}]{Piltz2016}%
  \BibitemOpen
  \bibfield  {author} {\bibinfo {author} {\bibfnamefont {C.}~\bibnamefont {Piltz}}, \bibinfo {author} {\bibfnamefont {T.}~\bibnamefont {Sriarunothai}}, \bibinfo {author} {\bibfnamefont {S.~S.}\ \bibnamefont {Ivanov}}, \bibinfo {author} {\bibfnamefont {S.}~\bibnamefont {W\"olk}},\ and\ \bibinfo {author} {\bibfnamefont {C.}~\bibnamefont {Wunderlich}},\ }\bibfield  {journal} {\bibinfo  {journal} {Science Advances}\ }\textbf {\bibinfo {volume} {2}},\ \href {https://doi.org/10.1126/sciadv.1600093} {10.1126/sciadv.1600093} (\bibinfo {year} {2016})\BibitemShut {NoStop}%
\bibitem [{\citenamefont {Ba{\ss}ler}\ \emph {et~al.}(2023)\citenamefont {Ba{\ss}ler}, \citenamefont {Zipper}, \citenamefont {Cedzich}, \citenamefont {Heinrich}, \citenamefont {Huber}, \citenamefont {Johanning},\ and\ \citenamefont {Kliesch}}]{Bassler2023}%
  \BibitemOpen
  \bibfield  {author} {\bibinfo {author} {\bibfnamefont {P.}~\bibnamefont {Ba{\ss}ler}}, \bibinfo {author} {\bibfnamefont {M.}~\bibnamefont {Zipper}}, \bibinfo {author} {\bibfnamefont {C.}~\bibnamefont {Cedzich}}, \bibinfo {author} {\bibfnamefont {M.}~\bibnamefont {Heinrich}}, \bibinfo {author} {\bibfnamefont {P.~H.}\ \bibnamefont {Huber}}, \bibinfo {author} {\bibfnamefont {M.}~\bibnamefont {Johanning}},\ and\ \bibinfo {author} {\bibfnamefont {M.}~\bibnamefont {Kliesch}},\ }\href {https://doi.org/10.22331/q-2023-04-20-984} {\bibfield  {journal} {\bibinfo  {journal} {Quantum}\ }\textbf {\bibinfo {volume} {7}},\ \bibinfo {pages} {984} (\bibinfo {year} {2023})}\BibitemShut {NoStop}%
\bibitem [{\citenamefont {Weidt}\ \emph {et~al.}(2016)\citenamefont {Weidt}, \citenamefont {Randall}, \citenamefont {Webster}, \citenamefont {Lake}, \citenamefont {Webb}, \citenamefont {Cohen}, \citenamefont {Navickas}, \citenamefont {Lekitsch}, \citenamefont {Retzker},\ and\ \citenamefont {Hensinger}}]{Weidt2016}%
  \BibitemOpen
  \bibfield  {author} {\bibinfo {author} {\bibfnamefont {S.}~\bibnamefont {Weidt}}, \bibinfo {author} {\bibfnamefont {J.}~\bibnamefont {Randall}}, \bibinfo {author} {\bibfnamefont {S.~C.}\ \bibnamefont {Webster}}, \bibinfo {author} {\bibfnamefont {K.}~\bibnamefont {Lake}}, \bibinfo {author} {\bibfnamefont {A.~E.}\ \bibnamefont {Webb}}, \bibinfo {author} {\bibfnamefont {I.}~\bibnamefont {Cohen}}, \bibinfo {author} {\bibfnamefont {T.}~\bibnamefont {Navickas}}, \bibinfo {author} {\bibfnamefont {B.}~\bibnamefont {Lekitsch}}, \bibinfo {author} {\bibfnamefont {A.}~\bibnamefont {Retzker}},\ and\ \bibinfo {author} {\bibfnamefont {W.~K.}\ \bibnamefont {Hensinger}},\ }\href {https://doi.org/10.1103/physrevlett.117.220501} {\bibfield  {journal} {\bibinfo  {journal} {Physical Review Letters}\ }\textbf {\bibinfo {volume} {117}},\ \bibinfo {pages} {220501} (\bibinfo {year} {2016})}\BibitemShut {NoStop}%
\bibitem [{\citenamefont {Arrazola}\ \emph {et~al.}(2018)\citenamefont {Arrazola}, \citenamefont {Casanova}, \citenamefont {Pedernales}, \citenamefont {Wang}, \citenamefont {Solano},\ and\ \citenamefont {Plenio}}]{Arrazola2018}%
  \BibitemOpen
  \bibfield  {author} {\bibinfo {author} {\bibfnamefont {I.}~\bibnamefont {Arrazola}}, \bibinfo {author} {\bibfnamefont {J.}~\bibnamefont {Casanova}}, \bibinfo {author} {\bibfnamefont {J.~S.}\ \bibnamefont {Pedernales}}, \bibinfo {author} {\bibfnamefont {Z.-Y.}\ \bibnamefont {Wang}}, \bibinfo {author} {\bibfnamefont {E.}~\bibnamefont {Solano}},\ and\ \bibinfo {author} {\bibfnamefont {M.~B.}\ \bibnamefont {Plenio}},\ }\href {https://doi.org/10.1103/physreva.97.052312} {\bibfield  {journal} {\bibinfo  {journal} {Physical Review A}\ }\textbf {\bibinfo {volume} {97}},\ \bibinfo {pages} {052312} (\bibinfo {year} {2018})}\BibitemShut {NoStop}%
\bibitem [{\citenamefont {Huber}\ \emph {et~al.}(2021)\citenamefont {Huber}, \citenamefont {Haber}, \citenamefont {Barthel}, \citenamefont {García-Ripoll}, \citenamefont {Torrontegui},\ and\ \citenamefont {Wunderlich}}]{Huber2021}%
  \BibitemOpen
  \bibfield  {author} {\bibinfo {author} {\bibfnamefont {P.}~\bibnamefont {Huber}}, \bibinfo {author} {\bibfnamefont {J.}~\bibnamefont {Haber}}, \bibinfo {author} {\bibfnamefont {P.}~\bibnamefont {Barthel}}, \bibinfo {author} {\bibfnamefont {J.~J.}\ \bibnamefont {García-Ripoll}}, \bibinfo {author} {\bibfnamefont {E.}~\bibnamefont {Torrontegui}},\ and\ \bibinfo {author} {\bibfnamefont {C.}~\bibnamefont {Wunderlich}},\ }\href {https://doi.org/10.48550/ARXIV.2111.08977} {\bibinfo {title} {Realization of a quantum perceptron gate with trapped ions}} (\bibinfo {year} {2021})\BibitemShut {NoStop}%
\bibitem [{\citenamefont {Leu}\ \emph {et~al.}(2023)\citenamefont {Leu}, \citenamefont {Gely}, \citenamefont {Weber}, \citenamefont {Smith}, \citenamefont {Nadlinger},\ and\ \citenamefont {Lucas}}]{Leu2023}%
  \BibitemOpen
  \bibfield  {author} {\bibinfo {author} {\bibfnamefont {A.~D.}\ \bibnamefont {Leu}}, \bibinfo {author} {\bibfnamefont {M.~F.}\ \bibnamefont {Gely}}, \bibinfo {author} {\bibfnamefont {M.~A.}\ \bibnamefont {Weber}}, \bibinfo {author} {\bibfnamefont {M.~C.}\ \bibnamefont {Smith}}, \bibinfo {author} {\bibfnamefont {D.~P.}\ \bibnamefont {Nadlinger}},\ and\ \bibinfo {author} {\bibfnamefont {D.~M.}\ \bibnamefont {Lucas}},\ }\href {https://doi.org/10.1103/physrevlett.131.120601} {\bibfield  {journal} {\bibinfo  {journal} {Physical Review Letters}\ }\textbf {\bibinfo {volume} {131}},\ \bibinfo {pages} {120601} (\bibinfo {year} {2023})}\BibitemShut {NoStop}%
\bibitem [{\citenamefont {Nagies}\ \emph {et~al.}(2025{\natexlab{a}})\citenamefont {Nagies}, \citenamefont {Geier}, \citenamefont {Akram}, \citenamefont {Okamoto}, \citenamefont {Bantounas}, \citenamefont {Wunderlich}, \citenamefont {Johanning},\ and\ \citenamefont {Hauke}}]{Nagies2025}%
  \BibitemOpen
  \bibfield  {author} {\bibinfo {author} {\bibfnamefont {S.}~\bibnamefont {Nagies}}, \bibinfo {author} {\bibfnamefont {K.~T.}\ \bibnamefont {Geier}}, \bibinfo {author} {\bibfnamefont {J.}~\bibnamefont {Akram}}, \bibinfo {author} {\bibfnamefont {J.}~\bibnamefont {Okamoto}}, \bibinfo {author} {\bibfnamefont {D.}~\bibnamefont {Bantounas}}, \bibinfo {author} {\bibfnamefont {C.}~\bibnamefont {Wunderlich}}, \bibinfo {author} {\bibfnamefont {M.}~\bibnamefont {Johanning}},\ and\ \bibinfo {author} {\bibfnamefont {P.}~\bibnamefont {Hauke}},\ }\href {https://doi.org/10.1088/2058-9565/adc1fe} {\bibfield  {journal} {\bibinfo  {journal} {Quantum Science and Technology}\ }\textbf {\bibinfo {volume} {10}},\ \bibinfo {pages} {025051} (\bibinfo {year} {2025}{\natexlab{a}})}\BibitemShut {NoStop}%
\bibitem [{\citenamefont {Viola}\ and\ \citenamefont {Lloyd}(1998)}]{Viola1998}%
  \BibitemOpen
  \bibfield  {author} {\bibinfo {author} {\bibfnamefont {L.}~\bibnamefont {Viola}}\ and\ \bibinfo {author} {\bibfnamefont {S.}~\bibnamefont {Lloyd}},\ }\href {https://doi.org/10.1103/physreva.58.2733} {\bibfield  {journal} {\bibinfo  {journal} {Physical Review A}\ }\textbf {\bibinfo {volume} {58}},\ \bibinfo {pages} {2733} (\bibinfo {year} {1998})}\BibitemShut {NoStop}%
\bibitem [{\citenamefont {Khodjasteh}\ and\ \citenamefont {Lidar}()}]{Khodjasteh}%
  \BibitemOpen
  \bibfield  {author} {\bibinfo {author} {\bibfnamefont {K.}~\bibnamefont {Khodjasteh}}\ and\ \bibinfo {author} {\bibfnamefont {D.}~\bibnamefont {Lidar}},\ }in\ \href {https://doi.org/10.1109/qels.2005.1548669} {\emph {\bibinfo {booktitle} {2005 Quantum Electronics and Laser Science Conference}}},\ \bibinfo {series and number} {QELS-05}\ (\bibinfo  {publisher} {IEEE})\BibitemShut {NoStop}%
\bibitem [{\citenamefont {Souza}\ \emph {et~al.}(2011)\citenamefont {Souza}, \citenamefont {\'alvarez},\ and\ \citenamefont {Suter}}]{Souza2011}%
  \BibitemOpen
  \bibfield  {author} {\bibinfo {author} {\bibfnamefont {A.~M.}\ \bibnamefont {Souza}}, \bibinfo {author} {\bibfnamefont {G.~A.}\ \bibnamefont {\'alvarez}},\ and\ \bibinfo {author} {\bibfnamefont {D.}~\bibnamefont {Suter}},\ }\href {https://doi.org/10.1103/physrevlett.106.240501} {\bibfield  {journal} {\bibinfo  {journal} {Physical Review Letters}\ }\textbf {\bibinfo {volume} {106}},\ \bibinfo {pages} {240501} (\bibinfo {year} {2011})}\BibitemShut {NoStop}%
\bibitem [{\citenamefont {Cai}\ \emph {et~al.}(2023)\citenamefont {Cai}, \citenamefont {Babbush}, \citenamefont {Benjamin}, \citenamefont {Endo}, \citenamefont {Huggins}, \citenamefont {Li}, \citenamefont {McClean},\ and\ \citenamefont {O’Brien}}]{Cai2023}%
  \BibitemOpen
  \bibfield  {author} {\bibinfo {author} {\bibfnamefont {Z.}~\bibnamefont {Cai}}, \bibinfo {author} {\bibfnamefont {R.}~\bibnamefont {Babbush}}, \bibinfo {author} {\bibfnamefont {S.~C.}\ \bibnamefont {Benjamin}}, \bibinfo {author} {\bibfnamefont {S.}~\bibnamefont {Endo}}, \bibinfo {author} {\bibfnamefont {W.~J.}\ \bibnamefont {Huggins}}, \bibinfo {author} {\bibfnamefont {Y.}~\bibnamefont {Li}}, \bibinfo {author} {\bibfnamefont {J.~R.}\ \bibnamefont {McClean}},\ and\ \bibinfo {author} {\bibfnamefont {T.~E.}\ \bibnamefont {O’Brien}},\ }\href {https://doi.org/10.1103/revmodphys.95.045005} {\bibfield  {journal} {\bibinfo  {journal} {Reviews of Modern Physics}\ }\textbf {\bibinfo {volume} {95}},\ \bibinfo {pages} {045005} (\bibinfo {year} {2023})}\BibitemShut {NoStop}%
\bibitem [{\citenamefont {Piltz}\ \emph {et~al.}(2013)\citenamefont {Piltz}, \citenamefont {Scharfenberger}, \citenamefont {Khromova}, \citenamefont {Varón},\ and\ \citenamefont {Wunderlich}}]{Piltz2013}%
  \BibitemOpen
  \bibfield  {author} {\bibinfo {author} {\bibfnamefont {C.}~\bibnamefont {Piltz}}, \bibinfo {author} {\bibfnamefont {B.}~\bibnamefont {Scharfenberger}}, \bibinfo {author} {\bibfnamefont {A.}~\bibnamefont {Khromova}}, \bibinfo {author} {\bibfnamefont {A.~F.}\ \bibnamefont {Varón}},\ and\ \bibinfo {author} {\bibfnamefont {C.}~\bibnamefont {Wunderlich}},\ }\href {https://doi.org/10.1103/physrevlett.110.200501} {\bibfield  {journal} {\bibinfo  {journal} {Physical Review Letters}\ }\textbf {\bibinfo {volume} {110}},\ \bibinfo {pages} {200501} (\bibinfo {year} {2013})}\BibitemShut {NoStop}%
\bibitem [{\citenamefont {Genov}\ \emph {et~al.}(2017)\citenamefont {Genov}, \citenamefont {Schraft}, \citenamefont {Vitanov},\ and\ \citenamefont {Halfmann}}]{Genov2017}%
  \BibitemOpen
  \bibfield  {author} {\bibinfo {author} {\bibfnamefont {G.~T.}\ \bibnamefont {Genov}}, \bibinfo {author} {\bibfnamefont {D.}~\bibnamefont {Schraft}}, \bibinfo {author} {\bibfnamefont {N.~V.}\ \bibnamefont {Vitanov}},\ and\ \bibinfo {author} {\bibfnamefont {T.}~\bibnamefont {Halfmann}},\ }\href {https://doi.org/10.1103/physrevlett.118.133202} {\bibfield  {journal} {\bibinfo  {journal} {Physical Review Letters}\ }\textbf {\bibinfo {volume} {118}},\ \bibinfo {pages} {133202} (\bibinfo {year} {2017})}\BibitemShut {NoStop}%
\bibitem [{\citenamefont {Valahu}\ \emph {et~al.}(2021)\citenamefont {Valahu}, \citenamefont {Lawrence}, \citenamefont {Weidt},\ and\ \citenamefont {Hensinger}}]{Valahu2021}%
  \BibitemOpen
  \bibfield  {author} {\bibinfo {author} {\bibfnamefont {C.~H.}\ \bibnamefont {Valahu}}, \bibinfo {author} {\bibfnamefont {A.~M.}\ \bibnamefont {Lawrence}}, \bibinfo {author} {\bibfnamefont {S.}~\bibnamefont {Weidt}},\ and\ \bibinfo {author} {\bibfnamefont {W.~K.}\ \bibnamefont {Hensinger}},\ }\href {https://doi.org/10.1088/1367-2630/ac320e} {\bibfield  {journal} {\bibinfo  {journal} {New Journal of Physics}\ }\textbf {\bibinfo {volume} {23}},\ \bibinfo {pages} {113012} (\bibinfo {year} {2021})}\BibitemShut {NoStop}%
\bibitem [{\citenamefont {Morong}\ \emph {et~al.}(2023)\citenamefont {Morong}, \citenamefont {Collins}, \citenamefont {De}, \citenamefont {Stavropoulos}, \citenamefont {You},\ and\ \citenamefont {Monroe}}]{Morong2023}%
  \BibitemOpen
  \bibfield  {author} {\bibinfo {author} {\bibfnamefont {W.}~\bibnamefont {Morong}}, \bibinfo {author} {\bibfnamefont {K.}~\bibnamefont {Collins}}, \bibinfo {author} {\bibfnamefont {A.}~\bibnamefont {De}}, \bibinfo {author} {\bibfnamefont {E.}~\bibnamefont {Stavropoulos}}, \bibinfo {author} {\bibfnamefont {T.}~\bibnamefont {You}},\ and\ \bibinfo {author} {\bibfnamefont {C.}~\bibnamefont {Monroe}},\ }\href {https://doi.org/10.1103/prxquantum.4.010334} {\bibfield  {journal} {\bibinfo  {journal} {PRX Quantum}\ }\textbf {\bibinfo {volume} {4}},\ \bibinfo {pages} {010334} (\bibinfo {year} {2023})}\BibitemShut {NoStop}%
\bibitem [{\citenamefont {Barthel}\ \emph {et~al.}(2023)\citenamefont {Barthel}, \citenamefont {Huber}, \citenamefont {Casanova}, \citenamefont {Arrazola}, \citenamefont {Niroomand}, \citenamefont {Sriarunothai}, \citenamefont {Plenio},\ and\ \citenamefont {Wunderlich}}]{Barthel2023}%
  \BibitemOpen
  \bibfield  {author} {\bibinfo {author} {\bibfnamefont {P.}~\bibnamefont {Barthel}}, \bibinfo {author} {\bibfnamefont {P.~H.}\ \bibnamefont {Huber}}, \bibinfo {author} {\bibfnamefont {J.}~\bibnamefont {Casanova}}, \bibinfo {author} {\bibfnamefont {I.}~\bibnamefont {Arrazola}}, \bibinfo {author} {\bibfnamefont {D.}~\bibnamefont {Niroomand}}, \bibinfo {author} {\bibfnamefont {T.}~\bibnamefont {Sriarunothai}}, \bibinfo {author} {\bibfnamefont {M.~B.}\ \bibnamefont {Plenio}},\ and\ \bibinfo {author} {\bibfnamefont {C.}~\bibnamefont {Wunderlich}},\ }\href {https://doi.org/10.1088/1367-2630/acd4db} {\bibfield  {journal} {\bibinfo  {journal} {New Journal of Physics}\ }\textbf {\bibinfo {volume} {25}},\ \bibinfo {pages} {063023} (\bibinfo {year} {2023})}\BibitemShut {NoStop}%
\bibitem [{\citenamefont {N\"unnerich}\ \emph {et~al.}(2024)\citenamefont {N\"unnerich}, \citenamefont {Cohen}, \citenamefont {Barthel}, \citenamefont {Huber}, \citenamefont {Niroomand}, \citenamefont {Retzker},\ and\ \citenamefont {Wunderlich}}]{Nuennerich2024}%
  \BibitemOpen
  \bibfield  {author} {\bibinfo {author} {\bibfnamefont {M.}~\bibnamefont {N\"unnerich}}, \bibinfo {author} {\bibfnamefont {D.}~\bibnamefont {Cohen}}, \bibinfo {author} {\bibfnamefont {P.}~\bibnamefont {Barthel}}, \bibinfo {author} {\bibfnamefont {P.~H.}\ \bibnamefont {Huber}}, \bibinfo {author} {\bibfnamefont {D.}~\bibnamefont {Niroomand}}, \bibinfo {author} {\bibfnamefont {A.}~\bibnamefont {Retzker}},\ and\ \bibinfo {author} {\bibfnamefont {C.}~\bibnamefont {Wunderlich}},\ }\href {https://doi.org/10.48550/ARXIV.2403.04730} {\bibinfo {title} {Fast, robust and laser-free universal entangling gates for trapped-ion quantum computing}} (\bibinfo {year} {2024})\BibitemShut {NoStop}%
\bibitem [{\citenamefont {Lidar}(2008)}]{Lidar2008}%
  \BibitemOpen
  \bibfield  {author} {\bibinfo {author} {\bibfnamefont {D.~A.}\ \bibnamefont {Lidar}},\ }\href {https://doi.org/10.1103/physrevlett.100.160506} {\bibfield  {journal} {\bibinfo  {journal} {Physical Review Letters}\ }\textbf {\bibinfo {volume} {100}},\ \bibinfo {pages} {160506} (\bibinfo {year} {2008})}\BibitemShut {NoStop}%
\bibitem [{\citenamefont {Quiroz}\ and\ \citenamefont {Lidar}(2012)}]{Quiroz2012}%
  \BibitemOpen
  \bibfield  {author} {\bibinfo {author} {\bibfnamefont {G.}~\bibnamefont {Quiroz}}\ and\ \bibinfo {author} {\bibfnamefont {D.~A.}\ \bibnamefont {Lidar}},\ }\href {https://doi.org/10.1103/physreva.86.042333} {\bibfield  {journal} {\bibinfo  {journal} {Physical Review A}\ }\textbf {\bibinfo {volume} {86}},\ \bibinfo {pages} {042333} (\bibinfo {year} {2012})}\BibitemShut {NoStop}%
\bibitem [{\citenamefont {Zaech}\ \emph {et~al.}(2022)\citenamefont {Zaech}, \citenamefont {Liniger}, \citenamefont {Danelljan}, \citenamefont {Dai},\ and\ \citenamefont {Van~Gool}}]{Zaech2022}%
  \BibitemOpen
  \bibfield  {author} {\bibinfo {author} {\bibfnamefont {J.-N.}\ \bibnamefont {Zaech}}, \bibinfo {author} {\bibfnamefont {A.}~\bibnamefont {Liniger}}, \bibinfo {author} {\bibfnamefont {M.}~\bibnamefont {Danelljan}}, \bibinfo {author} {\bibfnamefont {D.}~\bibnamefont {Dai}},\ and\ \bibinfo {author} {\bibfnamefont {L.}~\bibnamefont {Van~Gool}},\ }in\ \href {https://doi.org/10.1109/cvpr52688.2022.00861} {\emph {\bibinfo {booktitle} {2022 IEEE/CVF Conference on Computer Vision and Pattern Recognition (CVPR)}}}\ (\bibinfo  {publisher} {IEEE},\ \bibinfo {year} {2022})\BibitemShut {NoStop}%
\bibitem [{\citenamefont {Sherrington}\ and\ \citenamefont {Kirkpatrick}(1975)}]{Sherrington1975}%
  \BibitemOpen
  \bibfield  {author} {\bibinfo {author} {\bibfnamefont {D.}~\bibnamefont {Sherrington}}\ and\ \bibinfo {author} {\bibfnamefont {S.}~\bibnamefont {Kirkpatrick}},\ }\href {https://doi.org/10.1103/physrevlett.35.1792} {\bibfield  {journal} {\bibinfo  {journal} {Physical Review Letters}\ }\textbf {\bibinfo {volume} {35}},\ \bibinfo {pages} {1792} (\bibinfo {year} {1975})}\BibitemShut {NoStop}%
\bibitem [{\citenamefont {Panchenko}(2013)}]{Panchenko2013}%
  \BibitemOpen
  \bibfield  {author} {\bibinfo {author} {\bibfnamefont {D.}~\bibnamefont {Panchenko}},\ }\href {https://doi.org/10.1007/978-1-4614-6289-7} {\emph {\bibinfo {title} {The Sherrington-Kirkpatrick Model}}}\ (\bibinfo  {publisher} {Springer New York},\ \bibinfo {year} {2013})\BibitemShut {NoStop}%
\bibitem [{\citenamefont {Leal-Taix{\'e}}\ \emph {et~al.}(2015)\citenamefont {Leal-Taix{\'e}}, \citenamefont {Milan}, \citenamefont {Reid}, \citenamefont {Roth},\ and\ \citenamefont {Schindler}}]{leal2015motchallenge}%
  \BibitemOpen
  \bibfield  {author} {\bibinfo {author} {\bibfnamefont {L.}~\bibnamefont {Leal-Taix{\'e}}}, \bibinfo {author} {\bibfnamefont {A.}~\bibnamefont {Milan}}, \bibinfo {author} {\bibfnamefont {I.}~\bibnamefont {Reid}}, \bibinfo {author} {\bibfnamefont {S.}~\bibnamefont {Roth}},\ and\ \bibinfo {author} {\bibfnamefont {K.}~\bibnamefont {Schindler}},\ }\href {https://doi.org/https://doi.org/10.48550/arXiv.1504.01942} {\bibinfo {title} {Motchallenge 2015: Towards a benchmark for multi-target tracking}} (\bibinfo {year} {2015})\BibitemShut {NoStop}%
\bibitem [{\citenamefont {Jansen}\ \emph {et~al.}(2007)\citenamefont {Jansen}, \citenamefont {Ruskai},\ and\ \citenamefont {Seiler}}]{Jansen2007}%
  \BibitemOpen
  \bibfield  {author} {\bibinfo {author} {\bibfnamefont {S.}~\bibnamefont {Jansen}}, \bibinfo {author} {\bibfnamefont {M.-B.}\ \bibnamefont {Ruskai}},\ and\ \bibinfo {author} {\bibfnamefont {R.}~\bibnamefont {Seiler}},\ }\href {https://doi.org/10.1063/1.2798382} {\bibfield  {journal} {\bibinfo  {journal} {Journal of Mathematical Physics}\ }\textbf {\bibinfo {volume} {48}},\ \bibinfo {pages} {102111} (\bibinfo {year} {2007})}\BibitemShut {NoStop}%
\bibitem [{\citenamefont {Lidar}\ \emph {et~al.}(2009)\citenamefont {Lidar}, \citenamefont {Rezakhani},\ and\ \citenamefont {Hamma}}]{Lidar2009}%
  \BibitemOpen
  \bibfield  {author} {\bibinfo {author} {\bibfnamefont {D.~A.}\ \bibnamefont {Lidar}}, \bibinfo {author} {\bibfnamefont {A.~T.}\ \bibnamefont {Rezakhani}},\ and\ \bibinfo {author} {\bibfnamefont {A.}~\bibnamefont {Hamma}},\ }\bibfield  {journal} {\bibinfo  {journal} {Journal of Mathematical Physics}\ }\textbf {\bibinfo {volume} {50}},\ \href {https://doi.org/10.1063/1.3236685} {10.1063/1.3236685} (\bibinfo {year} {2009})\BibitemShut {NoStop}%
\bibitem [{\citenamefont {Amin}(2009)}]{Amin2009}%
  \BibitemOpen
  \bibfield  {author} {\bibinfo {author} {\bibfnamefont {M.~H.~S.}\ \bibnamefont {Amin}},\ }\href {https://doi.org/10.1103/physrevlett.102.220401} {\bibfield  {journal} {\bibinfo  {journal} {Physical Review Letters}\ }\textbf {\bibinfo {volume} {102}},\ \bibinfo {pages} {220401} (\bibinfo {year} {2009})}\BibitemShut {NoStop}%
\bibitem [{\citenamefont {Cheung}\ \emph {et~al.}(2011)\citenamefont {Cheung}, \citenamefont {Høyer},\ and\ \citenamefont {Wiebe}}]{Cheung2011}%
  \BibitemOpen
  \bibfield  {author} {\bibinfo {author} {\bibfnamefont {D.}~\bibnamefont {Cheung}}, \bibinfo {author} {\bibfnamefont {P.}~\bibnamefont {Høyer}},\ and\ \bibinfo {author} {\bibfnamefont {N.}~\bibnamefont {Wiebe}},\ }\href {https://doi.org/10.1088/1751-8113/44/41/415302} {\bibfield  {journal} {\bibinfo  {journal} {Journal of Physics A: Mathematical and Theoretical}\ }\textbf {\bibinfo {volume} {44}},\ \bibinfo {pages} {415302} (\bibinfo {year} {2011})}\BibitemShut {NoStop}%
\bibitem [{\citenamefont {{\v{C}}epaitė}\ \emph {et~al.}(2023)\citenamefont {{\v{C}}epaitė}, \citenamefont {Polkovnikov}, \citenamefont {Daley},\ and\ \citenamefont {Duncan}}]{Cepaite2023}%
  \BibitemOpen
  \bibfield  {author} {\bibinfo {author} {\bibfnamefont {I.}~\bibnamefont {{\v{C}}epaitė}}, \bibinfo {author} {\bibfnamefont {A.}~\bibnamefont {Polkovnikov}}, \bibinfo {author} {\bibfnamefont {A.~J.}\ \bibnamefont {Daley}},\ and\ \bibinfo {author} {\bibfnamefont {C.~W.}\ \bibnamefont {Duncan}},\ }\href {https://doi.org/10.1103/prxquantum.4.010312} {\bibfield  {journal} {\bibinfo  {journal} {PRX Quantum}\ }\textbf {\bibinfo {volume} {4}},\ \bibinfo {pages} {010312} (\bibinfo {year} {2023})}\BibitemShut {NoStop}%
\bibitem [{\citenamefont {García-Pintos}\ \emph {et~al.}(2024)\citenamefont {García-Pintos}, \citenamefont {Sahasrabudhe},\ and\ \citenamefont {Arenz}}]{GarciaPintos2024}%
  \BibitemOpen
  \bibfield  {author} {\bibinfo {author} {\bibfnamefont {L.~P.}\ \bibnamefont {García-Pintos}}, \bibinfo {author} {\bibfnamefont {M.}~\bibnamefont {Sahasrabudhe}},\ and\ \bibinfo {author} {\bibfnamefont {C.}~\bibnamefont {Arenz}},\ }\href@noop {} {\bibinfo {title} {Tighter lower bounds on quantum annealing times}} (\bibinfo {year} {2024}),\ \Eprint {https://arxiv.org/abs/2410.14779} {arXiv:2410.14779 [quant-ph]} \BibitemShut {NoStop}%
\bibitem [{\citenamefont {Bottarelli}\ \emph {et~al.}(2024)\citenamefont {Bottarelli}, \citenamefont {de~Andoin}, \citenamefont {Chandarana}, \citenamefont {Paul}, \citenamefont {Chen}, \citenamefont {Sanz},\ and\ \citenamefont {Hauke}}]{Bottarelli2024}%
  \BibitemOpen
  \bibfield  {author} {\bibinfo {author} {\bibfnamefont {A.}~\bibnamefont {Bottarelli}}, \bibinfo {author} {\bibfnamefont {M.~G.}\ \bibnamefont {de~Andoin}}, \bibinfo {author} {\bibfnamefont {P.}~\bibnamefont {Chandarana}}, \bibinfo {author} {\bibfnamefont {K.}~\bibnamefont {Paul}}, \bibinfo {author} {\bibfnamefont {X.}~\bibnamefont {Chen}}, \bibinfo {author} {\bibfnamefont {M.}~\bibnamefont {Sanz}},\ and\ \bibinfo {author} {\bibfnamefont {P.}~\bibnamefont {Hauke}},\ }\href@noop {} {\bibinfo {title} {Symmetry-enhanced counterdiabatic quantum algorithm for qudits}} (\bibinfo {year} {2024}),\ \Eprint {https://arxiv.org/abs/2410.06710} {arXiv:2410.06710 [quant-ph]} \BibitemShut {NoStop}%
\bibitem [{\citenamefont {Choi}(2021)}]{Choi2021}%
  \BibitemOpen
  \bibfield  {author} {\bibinfo {author} {\bibfnamefont {V.}~\bibnamefont {Choi}},\ }\href {https://doi.org/10.48550/ARXIV.2105.02110} {\bibinfo {title} {Essentiality of the non-stoquastic hamiltonians and driver graph design in quantum optimization annealing}} (\bibinfo {year} {2021})\BibitemShut {NoStop}%
\bibitem [{\citenamefont {Feinstein}\ \emph {et~al.}(2025)\citenamefont {Feinstein}, \citenamefont {Shalashilin}, \citenamefont {Bose},\ and\ \citenamefont {Warburton}}]{Feinstein2025}%
  \BibitemOpen
  \bibfield  {author} {\bibinfo {author} {\bibfnamefont {N.}~\bibnamefont {Feinstein}}, \bibinfo {author} {\bibfnamefont {I.}~\bibnamefont {Shalashilin}}, \bibinfo {author} {\bibfnamefont {S.}~\bibnamefont {Bose}},\ and\ \bibinfo {author} {\bibfnamefont {P.~A.}\ \bibnamefont {Warburton}},\ }\href {https://doi.org/10.1088/2058-9565/adab14} {\bibfield  {journal} {\bibinfo  {journal} {Quantum Science and Technology}\ }\textbf {\bibinfo {volume} {10}},\ \bibinfo {pages} {025011} (\bibinfo {year} {2025})}\BibitemShut {NoStop}%
\bibitem [{\citenamefont {Nagies}\ \emph {et~al.}(2025{\natexlab{b}})\citenamefont {Nagies}, \citenamefont {Geier}, \citenamefont {Akram}, \citenamefont {Bantounas}, \citenamefont {Johanning},\ and\ \citenamefont {Hauke}}]{Nagies2025a}%
  \BibitemOpen
  \bibfield  {author} {\bibinfo {author} {\bibfnamefont {S.}~\bibnamefont {Nagies}}, \bibinfo {author} {\bibfnamefont {K.~T.}\ \bibnamefont {Geier}}, \bibinfo {author} {\bibfnamefont {J.}~\bibnamefont {Akram}}, \bibinfo {author} {\bibfnamefont {D.}~\bibnamefont {Bantounas}}, \bibinfo {author} {\bibfnamefont {M.}~\bibnamefont {Johanning}},\ and\ \bibinfo {author} {\bibfnamefont {P.}~\bibnamefont {Hauke}},\ }\href {https://doi.org/10.1088/2058-9565/adcae6} {\bibfield  {journal} {\bibinfo  {journal} {Quantum Science and Technology}\ }\textbf {\bibinfo {volume} {10}},\ \bibinfo {pages} {035008} (\bibinfo {year} {2025}{\natexlab{b}})}\BibitemShut {NoStop}%
\bibitem [{\citenamefont {Luo}\ \emph {et~al.}(2021)\citenamefont {Luo}, \citenamefont {Xing}, \citenamefont {Milan}, \citenamefont {Zhang}, \citenamefont {Liu},\ and\ \citenamefont {Kim}}]{Luo2021}%
  \BibitemOpen
  \bibfield  {author} {\bibinfo {author} {\bibfnamefont {W.}~\bibnamefont {Luo}}, \bibinfo {author} {\bibfnamefont {J.}~\bibnamefont {Xing}}, \bibinfo {author} {\bibfnamefont {A.}~\bibnamefont {Milan}}, \bibinfo {author} {\bibfnamefont {X.}~\bibnamefont {Zhang}}, \bibinfo {author} {\bibfnamefont {W.}~\bibnamefont {Liu}},\ and\ \bibinfo {author} {\bibfnamefont {T.-K.}\ \bibnamefont {Kim}},\ }\href {https://doi.org/10.1016/j.artint.2020.103448} {\bibfield  {journal} {\bibinfo  {journal} {Artificial Intelligence}\ }\textbf {\bibinfo {volume} {293}},\ \bibinfo {pages} {103448} (\bibinfo {year} {2021})}\BibitemShut {NoStop}%
\bibitem [{\citenamefont {Zhao}\ \emph {et~al.}(2019)\citenamefont {Zhao}, \citenamefont {Zheng}, \citenamefont {Xu},\ and\ \citenamefont {Wu}}]{zhao2019object}%
  \BibitemOpen
  \bibfield  {author} {\bibinfo {author} {\bibfnamefont {Z.-Q.}\ \bibnamefont {Zhao}}, \bibinfo {author} {\bibfnamefont {P.}~\bibnamefont {Zheng}}, \bibinfo {author} {\bibfnamefont {S.-t.}\ \bibnamefont {Xu}},\ and\ \bibinfo {author} {\bibfnamefont {X.}~\bibnamefont {Wu}},\ }\href {https://doi.org/10.1109/TNNLS.2018.2876865} {\bibfield  {journal} {\bibinfo  {journal} {IEEE transactions on neural networks and learning systems}\ }\textbf {\bibinfo {volume} {30}},\ \bibinfo {pages} {3212} (\bibinfo {year} {2019})}\BibitemShut {NoStop}%
\bibitem [{\citenamefont {Ganian}\ \emph {et~al.}(2021)\citenamefont {Ganian}, \citenamefont {Hamm},\ and\ \citenamefont {Ordyniak}}]{ganian2021complexity}%
  \BibitemOpen
  \bibfield  {author} {\bibinfo {author} {\bibfnamefont {R.}~\bibnamefont {Ganian}}, \bibinfo {author} {\bibfnamefont {T.}~\bibnamefont {Hamm}},\ and\ \bibinfo {author} {\bibfnamefont {S.}~\bibnamefont {Ordyniak}},\ }in\ \href@noop {} {\emph {\bibinfo {booktitle} {Proceedings of the AAAI Conference on Artificial Intelligence}}},\ Vol.~\bibinfo {volume} {35}\ (\bibinfo {year} {2021})\ pp.\ \bibinfo {pages} {1388--1396}\BibitemShut {NoStop}%
\bibitem [{\citenamefont {Ihara}(2025)}]{ihara2025enhancing}%
  \BibitemOpen
  \bibfield  {author} {\bibinfo {author} {\bibfnamefont {Y.}~\bibnamefont {Ihara}},\ }\href {https://doi.org/10.1038/s41598-025-07492-7} {\bibfield  {journal} {\bibinfo  {journal} {Scientific Reports}\ }\textbf {\bibinfo {volume} {15}},\ \bibinfo {pages} {24294} (\bibinfo {year} {2025})}\BibitemShut {NoStop}%
\bibitem [{\citenamefont {Zheng}\ \emph {et~al.}(2019)\citenamefont {Zheng}, \citenamefont {Yang}, \citenamefont {Yu}, \citenamefont {Zheng}, \citenamefont {Yang},\ and\ \citenamefont {Kautz}}]{zheng2019joint}%
  \BibitemOpen
  \bibfield  {author} {\bibinfo {author} {\bibfnamefont {Z.}~\bibnamefont {Zheng}}, \bibinfo {author} {\bibfnamefont {X.}~\bibnamefont {Yang}}, \bibinfo {author} {\bibfnamefont {Z.}~\bibnamefont {Yu}}, \bibinfo {author} {\bibfnamefont {L.}~\bibnamefont {Zheng}}, \bibinfo {author} {\bibfnamefont {Y.}~\bibnamefont {Yang}},\ and\ \bibinfo {author} {\bibfnamefont {J.}~\bibnamefont {Kautz}},\ }in\ \href {https://doi.org/10.48550/arXiv.1904.07223} {\emph {\bibinfo {booktitle} {proceedings of the IEEE/CVF conference on computer vision and pattern recognition}}}\ (\bibinfo {year} {2019})\ pp.\ \bibinfo {pages} {2138--2147}\BibitemShut {NoStop}%
\bibitem [{\citenamefont {Hornakova}\ \emph {et~al.}(2021)\citenamefont {Hornakova}, \citenamefont {Kaiser}, \citenamefont {Swoboda}, \citenamefont {Rolinek}, \citenamefont {Rosenhahn},\ and\ \citenamefont {Henschel}}]{hornakova2021making}%
  \BibitemOpen
  \bibfield  {author} {\bibinfo {author} {\bibfnamefont {A.}~\bibnamefont {Hornakova}}, \bibinfo {author} {\bibfnamefont {T.}~\bibnamefont {Kaiser}}, \bibinfo {author} {\bibfnamefont {P.}~\bibnamefont {Swoboda}}, \bibinfo {author} {\bibfnamefont {M.}~\bibnamefont {Rolinek}}, \bibinfo {author} {\bibfnamefont {B.}~\bibnamefont {Rosenhahn}},\ and\ \bibinfo {author} {\bibfnamefont {R.}~\bibnamefont {Henschel}},\ }in\ \href {https://doi.org/10.48550/arXiv.2108.10606} {\emph {\bibinfo {booktitle} {Proceedings of the IEEE/CVF International Conference on Computer Vision}}}\ (\bibinfo {year} {2021})\ pp.\ \bibinfo {pages} {6330--6340}\BibitemShut {NoStop}%
\bibitem [{\citenamefont {Andriluka}\ \emph {et~al.}(2010)\citenamefont {Andriluka}, \citenamefont {Roth},\ and\ \citenamefont {Schiele}}]{andriluka2010monocular}%
  \BibitemOpen
  \bibfield  {author} {\bibinfo {author} {\bibfnamefont {M.}~\bibnamefont {Andriluka}}, \bibinfo {author} {\bibfnamefont {S.}~\bibnamefont {Roth}},\ and\ \bibinfo {author} {\bibfnamefont {B.}~\bibnamefont {Schiele}},\ }in\ \href {https://doi.org/10.1109/CVPR.2010.5540156} {\emph {\bibinfo {booktitle} {2010 IEEE Computer Society Conference on Computer Vision and Pattern Recognition}}}\ (\bibinfo {organization} {Ieee},\ \bibinfo {year} {2010})\ pp.\ \bibinfo {pages} {623--630}\BibitemShut {NoStop}%
\bibitem [{\citenamefont {Seelbach~Benkner}\ \emph {et~al.}(2020)\citenamefont {Seelbach~Benkner}, \citenamefont {Golyanik}, \citenamefont {Theobalt},\ and\ \citenamefont {Moeller}}]{benkner2020adiabatic}%
  \BibitemOpen
  \bibfield  {author} {\bibinfo {author} {\bibfnamefont {M.}~\bibnamefont {Seelbach~Benkner}}, \bibinfo {author} {\bibfnamefont {V.}~\bibnamefont {Golyanik}}, \bibinfo {author} {\bibfnamefont {C.}~\bibnamefont {Theobalt}},\ and\ \bibinfo {author} {\bibfnamefont {M.}~\bibnamefont {Moeller}},\ }in\ \href {https://doi.org/10.48550/arXiv.2107.04032} {\emph {\bibinfo {booktitle} {2020 International conference on 3D vision (3DV)}}}\ (\bibinfo {organization} {IEEE},\ \bibinfo {year} {2020})\ pp.\ \bibinfo {pages} {583--592}\BibitemShut {NoStop}%
\bibitem [{\citenamefont {Alessandroni}\ \emph {et~al.}(2023)\citenamefont {Alessandroni}, \citenamefont {Ramos-Calderer}, \citenamefont {Roth}, \citenamefont {Traversi},\ and\ \citenamefont {Aolita}}]{Alessandroni2023}%
  \BibitemOpen
  \bibfield  {author} {\bibinfo {author} {\bibfnamefont {E.}~\bibnamefont {Alessandroni}}, \bibinfo {author} {\bibfnamefont {S.}~\bibnamefont {Ramos-Calderer}}, \bibinfo {author} {\bibfnamefont {I.}~\bibnamefont {Roth}}, \bibinfo {author} {\bibfnamefont {E.}~\bibnamefont {Traversi}},\ and\ \bibinfo {author} {\bibfnamefont {L.}~\bibnamefont {Aolita}},\ }\href@noop {} {\bibinfo {title} {Alleviating the quantum big-$m$ problem}} (\bibinfo {year} {2023}),\ \Eprint {https://arxiv.org/abs/2307.10379} {arXiv:2307.10379 [quant-ph]} \BibitemShut {NoStop}%
\bibitem [{\citenamefont {Hauke}\ \emph {et~al.}(2015)\citenamefont {Hauke}, \citenamefont {Bonnes}, \citenamefont {Heyl},\ and\ \citenamefont {Lechner}}]{Hauke2015}%
  \BibitemOpen
  \bibfield  {author} {\bibinfo {author} {\bibfnamefont {P.}~\bibnamefont {Hauke}}, \bibinfo {author} {\bibfnamefont {L.}~\bibnamefont {Bonnes}}, \bibinfo {author} {\bibfnamefont {M.}~\bibnamefont {Heyl}},\ and\ \bibinfo {author} {\bibfnamefont {W.}~\bibnamefont {Lechner}},\ }\bibfield  {journal} {\bibinfo  {journal} {Frontiers in Physics}\ }\textbf {\bibinfo {volume} {3}},\ \href {https://doi.org/10.3389/fphy.2015.00021} {10.3389/fphy.2015.00021} (\bibinfo {year} {2015})\BibitemShut {NoStop}%
\bibitem [{\citenamefont {M{\o}lmer}\ and\ \citenamefont {S{\o}rensen}(1999)}]{Moelmer1999}%
  \BibitemOpen
  \bibfield  {author} {\bibinfo {author} {\bibfnamefont {K.}~\bibnamefont {M{\o}lmer}}\ and\ \bibinfo {author} {\bibfnamefont {A.}~\bibnamefont {S{\o}rensen}},\ }\href {https://doi.org/10.1103/physrevlett.82.1835} {\bibfield  {journal} {\bibinfo  {journal} {Physical Review Letters}\ }\textbf {\bibinfo {volume} {82}},\ \bibinfo {pages} {1835} (\bibinfo {year} {1999})}\BibitemShut {NoStop}%
\bibitem [{\citenamefont {S{\o}rensen}\ and\ \citenamefont {M{\o}lmer}(1999)}]{Soerensen1999}%
  \BibitemOpen
  \bibfield  {author} {\bibinfo {author} {\bibfnamefont {A.}~\bibnamefont {S{\o}rensen}}\ and\ \bibinfo {author} {\bibfnamefont {K.}~\bibnamefont {M{\o}lmer}},\ }\href {https://doi.org/10.1103/physrevlett.82.1971} {\bibfield  {journal} {\bibinfo  {journal} {Physical Review Letters}\ }\textbf {\bibinfo {volume} {82}},\ \bibinfo {pages} {1971} (\bibinfo {year} {1999})}\BibitemShut {NoStop}%
\bibitem [{\citenamefont {Khromova}\ \emph {et~al.}(2012)\citenamefont {Khromova}, \citenamefont {Piltz}, \citenamefont {Scharfenberger}, \citenamefont {Gloger}, \citenamefont {Johanning}, \citenamefont {Varón},\ and\ \citenamefont {Wunderlich}}]{Khromova2012}%
  \BibitemOpen
  \bibfield  {author} {\bibinfo {author} {\bibfnamefont {A.}~\bibnamefont {Khromova}}, \bibinfo {author} {\bibfnamefont {C.}~\bibnamefont {Piltz}}, \bibinfo {author} {\bibfnamefont {B.}~\bibnamefont {Scharfenberger}}, \bibinfo {author} {\bibfnamefont {T.~F.}\ \bibnamefont {Gloger}}, \bibinfo {author} {\bibfnamefont {M.}~\bibnamefont {Johanning}}, \bibinfo {author} {\bibfnamefont {A.~F.}\ \bibnamefont {Varón}},\ and\ \bibinfo {author} {\bibfnamefont {C.}~\bibnamefont {Wunderlich}},\ }\href {https://doi.org/10.1103/physrevlett.108.220502} {\bibfield  {journal} {\bibinfo  {journal} {Physical Review Letters}\ }\textbf {\bibinfo {volume} {108}},\ \bibinfo {pages} {220502} (\bibinfo {year} {2012})}\BibitemShut {NoStop}%
\bibitem [{\citenamefont {Balzer}\ \emph {et~al.}(2006)\citenamefont {Balzer}, \citenamefont {Braun}, \citenamefont {Hannemann}, \citenamefont {Paape}, \citenamefont {Ettler}, \citenamefont {Neuhauser},\ and\ \citenamefont {Wunderlich}}]{Balzer2006}%
  \BibitemOpen
  \bibfield  {author} {\bibinfo {author} {\bibfnamefont {C.}~\bibnamefont {Balzer}}, \bibinfo {author} {\bibfnamefont {A.}~\bibnamefont {Braun}}, \bibinfo {author} {\bibfnamefont {T.}~\bibnamefont {Hannemann}}, \bibinfo {author} {\bibfnamefont {C.}~\bibnamefont {Paape}}, \bibinfo {author} {\bibfnamefont {M.}~\bibnamefont {Ettler}}, \bibinfo {author} {\bibfnamefont {W.}~\bibnamefont {Neuhauser}},\ and\ \bibinfo {author} {\bibfnamefont {C.}~\bibnamefont {Wunderlich}},\ }\href {https://doi.org/10.1103/physreva.73.041407} {\bibfield  {journal} {\bibinfo  {journal} {Physical Review A}\ }\textbf {\bibinfo {volume} {73}},\ \bibinfo {pages} {041407} (\bibinfo {year} {2006})}\BibitemShut {NoStop}%
\bibitem [{\citenamefont {Schiffer}(1993)}]{Schiffer1993}%
  \BibitemOpen
  \bibfield  {author} {\bibinfo {author} {\bibfnamefont {J.~P.}\ \bibnamefont {Schiffer}},\ }\href {https://doi.org/10.1103/physrevlett.70.818} {\bibfield  {journal} {\bibinfo  {journal} {Physical Review Letters}\ }\textbf {\bibinfo {volume} {70}},\ \bibinfo {pages} {818} (\bibinfo {year} {1993})}\BibitemShut {NoStop}%
\bibitem [{\citenamefont {McKay}\ \emph {et~al.}(2017)\citenamefont {McKay}, \citenamefont {Wood}, \citenamefont {Sheldon}, \citenamefont {Chow},\ and\ \citenamefont {Gambetta}}]{McKay2017}%
  \BibitemOpen
  \bibfield  {author} {\bibinfo {author} {\bibfnamefont {D.~C.}\ \bibnamefont {McKay}}, \bibinfo {author} {\bibfnamefont {C.~J.}\ \bibnamefont {Wood}}, \bibinfo {author} {\bibfnamefont {S.}~\bibnamefont {Sheldon}}, \bibinfo {author} {\bibfnamefont {J.~M.}\ \bibnamefont {Chow}},\ and\ \bibinfo {author} {\bibfnamefont {J.~M.}\ \bibnamefont {Gambetta}},\ }\href {https://doi.org/10.1103/physreva.96.022330} {\bibfield  {journal} {\bibinfo  {journal} {Physical Review A}\ }\textbf {\bibinfo {volume} {96}},\ \bibinfo {pages} {022330} (\bibinfo {year} {2017})}\BibitemShut {NoStop}%
\bibitem [{\citenamefont {Sack}\ and\ \citenamefont {Serbyn}(2021)}]{Sack2021}%
  \BibitemOpen
  \bibfield  {author} {\bibinfo {author} {\bibfnamefont {S.~H.}\ \bibnamefont {Sack}}\ and\ \bibinfo {author} {\bibfnamefont {M.}~\bibnamefont {Serbyn}},\ }\href {https://doi.org/10.22331/q-2021-07-01-491} {\bibfield  {journal} {\bibinfo  {journal} {Quantum}\ }\textbf {\bibinfo {volume} {5}},\ \bibinfo {pages} {491} (\bibinfo {year} {2021})}\BibitemShut {NoStop}%
\bibitem [{\citenamefont {Merkel}\ \emph {et~al.}(2019)\citenamefont {Merkel}, \citenamefont {Thirumalai}, \citenamefont {Tarlton}, \citenamefont {Schäfer}, \citenamefont {Ballance}, \citenamefont {Harty},\ and\ \citenamefont {Lucas}}]{Merkel2019}%
  \BibitemOpen
  \bibfield  {author} {\bibinfo {author} {\bibfnamefont {B.}~\bibnamefont {Merkel}}, \bibinfo {author} {\bibfnamefont {K.}~\bibnamefont {Thirumalai}}, \bibinfo {author} {\bibfnamefont {J.~E.}\ \bibnamefont {Tarlton}}, \bibinfo {author} {\bibfnamefont {V.~M.}\ \bibnamefont {Schäfer}}, \bibinfo {author} {\bibfnamefont {C.~J.}\ \bibnamefont {Ballance}}, \bibinfo {author} {\bibfnamefont {T.~P.}\ \bibnamefont {Harty}},\ and\ \bibinfo {author} {\bibfnamefont {D.~M.}\ \bibnamefont {Lucas}},\ }\bibfield  {journal} {\bibinfo  {journal} {Review of Scientific Instruments}\ }\textbf {\bibinfo {volume} {90}},\ \href {https://doi.org/10.1063/1.5080093} {10.1063/1.5080093} (\bibinfo {year} {2019})\BibitemShut {NoStop}%
\bibitem [{\citenamefont {Ballance}\ \emph {et~al.}(2016)\citenamefont {Ballance}, \citenamefont {Harty}, \citenamefont {Linke}, \citenamefont {Sepiol},\ and\ \citenamefont {Lucas}}]{Ballance2016}%
  \BibitemOpen
  \bibfield  {author} {\bibinfo {author} {\bibfnamefont {C.~J.}\ \bibnamefont {Ballance}}, \bibinfo {author} {\bibfnamefont {T.~P.}\ \bibnamefont {Harty}}, \bibinfo {author} {\bibfnamefont {N.~M.}\ \bibnamefont {Linke}}, \bibinfo {author} {\bibfnamefont {M.~A.}\ \bibnamefont {Sepiol}},\ and\ \bibinfo {author} {\bibfnamefont {D.~M.}\ \bibnamefont {Lucas}},\ }\href {https://doi.org/10.1103/physrevlett.117.060504} {\bibfield  {journal} {\bibinfo  {journal} {Physical Review Letters}\ }\textbf {\bibinfo {volume} {117}},\ \bibinfo {pages} {060504} (\bibinfo {year} {2016})}\BibitemShut {NoStop}%
\bibitem [{\citenamefont {Blanes}\ \emph {et~al.}(2009)\citenamefont {Blanes}, \citenamefont {Casas}, \citenamefont {Oteo},\ and\ \citenamefont {Ros}}]{Blanes2009}%
  \BibitemOpen
  \bibfield  {author} {\bibinfo {author} {\bibfnamefont {S.}~\bibnamefont {Blanes}}, \bibinfo {author} {\bibfnamefont {F.}~\bibnamefont {Casas}}, \bibinfo {author} {\bibfnamefont {J.}~\bibnamefont {Oteo}},\ and\ \bibinfo {author} {\bibfnamefont {J.}~\bibnamefont {Ros}},\ }\href {https://doi.org/10.1016/j.physrep.2008.11.001} {\bibfield  {journal} {\bibinfo  {journal} {Physics Reports}\ }\textbf {\bibinfo {volume} {470}},\ \bibinfo {pages} {151} (\bibinfo {year} {2009})}\BibitemShut {NoStop}%
\bibitem [{\citenamefont {Rohde}\ \emph {et~al.}(2001)\citenamefont {Rohde}, \citenamefont {Gulde}, \citenamefont {Roos}, \citenamefont {Barton}, \citenamefont {Leibfried}, \citenamefont {Eschner}, \citenamefont {Schmidt-Kaler},\ and\ \citenamefont {Blatt}}]{Rohde2001}%
  \BibitemOpen
  \bibfield  {author} {\bibinfo {author} {\bibfnamefont {H.}~\bibnamefont {Rohde}}, \bibinfo {author} {\bibfnamefont {S.~T.}\ \bibnamefont {Gulde}}, \bibinfo {author} {\bibfnamefont {C.~F.}\ \bibnamefont {Roos}}, \bibinfo {author} {\bibfnamefont {P.~A.}\ \bibnamefont {Barton}}, \bibinfo {author} {\bibfnamefont {D.}~\bibnamefont {Leibfried}}, \bibinfo {author} {\bibfnamefont {J.}~\bibnamefont {Eschner}}, \bibinfo {author} {\bibfnamefont {F.}~\bibnamefont {Schmidt-Kaler}},\ and\ \bibinfo {author} {\bibfnamefont {R.}~\bibnamefont {Blatt}},\ }\href {https://doi.org/10.1088/1464-4266/3/1/357} {\bibfield  {journal} {\bibinfo  {journal} {Journal of Optics B: Quantum and Semiclassical Optics}\ }\textbf {\bibinfo {volume} {3}},\ \bibinfo {pages} {S34} (\bibinfo {year} {2001})}\BibitemShut {NoStop}%
\bibitem [{\citenamefont {Sriarunothai}\ \emph {et~al.}(2017)\citenamefont {Sriarunothai}, \citenamefont {Giri}, \citenamefont {W\"olk},\ and\ \citenamefont {Wunderlich}}]{Sriarunothai2017}%
  \BibitemOpen
  \bibfield  {author} {\bibinfo {author} {\bibfnamefont {T.}~\bibnamefont {Sriarunothai}}, \bibinfo {author} {\bibfnamefont {G.~S.}\ \bibnamefont {Giri}}, \bibinfo {author} {\bibfnamefont {S.}~\bibnamefont {W\"olk}},\ and\ \bibinfo {author} {\bibfnamefont {C.}~\bibnamefont {Wunderlich}},\ }\href {https://doi.org/10.1080/09500340.2017.1401137} {\bibfield  {journal} {\bibinfo  {journal} {Journal of Modern Optics}\ }\textbf {\bibinfo {volume} {65}},\ \bibinfo {pages} {560} (\bibinfo {year} {2017})}\BibitemShut {NoStop}%
\bibitem [{\citenamefont {Sosnova}\ \emph {et~al.}(2021)\citenamefont {Sosnova}, \citenamefont {Carter},\ and\ \citenamefont {Monroe}}]{Sosnova2021}%
  \BibitemOpen
  \bibfield  {author} {\bibinfo {author} {\bibfnamefont {K.}~\bibnamefont {Sosnova}}, \bibinfo {author} {\bibfnamefont {A.}~\bibnamefont {Carter}},\ and\ \bibinfo {author} {\bibfnamefont {C.}~\bibnamefont {Monroe}},\ }\href {https://doi.org/10.1103/physreva.103.012610} {\bibfield  {journal} {\bibinfo  {journal} {Physical Review A}\ }\textbf {\bibinfo {volume} {103}},\ \bibinfo {pages} {012610} (\bibinfo {year} {2021})}\BibitemShut {NoStop}%
\bibitem [{\citenamefont {Percival}(1992)}]{percival1992}%
  \BibitemOpen
  \bibfield  {author} {\bibinfo {author} {\bibfnamefont {D.~B.}\ \bibnamefont {Percival}},\ }\href@noop {} {\bibfield  {journal} {\bibinfo  {journal} {Computing Science and Statistics, 24, pp. 534–538}\ } (\bibinfo {year} {1992})}\BibitemShut {NoStop}%
\bibitem [{\citenamefont {Siegele-Brown}\ \emph {et~al.}(2022)\citenamefont {Siegele-Brown}, \citenamefont {Hong}, \citenamefont {Lebrun-Gallagher}, \citenamefont {Hile}, \citenamefont {Weidt},\ and\ \citenamefont {Hensinger}}]{SiegeleBrown2022}%
  \BibitemOpen
  \bibfield  {author} {\bibinfo {author} {\bibfnamefont {M.}~\bibnamefont {Siegele-Brown}}, \bibinfo {author} {\bibfnamefont {S.}~\bibnamefont {Hong}}, \bibinfo {author} {\bibfnamefont {F.~R.}\ \bibnamefont {Lebrun-Gallagher}}, \bibinfo {author} {\bibfnamefont {S.~J.}\ \bibnamefont {Hile}}, \bibinfo {author} {\bibfnamefont {S.}~\bibnamefont {Weidt}},\ and\ \bibinfo {author} {\bibfnamefont {W.~K.}\ \bibnamefont {Hensinger}},\ }\href {https://doi.org/10.1088/2058-9565/ac66fc} {\bibfield  {journal} {\bibinfo  {journal} {Quantum Science and Technology}\ }\textbf {\bibinfo {volume} {7}},\ \bibinfo {pages} {034003} (\bibinfo {year} {2022})}\BibitemShut {NoStop}%
\bibitem [{\citenamefont {Santra}\ \emph {et~al.}(2025)\citenamefont {Santra}, \citenamefont {Roy}, \citenamefont {Egger},\ and\ \citenamefont {Hauke}}]{Santra2025}%
  \BibitemOpen
  \bibfield  {author} {\bibinfo {author} {\bibfnamefont {G.~C.}\ \bibnamefont {Santra}}, \bibinfo {author} {\bibfnamefont {S.~S.}\ \bibnamefont {Roy}}, \bibinfo {author} {\bibfnamefont {D.~J.}\ \bibnamefont {Egger}},\ and\ \bibinfo {author} {\bibfnamefont {P.}~\bibnamefont {Hauke}},\ }\href {https://doi.org/10.1103/physreva.111.022434} {\bibfield  {journal} {\bibinfo  {journal} {Physical Review A}\ }\textbf {\bibinfo {volume} {111}},\ \bibinfo {pages} {022434} (\bibinfo {year} {2025})}\BibitemShut {NoStop}%
\bibitem [{\citenamefont {Capecci}\ \emph {et~al.}(2025)\citenamefont {Capecci}, \citenamefont {Santra}, \citenamefont {Bottarelli}, \citenamefont {Tirrito},\ and\ \citenamefont {Hauke}}]{Capecci2025}%
  \BibitemOpen
  \bibfield  {author} {\bibinfo {author} {\bibfnamefont {C.}~\bibnamefont {Capecci}}, \bibinfo {author} {\bibfnamefont {G.~C.}\ \bibnamefont {Santra}}, \bibinfo {author} {\bibfnamefont {A.}~\bibnamefont {Bottarelli}}, \bibinfo {author} {\bibfnamefont {E.}~\bibnamefont {Tirrito}},\ and\ \bibinfo {author} {\bibfnamefont {P.}~\bibnamefont {Hauke}},\ }\href {https://doi.org/10.48550/ARXIV.2505.17185} {\bibinfo {title} {Role of nonstabilizerness in quantum optimization}} (\bibinfo {year} {2025})\BibitemShut {NoStop}%
\bibitem [{mot()}]{motchallengeChallengeResults}%
  \BibitemOpen
  \href@noop {} {\bibinfo {title} {{M}{O}{T} {C}hallenge - {R}esults --- motchallenge.net}},\ \bibinfo {howpublished} {\url{https://motchallenge.net/results/MOT20Det/}},\ \bibinfo {note} {[Accessed 15-08-2025]}\BibitemShut {NoStop}%
\bibitem [{\citenamefont {Kantorovich}(1960)}]{kantorovich1960mathematical}%
  \BibitemOpen
  \bibfield  {author} {\bibinfo {author} {\bibfnamefont {L.~V.}\ \bibnamefont {Kantorovich}},\ }\href {https://doi.org/10.1287/mnsc.6.4.366} {\bibfield  {journal} {\bibinfo  {journal} {Management science}\ }\textbf {\bibinfo {volume} {6}},\ \bibinfo {pages} {366} (\bibinfo {year} {1960})}\BibitemShut {NoStop}%
\bibitem [{\citenamefont {Bontekoe}\ \emph {et~al.}(2023)\citenamefont {Bontekoe}, \citenamefont {Phillipson},\ and\ \citenamefont {Schoot}}]{bontekoe2023translating}%
  \BibitemOpen
  \bibfield  {author} {\bibinfo {author} {\bibfnamefont {T.}~\bibnamefont {Bontekoe}}, \bibinfo {author} {\bibfnamefont {F.}~\bibnamefont {Phillipson}},\ and\ \bibinfo {author} {\bibfnamefont {W.~v.~d.}\ \bibnamefont {Schoot}},\ }in\ \href {https://doi.org/10.1007/978-3-031-36030-5_8} {\emph {\bibinfo {booktitle} {International Conference on Computational Science}}}\ (\bibinfo {organization} {Springer},\ \bibinfo {year} {2023})\ pp.\ \bibinfo {pages} {90--107}\BibitemShut {NoStop}%
\bibitem [{\citenamefont {Ohno}\ \emph {et~al.}(2024)\citenamefont {Ohno}, \citenamefont {Shirai},\ and\ \citenamefont {Togawa}}]{ohno2024toward}%
  \BibitemOpen
  \bibfield  {author} {\bibinfo {author} {\bibfnamefont {K.}~\bibnamefont {Ohno}}, \bibinfo {author} {\bibfnamefont {T.}~\bibnamefont {Shirai}},\ and\ \bibinfo {author} {\bibfnamefont {N.}~\bibnamefont {Togawa}},\ }\href {https://doi.org/10.1109/access.2024.3425711} {\bibfield  {journal} {\bibinfo  {journal} {IEEE Access}\ }\textbf {\bibinfo {volume} {12}},\ \bibinfo {pages} {97678–97690} (\bibinfo {year} {2024})}\BibitemShut {NoStop}%
\bibitem [{\citenamefont {G{\"u}ney}\ \emph {et~al.}(2025)\citenamefont {G{\"u}ney}, \citenamefont {Ehrenthal},\ and\ \citenamefont {Hanne}}]{guney2025qubo}%
  \BibitemOpen
  \bibfield  {author} {\bibinfo {author} {\bibfnamefont {E.}~\bibnamefont {G{\"u}ney}}, \bibinfo {author} {\bibfnamefont {J.}~\bibnamefont {Ehrenthal}},\ and\ \bibinfo {author} {\bibfnamefont {T.}~\bibnamefont {Hanne}},\ }\href {https://doi.org/10.1109/ACCESS.2025.3550788} {\bibfield  {journal} {\bibinfo  {journal} {IEEE Access}\ }\textbf {\bibinfo {volume} {13}},\ \bibinfo {pages} {47086} (\bibinfo {year} {2025})}\BibitemShut {NoStop}%
\bibitem [{\citenamefont {Quintero}\ and\ \citenamefont {Zuluaga}(2021)}]{quintero2021characterizing}%
  \BibitemOpen
  \bibfield  {author} {\bibinfo {author} {\bibfnamefont {R.~A.}\ \bibnamefont {Quintero}}\ and\ \bibinfo {author} {\bibfnamefont {L.~F.}\ \bibnamefont {Zuluaga}},\ }\href@noop {} {\bibfield  {journal} {\bibinfo  {journal} {Technical Report. Department of Industrial and Systems Engineering, Lehigh University, Tech. Rep.}\ } (\bibinfo {year} {2021})}\BibitemShut {NoStop}%
\bibitem [{\citenamefont {Montanez-Barrera}\ \emph {et~al.}(2022)\citenamefont {Montanez-Barrera}, \citenamefont {Willsch}, \citenamefont {Maldonado-Romo},\ and\ \citenamefont {Michielsen}}]{montanez2024unbalanced}%
  \BibitemOpen
  \bibfield  {author} {\bibinfo {author} {\bibfnamefont {A.}~\bibnamefont {Montanez-Barrera}}, \bibinfo {author} {\bibfnamefont {D.}~\bibnamefont {Willsch}}, \bibinfo {author} {\bibfnamefont {A.}~\bibnamefont {Maldonado-Romo}},\ and\ \bibinfo {author} {\bibfnamefont {K.}~\bibnamefont {Michielsen}},\ }\bibfield  {journal} {\bibinfo  {journal} {Quantum Science and Technology, Volume 9, Number 2 (2024)}\ }\href {https://doi.org/10.1088/2058-9565/ad35e4} {10.1088/2058-9565/ad35e4} (\bibinfo {year} {2022}),\ \Eprint {https://arxiv.org/abs/2211.13914} {arXiv:2211.13914 [quant-ph]} \BibitemShut {NoStop}%
\bibitem [{\citenamefont {Bottarelli}\ \emph {et~al.}(2025)\citenamefont {Bottarelli}, \citenamefont {Schmitt},\ and\ \citenamefont {Hauke}}]{Bottarelli2025}%
  \BibitemOpen
  \bibfield  {author} {\bibinfo {author} {\bibfnamefont {A.}~\bibnamefont {Bottarelli}}, \bibinfo {author} {\bibfnamefont {S.}~\bibnamefont {Schmitt}},\ and\ \bibinfo {author} {\bibfnamefont {P.}~\bibnamefont {Hauke}},\ }\bibfield  {journal} {\bibinfo  {journal} {Physical Review Research}\ }\textbf {\bibinfo {volume} {7}},\ \href {https://doi.org/10.1103/3l96-41xf} {10.1103/3l96-41xf} (\bibinfo {year} {2025})\BibitemShut {NoStop}%
\bibitem [{\citenamefont {Glover}\ \emph {et~al.}(2019)\citenamefont {Glover}, \citenamefont {Kochenberger},\ and\ \citenamefont {Du}}]{glover2018tutorial}%
  \BibitemOpen
  \bibfield  {author} {\bibinfo {author} {\bibfnamefont {F.}~\bibnamefont {Glover}}, \bibinfo {author} {\bibfnamefont {G.}~\bibnamefont {Kochenberger}},\ and\ \bibinfo {author} {\bibfnamefont {Y.}~\bibnamefont {Du}},\ }\href {https://arxiv.org/abs/1811.11538} {\bibinfo {title} {A tutorial on formulating and using qubo models}} (\bibinfo {year} {2019}),\ \Eprint {https://arxiv.org/abs/1811.11538} {arXiv:1811.11538} \BibitemShut {NoStop}%
\bibitem [{\citenamefont {Huber}(2024)}]{Patrick2024}%
  \BibitemOpen
  \bibfield  {author} {\bibinfo {author} {\bibfnamefont {P.}~\bibnamefont {Huber}},\ }\href {https://d-nb.info/1365040550/34} {\bibinfo {title} {Multi-qubit gates in a trapped-ion quantum computer}} (\bibinfo {year} {2024})\BibitemShut {NoStop}%
\end{thebibliography}%
